\pdfoutput = 1
\documentclass[aps,prx,reprint,superscriptaddress,floatfix,amsmath,amssymb]{revtex4-2}
\usepackage[tworuled,vlined]{algorithm2e}
\usepackage[english]{babel}
\usepackage{booktabs}
\usepackage{comment}
\usepackage{color}
\usepackage{enumitem}
\usepackage{graphicx}
\usepackage{hyperref}

\usepackage{physics}
\usepackage{bm}
\usepackage[normalem]{ulem}

\hypersetup{colorlinks=true,urlcolor=blue,citecolor=blue,linkcolor=blue,breaklinks=true}

\definecolor{purple}{rgb}{0.6, 0.2, 0.8}
\definecolor{orange}{rgb}{0.91, 0.41, 0.17}

\newcommand\ee{\mathrm{e}}%
\newcommand\ii{\mathrm{i}}%
\newcommand\iv{\ii\nu}%
\newcommand\iw{\ii\omega}%

\newcommand\DFT{\operatorname{FT}}

\newcommand\PH{{\mathrm{ph}}}

\begin{document}

\hyphenation{par-ti-cle--hole Hart-ree--Fock}

\makeatletter
\def\bbl@set@language#1{%
  \edef\languagename{%
    \ifnum\escapechar=\expandafter`\string#1\@empty
    \else\string#1\@empty\fi}%
  \@ifundefined{babel@language@alias@\languagename}{}{%
    \edef\languagename{\@nameuse{babel@language@alias@\languagename}}%
  }%
  \select@language{\languagename}%
  \expandafter\ifx\csname date\languagename\endcsname\relax\else
    \if@filesw
      \protected@write\@auxout{}{\string\select@language{\languagename}}%
      \bbl@for\bbl@tempa\BabelContentsFiles{%
        \addtocontents{\bbl@tempa}{\xstring\select@language{\languagename}}}%
      \bbl@usehooks{write}{}%
    \fi
  \fi}
\newcommand{\DeclareLanguageAlias}[2]{%
  \global\@namedef{babel@language@alias@#1}{#2}%
}
\makeatother
\DeclareLanguageAlias{en}{english}

\allowdisplaybreaks
\title{
Multiscale space-time ansatz for correlation functions of quantum systems based on quantics tensor trains
}

\author{Hiroshi Shinaoka}
\affiliation{Department of Physics, Saitama University, Saitama 338-8570, Japan}
\affiliation{JST, PRESTO, 4-1-8 Honcho, Kawaguchi, Saitama 332-0012, Japan}

\author{Markus Wallerberger}
\affiliation{Institute of Solid State Physics, TU Wien, 1040 Vienna, Austria}

\author{Yuta Murakami}
\affiliation{Center for Emergent Matter Science, RIKEN, Wako, Saitama 351-0198, Japan}

\author{Kosuke Nogaki}
\affiliation{Department of Physics, Kyoto University, Kyoto 606-8502, Japan}

\author{Rihito Sakurai}
\affiliation{Department of Physics, Saitama University, Saitama 338-8570, Japan}

\author{Philipp Werner}
\affiliation{Department of Physics, University of Fribourg, 1700 Fribourg, Switzerland}

\author{Anna Kauch}
\affiliation{Institute of Solid State Physics, TU Wien, 1040 Vienna, Austria}

\begin{abstract}
Correlation functions of quantum systems---central objects in quantum field theories---are defined in high-dimensional space-time domains. Their numerical treatment thus suffers from the curse of dimensionality, which hinders the application of sophisticated many-body theories to interesting problems.
{Here, we propose a multi-scale space-time ansatz for correlation functions of quantum systems based on quantics tensor trains (QTT),
{``qubits''}
} describing exponentially different length scales.  The ansatz then assumes a separation of length scales by 
{decomposing the resulting high-dimensional tensors into tensor trains (known also as matrix product states).}
We numerically verify the ansatz for various equilibrium and nonequilibrium systems and demonstrate compression rates of several orders of magnitude for challenging cases.
Essential building blocks of diagrammatic equations, such as convolutions or Fourier transforms are formulated in the compressed form.
We numerically demonstrate the stability and efficiency of the proposed methods for the Dyson and Bethe-Salpeter equations.
{The QTT representation} provides a unified framework for implementing efficient computations of quantum field theories.
\end{abstract}
\maketitle

\section{Introduction}
Correlation functions are central building blocks of quantum field theories for many-body and first-principles calculations~\cite{Mahan}.
A typical example is the Matsubara or nonequilibrium Green's functions.
These correlation functions are high-dimensional space-time objects,
which creates a severe challenge for numerical calculations. A long-standing and fundamental problem of great practical importance is thus the search for compact representations of correlation functions. 

Notable theoretical developments have been made  in the Matsubara-frequency domain for the one-particle (1P) Green's function, for which  compact representations, such as Legendre~\cite{Boehnke2011-rt,Dong2020-tr} and Chebyshev~\cite{Gull2018-ci} bases, were constructed. The 1P Green's function is   related to a spectral function through the ill-conditioned analytic continuation kernel.
This prior knowledge was recently employed to construct the intermediate representation (IR) ~\cite{Shinaoka2017-ah, Otsuki2017-qf} and the sparse sampling method~\cite{Li2020-kb,Shinaoka2022-xv}, the minimax method~\cite{Kaltak2020-hu}, and the discrete Lehmann representation (DLR)~\cite{Kaye2022-ad}.
These methods are all based on the same prior knowledge and allow to treat a wide range of energy scales, from the bandwidth to low-temperature phenomena, 
for one-dimensional (1D) objects in the Matsubara frequency domain.
Their application enabled first-principles calculations of correlation functions for unconventional~\cite{Witt2021} and phonon-mediated superconductors~\cite{Wang2020-uz} with low $T_\mathrm{c}$, as well as recent studies of transition metal oxides~\cite{Wang2020-uz, Nomoto2020-zb, Nomoto2020-qa, Nomura2020-an,Yeh2021-fu, Iskakov2020-nt}.

Extending these developments to other space-time domains, particularly to higher order correlation functions,  has been a central challenge in many different fields of computational physics.
For the numerical renormalization group (NRG) and diagrammatic calculations at the two-particle (2P) level~\cite{Rohringer2018-gt,Iskakov2018-bi, Vucicevic2017-jy,Kitatani2022-xh}, compact representations for the Matsubara-frequency dependence of 2P quantities have been proposed.
{Recently, the analytic structure of arbitrary correlation functions
has been clarified~\cite{Kugler2021-gq}; this structure can be
leveraged in NRG calculations~\cite{Lee2021-ho}, compression of 2P quantities~\cite{Shinaoka2018-qz,Shinaoka2020-rp}, and associated diagrammatic equations~\cite{Wallerberger2021-kv}.
}

An efficient description of the three-momentum dependence of 2P quantities is another actively pursued direction, relevant for diagrammatic calculations at the 2P level and the functional renormalization group (fRG)~\cite{Metzner2012-dq}.
Examples of such efforts include the truncated unity approach based on a truncated form-factor basis ~\cite{Eckhardt2018, Eckhardt2020} and a machine-learning approach~\cite{Di_Sante2022-su}.
There is also an increasing demand for efficient treatment of 2P quantities in \textit{ab initio} calculations for such as the inclusion of vertex corrections in \textit{GW}~\cite{Maggio2017} and the Migdal--Eliashberg theory~\cite{Schrodi2020-le}.
For nonequilibrium systems, a hierarchical low-rank data structure has been proposed~\cite{Kaye2021-ly} for the real-time 1P Green's function with two time arguments.

Despite these extensive efforts, a generic and  efficient treatment of high-dimensional space-time objects has not yet been established. 
The difficulty can be attributed to the absence of a common and general ansatz for different space-time domains.
A promising ansatz requires (1) an accurate treatment of a wide range of length scales in space-time, (2) systematic control over the truncation error, (3) the possibility of efficient computations in the compressed form and (4) straightforward and robust implementations as computer code.

In this paper, we propose the multi-scale space-time ansatz based on {quantics tensor trains (QTTs)~\cite{Oseledets09,Khoromskij11} as a universal solution.}
The space-time dependence is described by auxiliary {bits, which we call ``qubits'' in the present study,} representing exponentially different length scales in space-time. The resultant high-dimensional object in the qubit space is decomposed into tensor trains {(TTs)}, to the physics community better known as matrix product states (MPS), based on the assumption of length scale separation.
{The QTT representation} allows us to describe the space-time dependence of correlation functions in exponentially wide scales 
using memory and computational resources which scale linearly, and thus essentially removes a major bottleneck for numerical many-body calculations.
Basic operations such as the Fourier transform can be formulated in compressed form and
the methods can be implemented straightforwardly using standard MPS libraries.
We numerically verify the ansatz for various equilibrium and nonequilibrium systems: from 1P and 2P Matsubara and real-time Green's functions.
Compression rates of several orders of magnitude are demonstrated for challenging cases.
We also numerically show the stability and efficiency of the proposed methods for the Dyson and Bethe-Salpeter equations (BSEs).

Recently, related quantum-inspired algorithms using the qubit mapping have been proposed for image compression~\cite{Latorre2005-eo}, and for solving Navier-Stokes equations for turbulent flows~\cite{Gourianov2022-vn} or the Vlasov-Poisson equations for collisionless plasmas~\cite{Ye2022-ep}.
A low-rank tensor train approximation has been applied to the numerical integration of high-order perturbation series of quantum systems without the multi-scale ansatz~\cite{Fernandez2022}.
{The quantics represenation was used to represent spectral functions in combination with a Boltzmann machine~\cite{Yamaji2021-zh}.}
In this paper, we clarify the fundamental question how such a multi-scale ansatz performs in the context of quantum field theories.
{The QTT representation has the potential} not only to change the way in which numerical many-body calculations will be performed, but also to bridge the fields of quantum information theory and quantum field theory.

The paper is organized as follows:
In Sec.~\ref{sec:mssta}, we introduce the QTT representation.  We detail common operations performed with this ansatz
in Sec.~\ref{sec:ops}.
In Sec.~\ref{sec:compression}, we show the performance of the QTT representation in encoding the imaginary-time or Matsubara-frequency, momentum, and real-time dependence of correlation functions in a variety of equilibrium and non-equilibrium systems. 
Sec.~\ref{sec:computation} is devoted to the demonstration of the computation of correlation functions.
We summarize the main results of the paper in Sec.~\ref{sec:conclusions}.
Appendices are devoted to technical discussions on (\ref{sec:mps}) matrix product states,
(\ref{sec:mpo}) matrix product operators, (\ref{app:fft}) Fourier transforms,
and (\ref{app:mesh}) frequency meshes.

{\em Note on nomenclature.} QTT representation is based on concepts already known in literature as quantics tensor trains and we adopt this name also here. However, in the physics community tensor trains are better known as matrix product states (MPS) and matrix product operators (MPO) and in the technical parts  of the paper we use these names.

\section{Multi-scale space--time ansatz}
\label{sec:mssta}

\begin{figure}
    \centering
    \includegraphics{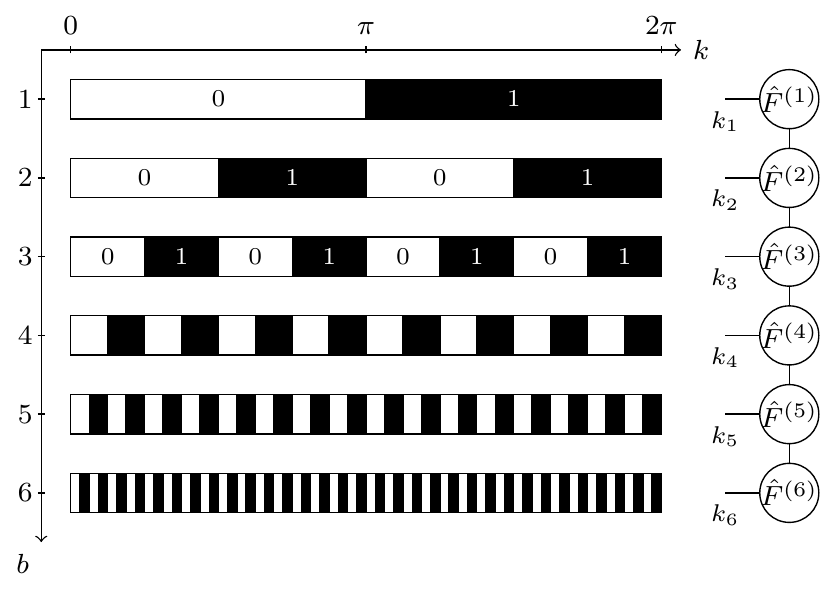}
    \caption{Multi-scale ansatz for momentum space. Each row, numbered by the bond index $b$, corresponds to a different level of discretization of the 1D momentum $k$ (different length scale). In this way, $k$ can be represented by a set of bits $k_1, \cdots, k_R$ (see text). On the right, the {QTT} representation of a momentum dependent function, Eq.~\eqref{eq:fhat-mps}, is shown.}
    \label{fig:scales}
\end{figure}

\begin{figure}
    \centering
    \includegraphics[width=0.99\columnwidth]{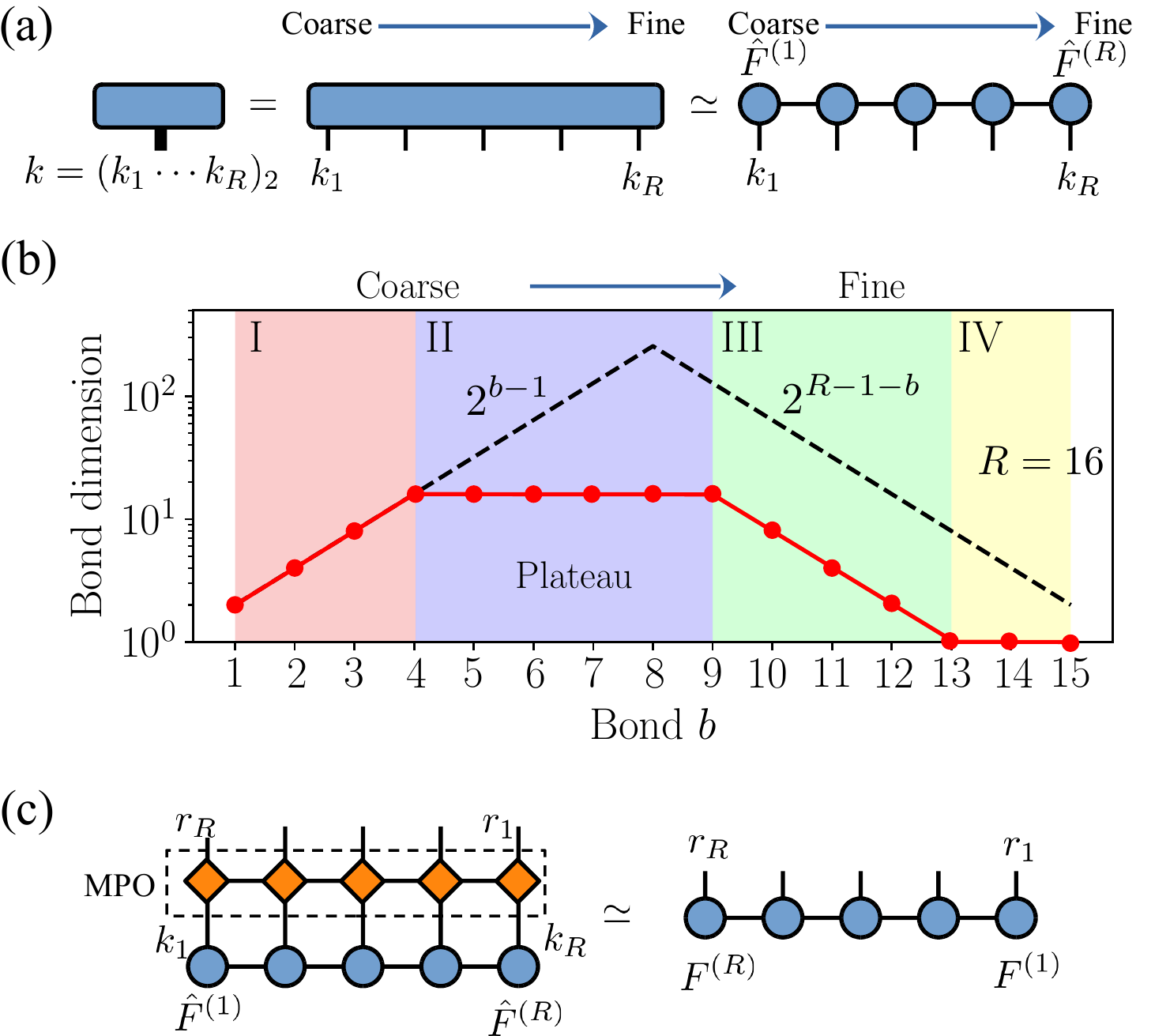}
    \caption{
        (a) QTT representation in momentum space.
        The rightmost {bits} (indices) represent fine structures in momentum space.
        Low entanglement structures are assumed between different length scales.
        (b) Schematic illustration of the bond dimensions along the chain representing the momentum dependence. The dashed line indicates the maximum bond dimensions in maximally entangled cases.
        (c) Fourier transform from momentum space to real space by applying a matrix product operator (MPO).
        The orange diamonds represent the MPO tensors.
        The structure of the MPO is illustrated in Fig.~\ref{fig:fft-tn} in Appendix~\ref{app:fft}.
    }
    \label{fig:mssta}
\end{figure}

In this section, we explain the multi-scale space-time ansatz based on {QTT}.
The essence of the ansatz is to introduce multiple indices to describe different space-time length scales, and to assume low entanglement structures between different scales (see Fig.~\ref{fig:scales}).
We focus on the momentum space and the associated real space as the first examples.

\subsection{Momentum space}
Let us consider first a function $f(k)$ in momentum space, where $k\in [0, 2\pi)$ is the (for
now one-dimensional) momentum.  Usually, we discretize $f(k)$ on an equidistant grid
of size, e.g., $2^R$.
This technique is straight-forward
to implement, but comes with a series of drawbacks: many-body propagators have sharp and intricate 
structures in momentum space, which means that the precision of the approximation 
only improves slowly with $2^R$.

Instead of considering a ``flat'' discretization into a vector of $2^R$ momenta, 
in the multi-scale ansatz, we first separate out $R$ distinct scales $k_1, \ldots, k_R$:
\begin{align}
    f(k) \approx f\left(k_1\pi + k_2\frac{\pi}{2} + \cdots+ k_R\frac{2\pi}{2^R}\right) = f(k_1, \cdots, k_R),
    \label{eq:fhat-scales}
\end{align}
where each $k_b$, $1\le b\le R$, now only takes two values: zero or one. Put
differently, $k_1, \ldots, k_R$ are the bits of $2^Rk/(2\pi)$, i.e.,
$k = 2\pi (k_1\cdots k_R)_2/2^R$. In this notation $k=0$ corresponds to $(00\cdots0)_2$, $k = 2\pi/2^R$ corresponds to $(00\cdots 1)_2$, and so forth.

We can
interpret $f(k_1, \cdots, k_R)$ in a couple of ways. In terms of physics, we have separated out different scales of the problem, as illustrated in Fig.~\ref{fig:scales}: $k_1$ partitions the Brillouin zone into two coarse regions, $[0,\pi)$ and $[\pi,2\pi)$, and as we move towards
$k_R$, features on finer and finer scales are captured.  In terms of quantum information theory, $f(k_1, \cdots, k_R)$ can be regarded as an (unnormalized) wavefunction in the Hilbert space of dimension $2^R$ spanned by $S=1/2$ spins or \textit{qubits}. In terms of linear algebra,
we have simply reinterpreted the $2^R$-vector of momenta as $2 \times \cdots \times 2$ ($R$-way) tensor.   

Since up to this point, we have merely reshaped our data from a vector to a tensor, no information of the original discretization is lost.
The main idea of the multi-scale ansatz 
is to express the single $R$-way tensor $f$ by a tensor train,
a contraction of $R$ three-way tensors
$\hat F^{(1)}, \ldots, \hat F^{(R)}$:
\begin{align}
    f(k_1, \cdots, k_R) &\approx \sum_{\alpha_1=1}^{D_1} \cdots \sum_{\alpha_{R-1}=1}^{D_{R-1}}
    \hat F^{(1)}_{k_1,1\alpha_1} \cdots \hat F^{(R)}_{k_R,\alpha_{R-1}1}
    \nonumber\\
    & \equiv
    \hat F^{(1)}_{k_1} \cdot \hat F^{(2)}_{k_2} 
    \cdot \ldots \cdot \hat F^{(R)}_{k_R},
    \label{eq:fhat-mps}
\end{align}
where $\hat F^{(b)}$ is an auxiliary $2\times D_{b-1}\times D_b$ tensor,  
$\alpha_1, \ldots, \alpha_{R-1}$ forming bonds between neighbouring tensors, 
and $D_b$ is the bond dimension of the $b$-th bond. 
$D = \max_b D_b$ is the bond dimension of the whole MPS. (We refer the reader
to Appendix~\ref{sec:mps} for more details.) We illustrate Eq.~(\ref{eq:fhat-mps}) in Fig.~\ref{fig:mssta}(a).

The ansatz (\ref{eq:fhat-mps}) is still exact if the bond dimension is very large, $D \sim 2^R$;
it becomes approximate if the bonds are truncated to the most important contribution.
The core insight is that for many functions, including, as we shall show, the
propagators in momentum space, the bond dimension needed to approximate
the original tensor grows only modestly with the desired accuracy $\epsilon$ {measured by the Frobenius norm [see Eq.~\eqref{eq:epsilon}]},
thus allowing us to compress the function significantly.

More specifically, Fig.~\ref{fig:mssta}(b) illustrates how the bond dimension typically varies along the chain when the MPS is truncated with a certain cutoff $\epsilon$.
First, the bond dimension increases exponentially in the region I, where coarse global structures are not compressible.
This is followed by region II (plateau region), where different length scales are not strongly entangled (separation in length scale).
In region III, the bond dimension decreases but there is still a finite entanglement that is important for a quantitative description of the $k$ dependence within the given $\epsilon$.
In region IV, the bond dimension is 1 and the tensor train can be truncated without sacrificing any accuracy. {The efficiency of the QTT representation} relies on the existence of the plateau.

\subsection{Real space}
We construct a similar representation for the real space that is associated with the momentum space by Fourier transform.
In the case of a regular lattice, the lattice points are labeled by natural numbers, $r/a=0, 1, \cdots, 2^R-1$, where $a$ is the lattice constant.
As we did for $k$, we map natural integers to binary numbers as $r/a =(r_1 \cdots r_R)_2$.
Note that $r_i$ and $k_{R+1-i}$ correspond to the same length scale.
We represent the Fourier transformed function $f(r)$ in the $r$ space
as an MPS:
\begin{align}
    f(r) &= \sum_k e^{-\ii k r} f(k)\nonumber \\
    &\equiv \frac{1}{2^R} \sum_{k_1,\cdots,k_R}\!\! e^{-2\pi\ii \left (k_1/2 +\; \cdots\; +  k_R/2^R \right) r} f(k_1,\cdots,k_R) \nonumber \\
    &\approx \sum_{\alpha_1=1}^{r_1} \cdots \sum_{\alpha_{R-1}=1}^{r_{R-1}} 
    F^{(R)}_{r_R,1\alpha_1} \cdots F^{(1)}_{r_1,\alpha_{R-1}1} \nonumber \\
    &= F^{(R)}_{r_R} \cdot F^{(R-1)}_{r_{R-1}} \cdot (\ldots) \cdot F^{(1)}_{r_1}.
    \label{eq:ft} 
\end{align}

\subsection{Other spaces}
The aforementioned representation can be applied to other variables.
However, special care is needed for imaginary-time and Matsubara-frequency spaces.
An imaginary time $\tau$ is represented as
$2^R \tau/\beta = (\tau_1 \cdots \tau_R)_2$.
A Matsubara frequency is represented as $\nu = (2 (n - 2^{R-1}) + \xi)\pi/\beta$ using $n=0,1,2,\cdots,2^R-1$ ($\xi=0, 1$ for bosons and fermions, respectively).
Note that the artificial periodic boundary condition maps negative Matsubara frequencies as $-n = 2^R - n$ to the higher half of the positive part.
This allows to treat momentum space and frequency space consistently in the implementation.
The boundary effects of this periodic boundary condition vanish exponentially with increasing $R$ and do not matter in practice.
A real time $t$ ($0 \le t < t_\mathrm{max}$) is represented by a natural number, $2^R t/t_\mathrm{max}$.
{In a similar manner,  a real frequency $\omega$ ($-W \le \omega < W$) is represented by a natural number, $2^R (\omega+W)/(2W)$.}

\subsection{Higher dimensions}\label{sec:higher-dim}
\begin{figure}
    \centering
    \includegraphics[width=0.8\columnwidth]{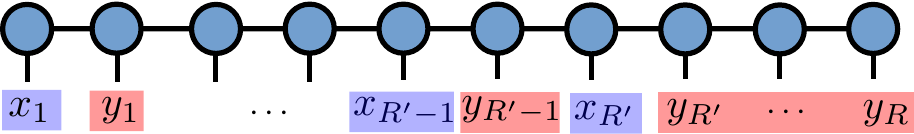}
    \caption{
    Matrix product states representing a 2D space spanned by $x$ and $y$. The expansion can be truncated at the right edge.
    }
    \label{fig:higherdim}
\end{figure}

It is easy to construct 
QTT representations for higher dimensional objects spanned by multiple space-time axes.
As an example, let us consider a 2D space spanned by the two variables $x=(x_1 \cdots x_{R'})_2$ and $y=(y_1 \cdots y_{R})_2$ ($R'<R$).
We assume that we are going to truncate the expansion at $x_{R'}$ and $y_R$.
In other words, the right qubits correspond to fine resolution for both $x$ and $y$.
In the present study,  we use the MPS structure shown in Fig.~\ref{fig:higherdim}.
An important point is that two qubits/tensors corresponding to the same length scale are next to each other because they are expected to be strongly entangled.

\section{Operations in the QTT representation}
\label{sec:ops}

\subsection{Fourier transform}\label{sec:fft}
The discrete Fourier transform (DFT) in Eq.~\eqref{eq:ft} can be represented as a matrix product operator (MPO), with a small bond dimension. (We refer the reader to Appendix~\ref{sec:mpo} for more details.)
This can be intuitively understood by the fact that two space-time indices $r_{R+1-i}$ and $k_i$ at the same position ($i=1, \cdots, R$) in Fig.~\ref{fig:mssta}(c) correspond to the same length scale.
The small bond dimension of the MPO was shown numerically in 2017~\cite{Woolfe2017-yy}.

\begin{figure}
    \includegraphics[width=0.9\columnwidth]{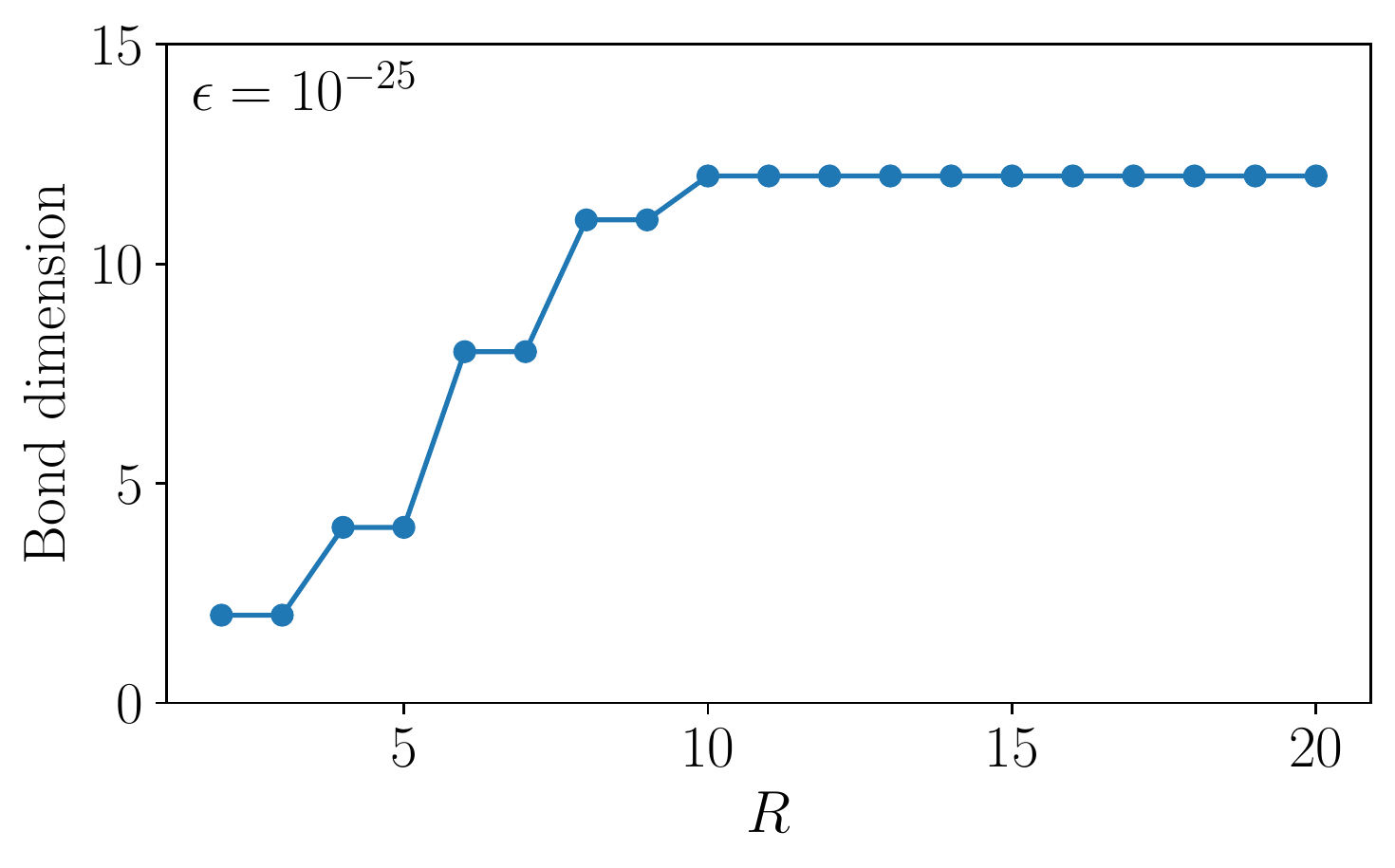}
    \caption{
        Bond dimensions of the MPO for the discrete Fourier transform recursively constructed with truncation cutoff $\epsilon=10^{-25}$.
    }
    \label{fig:fft-bonddims}
\end{figure}

As detailed in Appendix~\ref{appendix:fft}, one can construct MPOs recursively for $R=1, 2, 3, \cdots$.
Figure~\ref{fig:fft-bonddims} shows the bond dimensions of the numerically constructed MPOs with $\epsilon=10^{-25}$.
One can clearly see that the bond dimension weakly depends on $R$ and becomes saturated for $R>10$.
This crossover point shifts to larger $R$ as the cutoff $\epsilon$ is reduced.
This result indicates that the Fourier transform can be performed efficiently, with a computational time $\mathcal{O}(R)$ for fixed target accuracy.

\subsection{Element-wise product}\label{sec:elementwiseprod}
\begin{figure}
    \centering
    \includegraphics[width=0.99\columnwidth]{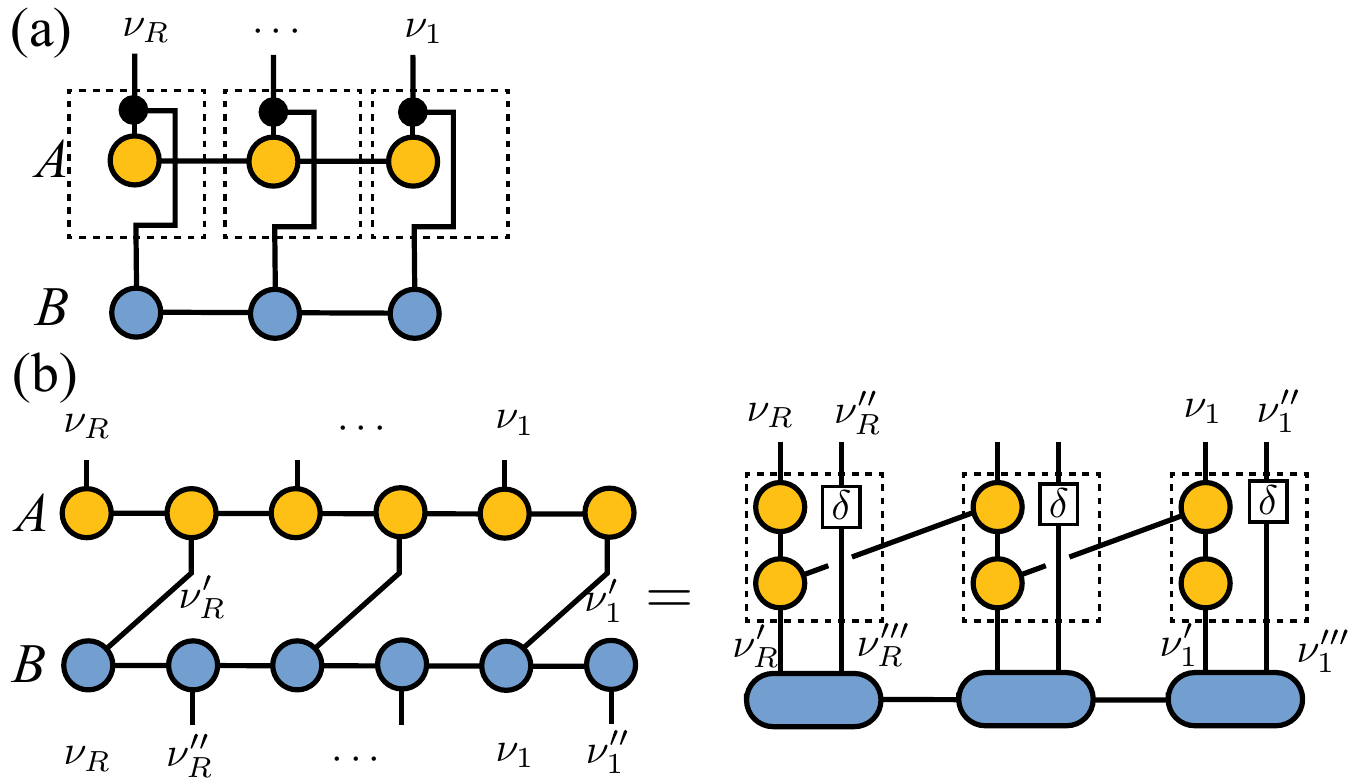}
    \caption{
    Tensor contraction for the element-wise product [(a)] and matrix product [(b)] of two MPSs $A$ and $B$.
    The filled circles in (a) denote a superdiagonal tensor whose nonzero entries are one. The dashed squares denote the tensors of the auxiliary MPOs.
    }
    \label{fig:mul}
\end{figure}

Solving the Dyson equation requires the computation of the element-wise product of two MPSs, $A$ and $B$:
\begin{align}
C(\iv) &= A(\iv) B(\iv).
\end{align}
To be precise, for given MPSs $A$ and $B$, one needs to compute an MPS for the product $C$.
In the compressed form, the product can be expressed as
\begin{align}
&C(\nu_R, \cdots, \nu_1)\nonumber\\
&= A(\nu_R, \cdots, \nu_1) B(\nu_R, \cdots, \nu_1)\nonumber\\
&= \sum_{\nu'_1,\cdots, \nu'_R}
A^{\nu_R, \cdots, \nu_1}_{\nu'_R, \cdots, \nu'_1}
B(\nu_R', \cdots, \nu_1'),
\end{align}
where
\begin{align}
A^{\nu_R, \cdots, \nu_1}_{\nu'_R, \cdots, \nu'_1}
&\equiv 
A(\nu_R, \cdots, \nu_1)\delta_{\nu^{\vphantom\prime}_R,\nu_R'}\cdots \delta_{\nu^{\vphantom\prime}_1,\nu_1'}.
\end{align}
An MPO for the auxiliary linear operator $A$ can be constructed from the MPS tensors of $A$ as
\begin{align}
    &
    (A^{\nu_R}_{1,a_1} \delta_{\nu_R,\nu_R'})
    \cdots
    (A^{\nu_1}_{a_R,1} \delta_{\nu_1,\nu_1'}).
\end{align}
The MPO is illustrated in Fig.~\ref{fig:mul}(a).
This allows to use an efficient implementation of
an MPO--MPS multiplication.

\subsection{Matrix multiplication for two-frequency objects}\label{sec:matmul}
To solve the BSE or the Dyson equation for the non-equilibrium Green's function, one needs to multiply two-frequency quantities:
\begin{align}
C(\iv, \iv'') &= \sum_{\nu'} A(\iv, \iv') B(\iv', \iv''),
\end{align}
where we perform the summation on the mesh of size $2^R$.
This can be expressed as an MPO-MPS product:
\begin{align}
&C(\nu_R, \nu''_R, \cdots, \nu_1, \nu''_1) \nonumber\\
&= \sum_{(\nu'_1\nu'''_1) \cdots, (\nu'_R \nu'''_R)} A^{(\nu_R \nu''_R), \cdots, (\nu_1 \nu''_1)}_{(\nu'_R \nu'''_R), \cdots, (\nu'_1 \nu'''_1)}
B((\nu'_R \nu'''_R), \cdots, (\nu'_1 \nu'''_1)).
\label{eq:mulmps}
\end{align}
Here, we introduced a combined index of dimension $4~(=2^2)$, and an auxiliary MPO, $A$, which is illustrated in Fig.~\ref{fig:mul}(b).

\subsection{Linear transformation of arguments of multidimensional objects}\label{sec:argument}
Another typical operation required for solving a diagrammatic equation is the linear transformation of arguments of multidimensional objects.
As as an example, we consider a function with two time arguments, $f(t, t')$.
We want to transform this to a function $g(t_1, t_2) = f((t-t')/2, (t+t')/2)$ which depends on the relative and average times. 
This linear transformation can be represented by an MPO with a small bond dimension of $\mathcal{O}(1)$ because the linear transformation can be performed almost independently at different length scales.
Indeed, the MPOs can be constructed using adders or subtractors of binary numbers.

\section{Compression}
\label{sec:compression}

\subsection{Imaginary-time/Matsubara-frequency Green's function}\label{sec:1P}
As the minimum example, we consider the imaginary-time and Matsubara-frequency dependence of the fermionic Green's function generated by a few poles.
The Green's function reads
\begin{align}
   G(\iv) &= \int \dd\omega \frac{\rho(\omega)}{\iv - \omega} = \sum_{i=1}^{N_\mathrm{P}} \frac{c_i}{\iv - \omega_i}, \\
   G(\tau) &= - \sum_{i=1}^{N_\mathrm{P}} \frac{c_i e^{-\tau \omega_i}}{1 + e^{-\beta \omega_i}},
\end{align}
with
\begin{align}
   \rho(\omega) &= \sum_{i=1}^{N_\mathrm{P}} c_i \delta(\omega - \omega_i),
\end{align}
where $\omega_i$ and $c_i$ are the positions of the poles and the associated coefficients, respectively.
The $G(\iv)$ decay asymptotically as $\mathcal{O}(1/\iv)$ for large Matsubara frequencies (high frequency \textit{tail}).
Since we know the normalization factor $\sum_{i=1}^{N_\mathrm{P}} c_i$ a priori from the commutation relation of the operators, 
this contribution can be subtracted as
\begin{align}
   \tilde G(\iv) &\equiv G(\iv) - \frac{\sum_{i=1}^{N_\mathrm{P}} c_i}{\iv},
\end{align}
where $\tilde G(\iv)$ decays faster than $\mathcal{O}(1/(\iv)^2)$.
As we will see later, this subtraction slightly suppresses the bond dimension at high temperatures.

For $N_\mathrm{P}=1$, $G(\tau)$ can be represented as an MPS of bond dimension 1:
\begin{align}
   G(\tau) &= - \frac{c_1}{1 + e^{-\beta \omega_1}} \prod_{t=1}^R e^{- \tau_t 2^{-t} \beta \omega_1},\nonumber \\
   &= - \frac{c_1}{1 + e^{-\beta \omega_1}} G^{(1)} \cdot (\cdots) \cdot G^{(R)}
   \label{eq:pole-mps}
\end{align}
{
with the $t$-th TT tensor 
\begin{align}
    G_{\alpha_t, \alpha_{t+1}}^{(t)} &\equiv e^{- \tau_t 2^{-t} \beta \omega_1} \delta_{\alpha_t, \alpha_{t+1}},
\end{align}
where $\tau/\beta = (0.\tau_1 \tau_2 \cdots \tau_R)_2$ and $t=1,2,\cdots,R$, while $\alpha_{t}$ and $\alpha_{t+1}$ are indices of the virtual bonds.}
The coefficient in Eq.~\eqref{eq:pole-mps} can be absorbed into one of the tensors.

For $N_\mathrm{P} > 1$, the bond dimension of the natural MPS of $G(\tau)$ is bounded from above: $D \le N_\mathrm{P}$. This explicitly constructed MPS is highly compressible as we will demonstrate below.

We investigate the compactness of the representation for a model with $N_\mathrm{P}=100$ where the position and coefficients of the poles are chosen randomly according to the normal Gaussian distribution.
We use the truncation parameter $\epsilon = 10^{-20}$.

\begin{figure}
    \centering
    \includegraphics[width=0.99\columnwidth]{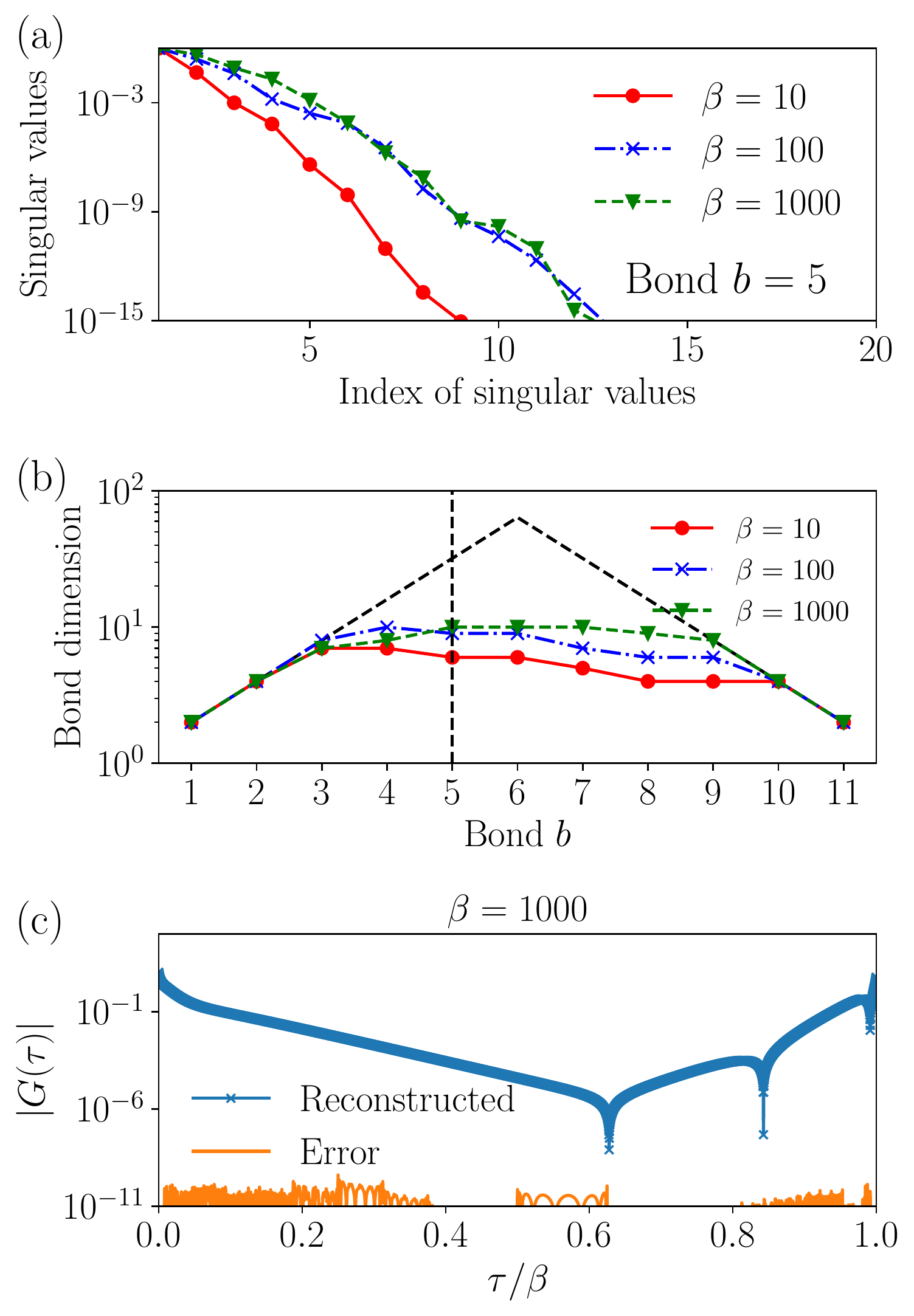}
    \caption{
    Compression of $G(\tau)$ generated by randomly chosen 100 poles with $R=12$ (see the text).
    (a) Singular values at $b=5$, (b) bond dimensions for $\epsilon=10^{-20}$, (c) comparison between the exact and reconstructed data.
    $G(\tau)$ has three sign changes.
    In (b), the dashed line indicates the maximum bond dimensions in maximally entangled cases.
    }
    \label{fig:randompoles-gtau}
\end{figure}

Figure~\ref{fig:randompoles-gtau} presents the results for $G(\tau)$.
As shown in Fig.~\ref{fig:randompoles-gtau}(a), the singular values decay exponentially.
As one can see in Fig.~\ref{fig:randompoles-gtau}(b), the number of relevant singular values, i.e., the bond dimensions, only mildly depend on $\beta$ and become converged at $\beta=1000$.
This convergence may reflect the fact that $G(\tau)$ has limited information, i.e., a nonzero lower bound for excitation energies,  due to the finite number of poles.
The bond dimensions slowly vanish after the first few bonds, indicating that one can increase the grid size exponentially with respect to the memory size, i.e., the number of tensors.
The error in the reconstructed data is almost constant in amplitude over $\beta$ [Fig.~\ref{fig:randompoles-gtau}(c)].

\begin{figure}
    \centering
    \includegraphics[width=0.99\columnwidth]{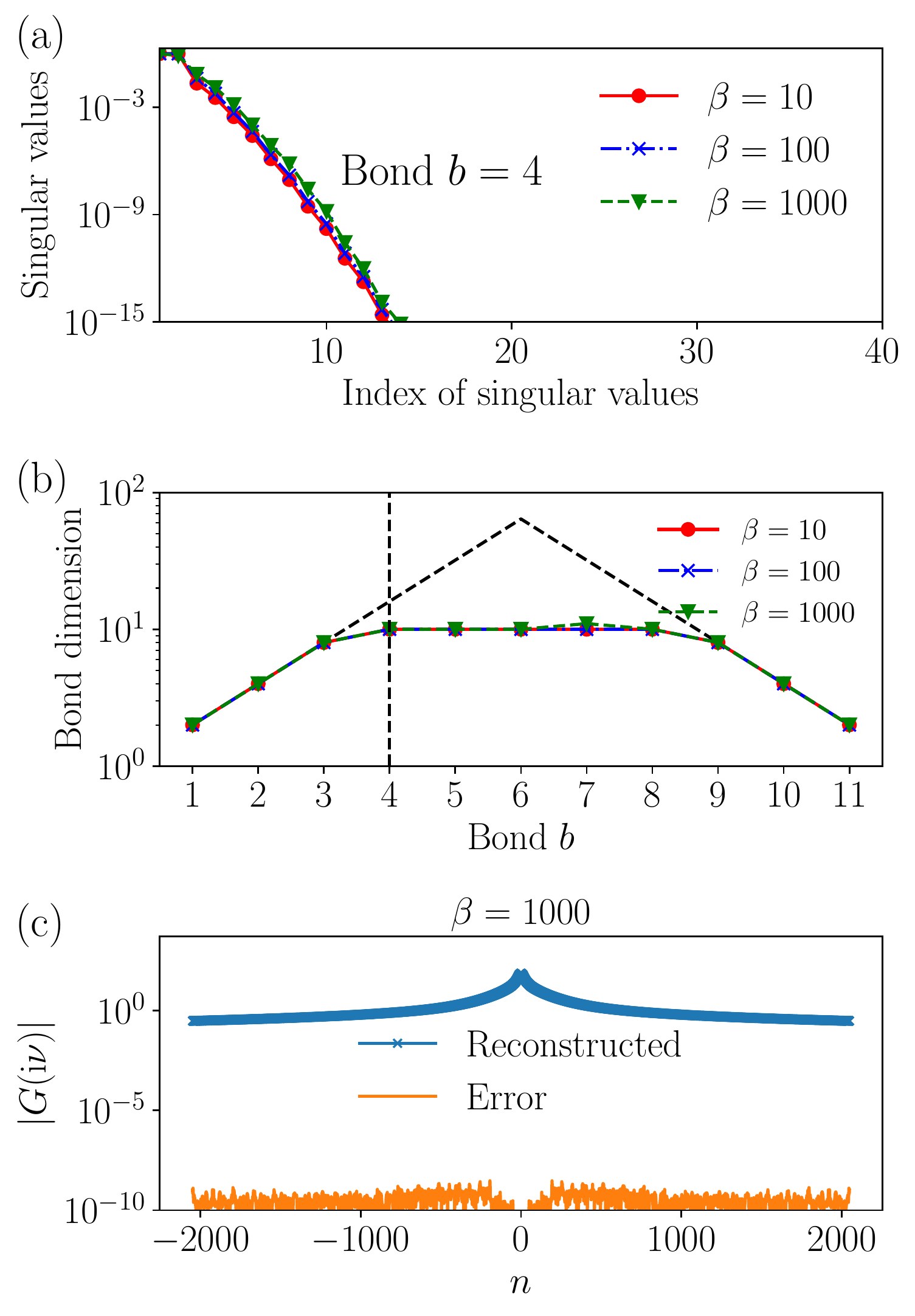}
    \caption{
    Compression of $G(\iv)$ for the same model as in Fig.~\ref{fig:randompoles-gtau} with $R=12$.
    (a) Singular values at $b=4$, (b) bond dimensions, (c) comparison between the exact and reconstructed data.
    In (b), the dashed line indicates the maximum bond dimensions in maximally entangled cases.
    }
    \label{fig:randompoles-giv}
\end{figure}

Figure~\ref{fig:randompoles-giv} shows the results of the decomposition of $G(\iv)$.
As seen in Figs.~\ref{fig:randompoles-giv}(b) and (c), in contrast to $G(\tau)$, the singular values and the bond dimensions are almost independent of $\beta$.
The bond dimensions are close to the maximum bond dimensions of $G(\tau)$, which is reasonable because the two objects contain the same amount of information. 

\begin{figure}
    \centering
    \includegraphics[width=0.99\columnwidth]{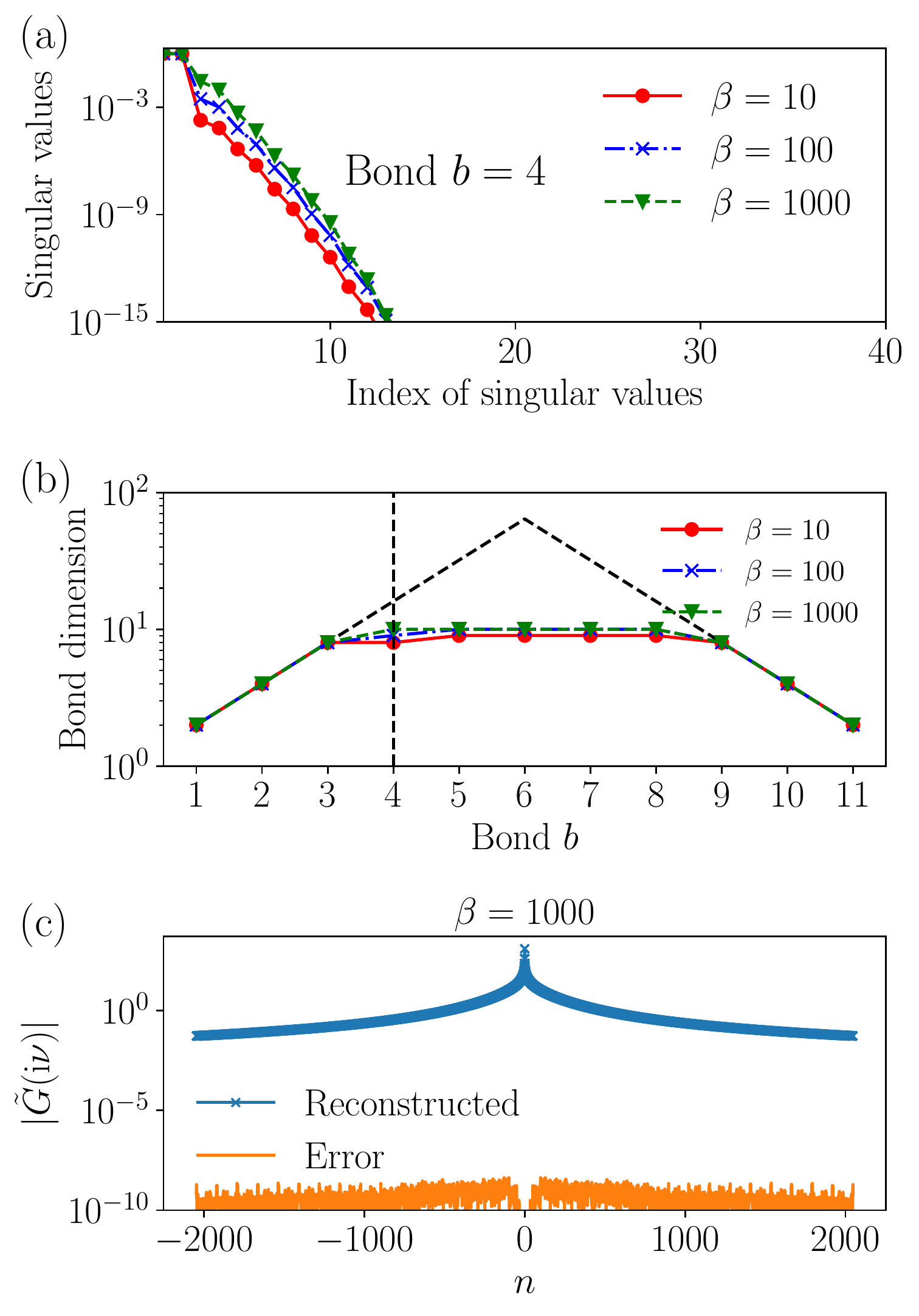}
    \caption{
    Compression of $\tilde G(\iv)$ for the same model as in Fig.~\ref{fig:randompoles-gtau} with $R=12$.
    Note that the tail is subtracted from the data before the compression.
    (a) Singular values at $b=4$, (b) bond dimensions, (c) comparison between the exact and reconstructed data.
    In (b), the dashed line indicates the maximum bond dimensions in maximally entangled cases.
    }
    \label{fig:randompoles-giv-tail}
\end{figure}

We now analyze the results for $\tilde G(\iv)$ to get an insight into the insensitivity of the bond dimensions.
Figure~\ref{fig:randompoles-giv-tail} shows the results for $\tilde G(\iv)$.
Comparing Figs.~\ref{fig:randompoles-giv}(a) and \ref{fig:randompoles-giv-tail}(a) reveals that subtracting the tail slightly reduces the number of relevant singular values for small $\beta$.
This indicates that fitting the trivial high-frequency asymptotic behavior requires some bond dimensions.
As seen in Fig.~\ref{fig:randompoles-giv-tail}(b), the subtraction of the tail enhances the $\beta$ dependence of the singular values and the bond dimensions as expected.
At low temperatures, the subtraction of the tail does not change the bond dimensions significantly.
In practical calculations, thus, such treatment of the tail may not be necessary.

It should be noted that the size of a compressed object scales as $\mathcal{O}(D^2)$. The new representation is less compact than IR~\cite{Shinaoka2017-ah} and DLR~\cite{Kaye2022-ad} but can be naturally generalized to higher dimensions, e.g., four-point functions, as we will see later in Sec.~\ref{sec:hubbardatom}.

\subsection{Momentum dependence: 1D case}\label{sec:1P-mom-1D}
\begin{figure}
    \centering
    \includegraphics[width=0.95\columnwidth]{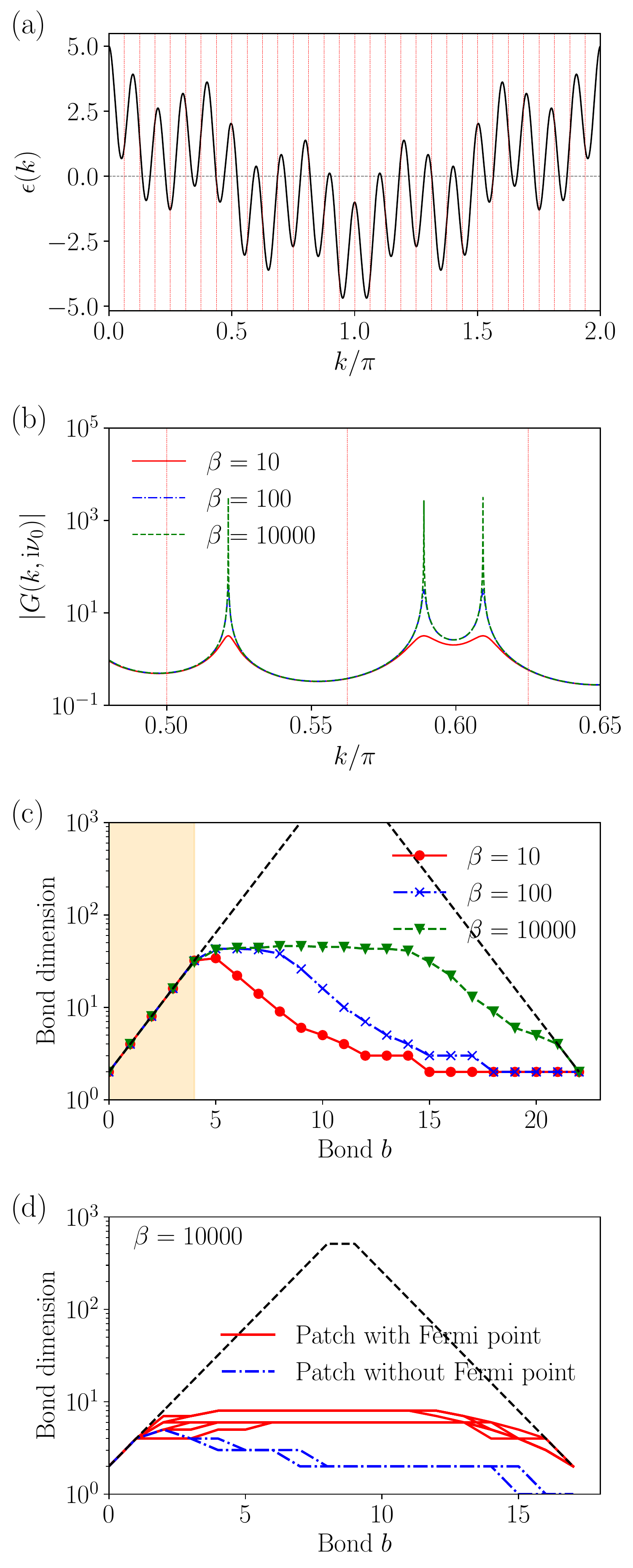}
    \caption{
        Momentum dependence of the Matsubara Green's function of the 1D tight-binding model \eqref{eq:1d-tb}.
        (a) Band dispersion, (b) Green's function at the lowest positive Matsubara frequency, (c) bond dimensions of an MPS for the full BZ, (d) bond dimensions of patch-wise constructed MPSs.
        We used $\epsilon = 10^{-10}$ and $2^5$ (=32) patches in (d).
        The vertical lines in (a) denote the boundaries of the patches.
        $\epsilon(k)$ has 34 roots.
        The shaded region in (c) corresponds to the patch size.
        Two out of the 32 patches do not contain a Fermi point.
        In (d), the dashed line indicates the maximum bond dimensions in maximally entangled cases.
    }
    \label{fig:1P-mom}
\end{figure}

To demonstrate the first example of a momentum dependent object, we consider a 1D tight-binding model whose band dispersion is given by
\begin{align}
    \epsilon(k) &= 2\cos(k) + \cos(5k)+ 2\cos(20k).\label{eq:1d-tb}
\end{align}
The chemical potential is at zero energy.
As illustrated in Fig.~\ref{fig:1P-mom}(a), this model has 34 Fermi points, where the Green's function at low frequency has large values [see Fig.~\ref{fig:1P-mom}(b)].
Here, we consider the Matsubara Green's function at the lowest positive Matsubara frequency,
\begin{align}
    G(\iv_0, k) &= \frac{1}{\iv_0 - \epsilon(k)}.
\end{align}

Figure~\ref{fig:1P-mom}(c) shows the bond dimension of an MPS constructed to describe the momentum dependence in the full BZ with cutoff $\epsilon=10^{-10}$.
For the first few bonds, the bond dimension increases exponentially, indicating that the global structures of $G(k)$ (with large length scales in momentum space) are not compressible with {QTT}.
After the first few bonds, the bond dimension gets saturated and eventually decreases.
The maximum of the bond dimension does not strongly depend on $\beta$.
As we lower the temperature, the region with almost constant bond dimensions is enhanced.
As we will see in the next subsection, this behavior is specific to 1D cases.

Large bond dimensions should be avoided in practical calculations, because of the increase in the computational time.
{The bond dimensions can be significantly reduced by \textit{patching} shown in Fig.~\ref{fig:1P-mom}(a).
The patching means partitioning the full momentum space into several patches of the same length and representing the momentum dependence within each patch by a single MPS.
In Fig.~\ref{fig:1P-mom}(a), we constructed the patches so that each patch  contains only a few Fermi points (or none).}
Figure~\ref{fig:1P-mom}(d) shows the bond dimension required to represent the momentum dependence within each patch with the same cutoff.
One can clearly see that the bond dimensions are below 10.
Patches including Fermi points require larger bond dimensions, as expected.
It should be noted that such a patching is consistent with the fast Fourier transform and can be done adaptively in solving the Dyson equation in the QTT representation.

\subsection{Momentum dependence: 2D case}\label{sec:1P-mom-2D}
We now move on to 2D systems.
As a simple case, we consider a nearest-neighbor tight-binding model on the square lattice. At half filling, the Green's function can be expressed as
\begin{align}
    G(\iv, \bm{k}) &= \frac{1}{\iv - \epsilon(\bm{k})},
\end{align}
where $\epsilon(\bm{k}) = -2 \cos(k_x) - 2 \cos(k_y)$.
We consider the lowest positive Matsubara frequency.

\begin{figure}
    \centering
    \includegraphics[width=0.99\columnwidth]{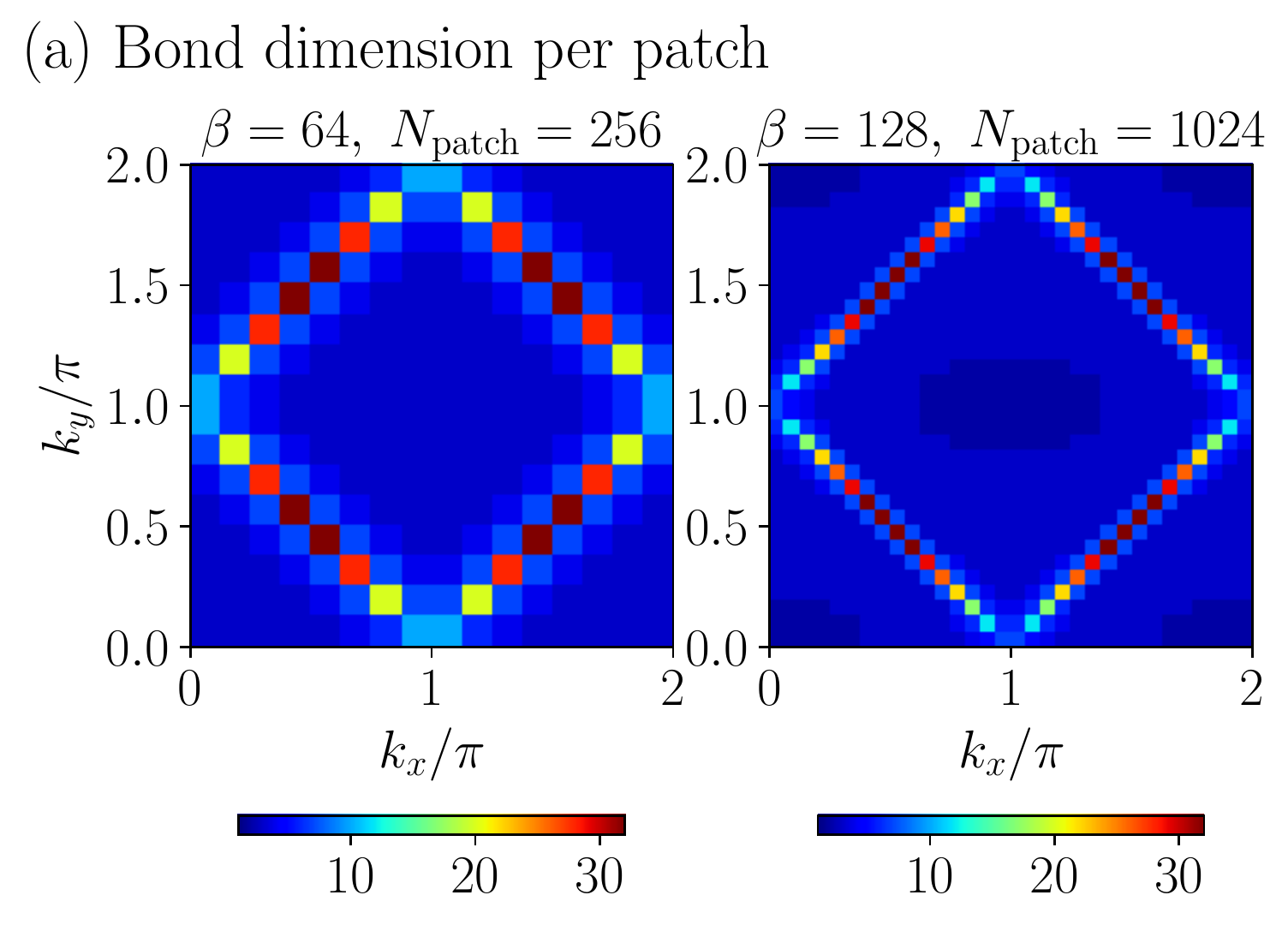}
    \includegraphics[width=0.99\columnwidth]{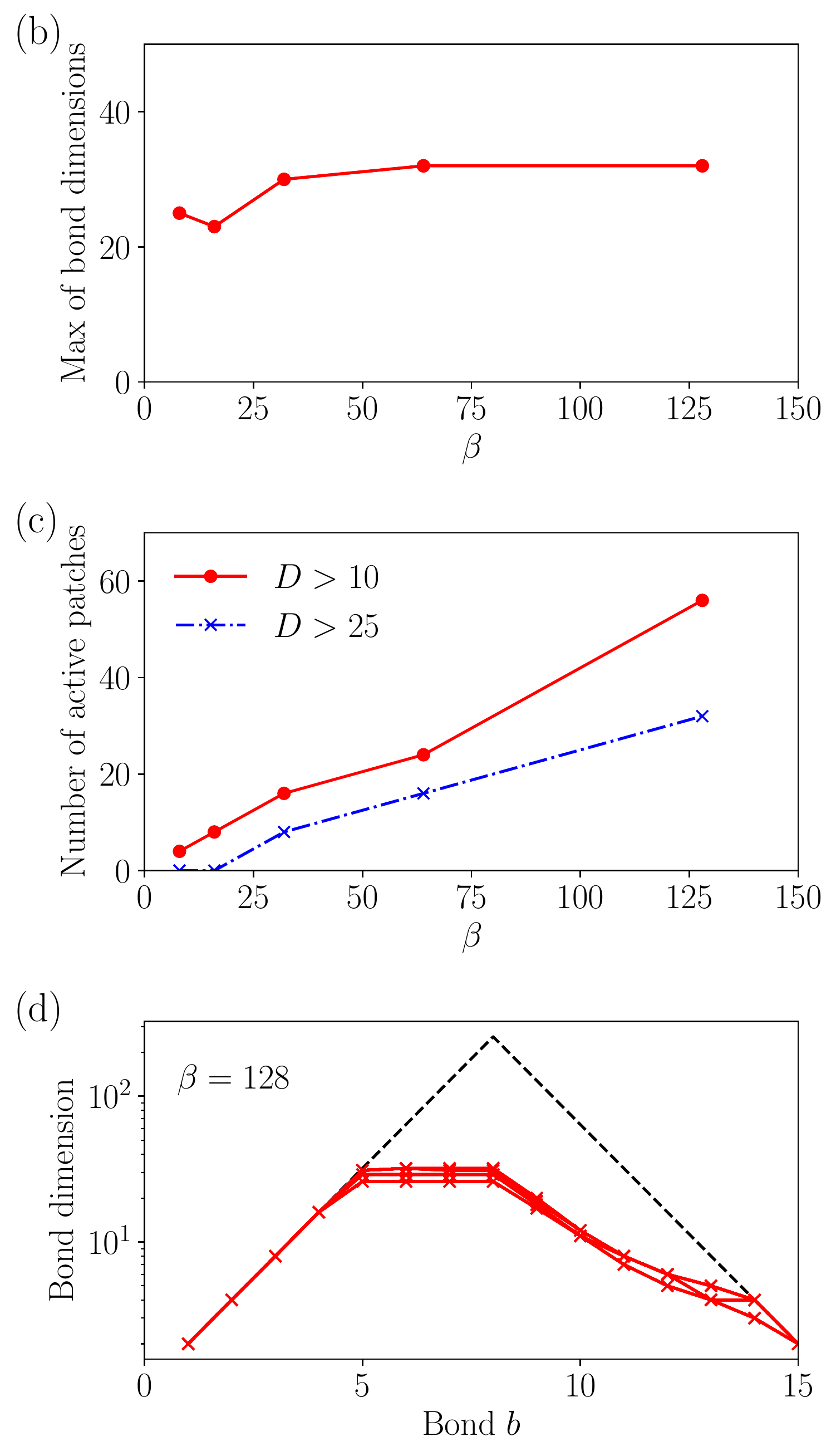}
    \caption{
        Momentum dependence of the Matsubara Green's function for the next-nearest tight-binding model on the square lattice. (a) Bond dimensions per patch, (b) maximum value of the bond dimensions, (c) number of patches with  large bond dimensions, (d) bond dimensions along the chain for patches with relatively large bond dimensions ($\ge 25$).
        In (d), the dashed line indicates the maximum bond dimensions in maximally entangled cases.
    }
    \label{fig:1P-mom-2D}
\end{figure}

At half filling, there is a large Fermi surface where the Green's function has large values.
The length scale of the structure in momentum space scales as $\mathcal{O}(T)$ at low temperatures.
Motivated by this fact, we divide the full BZ into $2^P \times 2^P$ patches. We take $2^P=\beta/4$ for $\beta=8, 16, 32, 64, 128$.
The momentum dependence within each patch is represented using a $256\times 256$ mesh.
We compress the momentum dependence by an MPS within each patch using the cutoff $\epsilon=10^{-10}$.
Figure~\ref{fig:1P-mom-2D}(a) shows the patches and bond dimension per patch computed at $\beta=64$ and 128.
One can see that relatively large bond dimensions are required only for a small number of patches near the Fermi surface.
As shown in Fig.~\ref{fig:1P-mom-2D}(b), the maximum of the bond dimensions stays constant at low temperatures.
Figure~\ref{fig:1P-mom-2D}(c) shows the number of \textit{active} patches with relatively large bond dimensions and indicates that the number of these active patches grows linearly with $\beta$.
We expect that the number of active patches grows as $\mathcal{O}(\beta^2)$ in 3D systems.

Let us look at patches with large bond dimensions for the lowest temperature $\beta=128$. As shown in Fig.~\ref{fig:1P-mom-2D}(d), after the first few bonds, the bond dimension exhibits a plateau, followed by an exponential decrease.
The existence of the plateau supports length scale separation.

These results indicate the importance of the combination of two different schemes for compressing the momentum dependence of non-interacting Green's functions. The overall structure of the momentum dependence is barely compressible with QTT because Fermi surfaces can appear anywhere. This leads to the need for patching.
However, this issue is specific to the non-interacting Green's function with sharp Fermi surfaces and it will be less significant in interacting models.
Since the momentum dependence within each patch has a simpler structure, it is compressible with QTT. This allows us to essentially eliminate discretization errors by using a large $R$.

\subsection{2D Hubbard model}
In this subsection, we discuss examples of interacting electrons.
We consider the momentum dependence in the case of the single-orbital Hubbard model on the square lattice at half-filling.

The Hamiltonian of the Hubbard model reads
\begin{align}
\mathcal{H} &= - \sum_{\langle ij\rangle, \sigma} \hat{c}^\dagger_{i\sigma} \hat{c}_{j\sigma} + U \sum_{i} \hat{n}_{i\uparrow} \hat{n}_{i\downarrow} - \mu \sum_{i,\sigma} \hat{n}_{i\sigma},
\label{eq:Hubbard}
\end{align}
where $\hat{c}^\dagger_{i\sigma}$ is the creation operator for an electron with spin $\sigma$ at site $i$, $\langle ij\rangle$ indicates a pair of neighboring sites, and 
$\hat{n}_{i\sigma} = \hat{c}^\dagger_{i\sigma} \hat{c}_{i\sigma}$.  $U$ is the onsite repulsion, $\mu = U/2$, and we have set the nearest-neighbour hopping amplitude to $1$.

We solve the model within the FLEX approximation~\cite{Bickers1989,Bickers1989_2,Bickers1991}.
FLEX is a conserving approximation in which several conservation laws are satisfied in the framework of the Luttinger-Ward theory~\cite{Luttinger1960,Luttinger1960_2,Baym1961,Baym1962}.
It is widely used to study unconventional superconductivity induced by spin fluctuations~\cite{Moriya2000,Yanase2003}.

In the FLEX approximation, for a paramagnetic state, the self-energy is approximated as
\begin{align}
   \Sigma(\bm{k},\iv) &= \frac{T}{N} \sum_{\bm{q},\iw} V(\bm{q},\iw) G(\bm{k}-\bm{q},\iv-\iw),
\end{align}
where $N$ is the size of the momentum grid, $\bm{q}$ is a bosonic momentum, and $\nu$ and   $\omega$ are Matsubara fermionic and bosonic frequencies, respectively.
The effective interaction $V$ is defined as
\begin{align}
   V(\bm{q},\iw) &= U^2\left(\frac{3}{2}\chi_\mathrm{s}(\bm{q},\iw)+\frac{1}{2}\chi_\mathrm{c}(\bm{q},\iw)-\chi_0(\bm{q},\iw)\right),
\end{align}
where we introduced the bare, spin, and charge susceptibility:
\begin{align}
    \chi_0(\bm{q},\iw) &= -\frac{T}{N} \sum_{\bm{k},\iv} G(\bm{k}+\bm{q},\iv+\iw) G(\bm{k},\iv), \\
   \chi_\mathrm{s}(\bm{q},\iw) &= \frac{\chi_0(\bm{q},\iw)}{1-\chi_0(\bm{q},\iw)U}, \\
   \chi_\mathrm{c}(\bm{q},\iw) &= \frac{\chi_0(\bm{q},\iw)}{1+\chi_0(\bm{q},\iw)U}.
\end{align}
The Green's function is given by
\begin{align}
    G(\iv, \bm{k}) &= \frac{1}{\iv - \epsilon(\bm{k}) + \mu - \Sigma(\iv, \bm{k})},
\end{align}
where $\epsilon(\bm{k}) = -2 \cos(k_x) - 2 \cos(k_y)$.
Using these equations, the Green's function, self-energy, and effective interaction are self-consistently determined.

We choose the temperature $T = 0.03$ and $U=1.1$, which is a critical region near the antiferromagnetic (AF) phase, where the spin susceptibility acquires a strong momentum dependence.
We use a $1024\times 1024$ grid of $k$ points in the full BZ.
We use the IR basis~\cite{Shinaoka2017-ah} and the sparse-sampling method~\cite{Li2020-kb} to perform efficient FLEX calculations~\cite{Witt2021}.

\begin{figure}
    \centering
    \includegraphics[width=1.0\columnwidth]{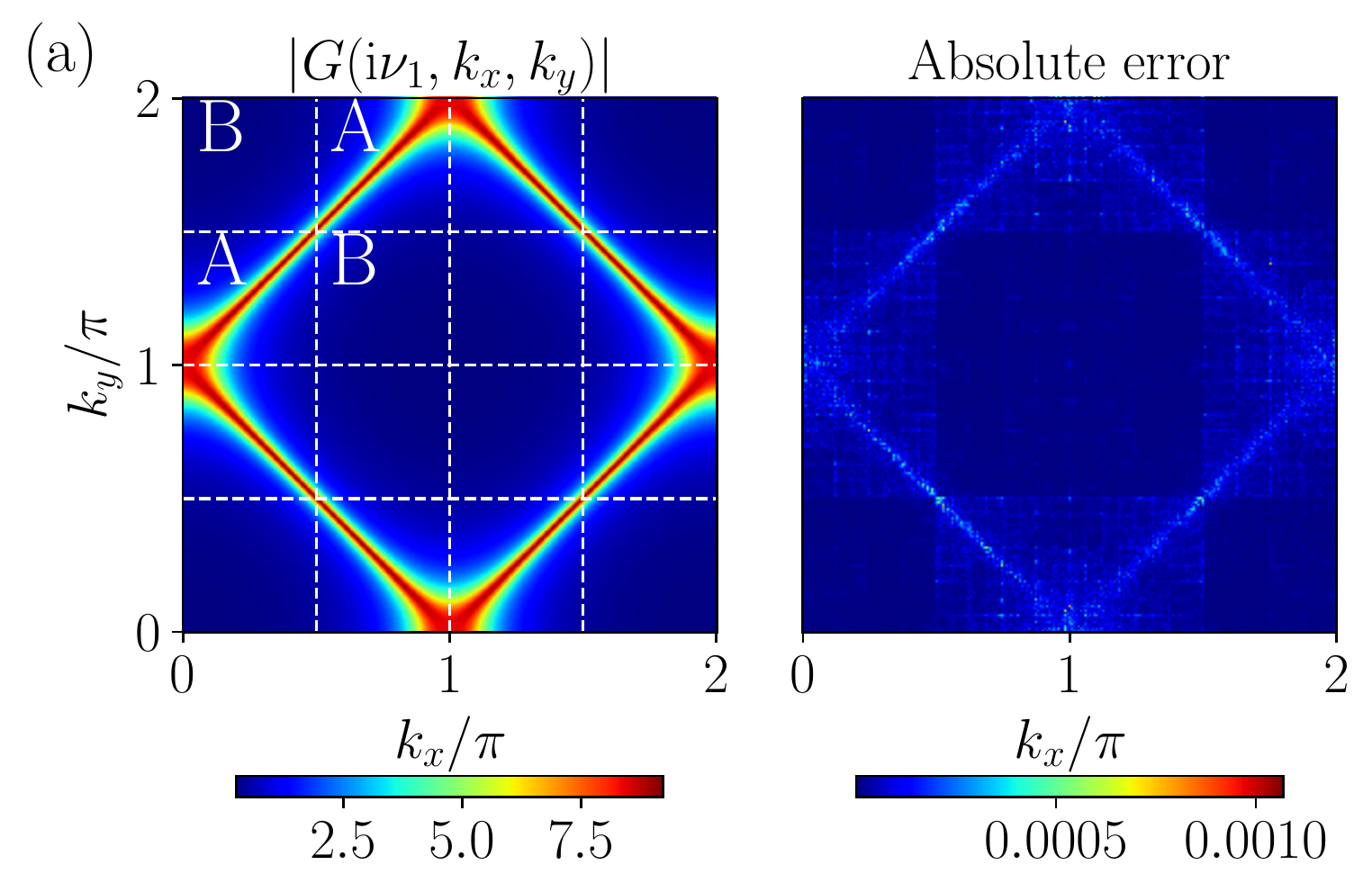}
    \includegraphics[width=1.0\columnwidth]{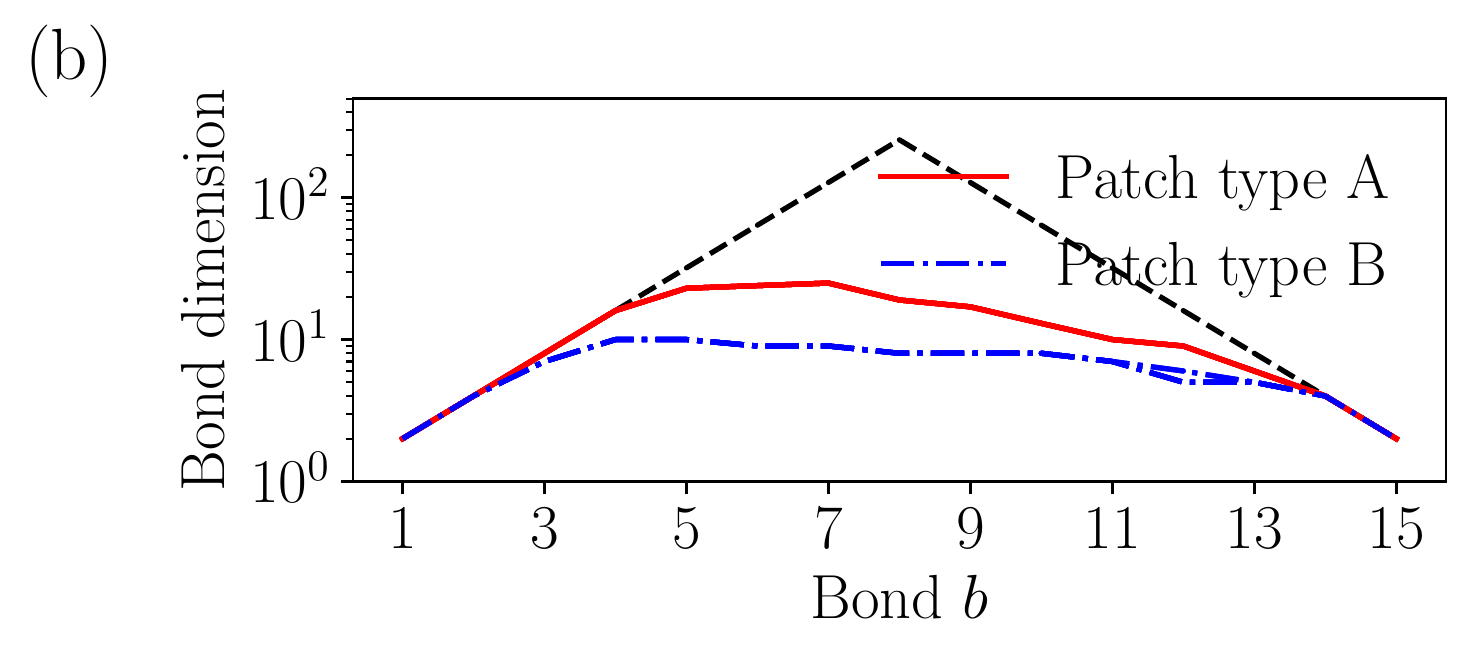}
    \caption{
    Green's function of the 2D Hubbard model solved within FLEX at the lowest positive Matsubara frequency. (a) Intensity map of the original data and the error in the reconstructed data. The full BZ is divided into 16 patches. (b) Bond dimensions for all 16 patches.
    The dashed line indicates the maximum bond dimensions in maximally entangled cases.
    {The compression rates are 10.86 and 45.20 for the patches A and B, respectively. Here, the compression rate is defined as the ratio between the number of elements in a TT and that of the original tensor.}
   }
    \label{fig:flex-givk}
\end{figure}

Figure~\ref{fig:flex-givk}(a) shows the intensity map of the Green's function at the lowest positive Matsubara frequency.
The large Fermi surface is broadened by finite-$T$ and correlation effects.
We thus need fewer patches than in the previous subsection without the self-energy.
We use $4\times 4=16$ patches, which are classified into two types: A and B.
The patches of type A contain more complex features in the BZ.
Within each patch, we expand the momentum dependence of the Green's function using an MPS.
This approach is natural because the Dyson equation can be solved patch-wise in the QTT representation.

We represented the momentum dependence within each patch using an MPS with $\epsilon=10^{-10}$.
Figure~\ref{fig:flex-givk}(a) shows the error in the reconstructed data, while Fig.~\ref{fig:flex-givk}(b) shows the bond dimensions for the two types of patches.
The absolute error is as small as $10^{-5}$, which is consistent with the square root of cutoff.
The bond dimensions are larger for the patches of type A, which is consistent with the complex momentum dependence within these patches.
For all the patches, the bond dimensions decrease after the first few bonds, indicating the validity of {the QTT representation}.
In practical solutions of the Dyson equation, one could partition the BZ adaptively by dividing patches with large bond dimensions.

\begin{figure}
          \centering
          \includegraphics[width=1.0\columnwidth]{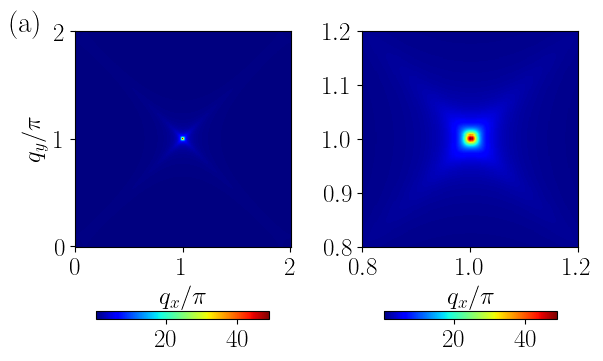}
          \includegraphics[width=1.0\columnwidth]{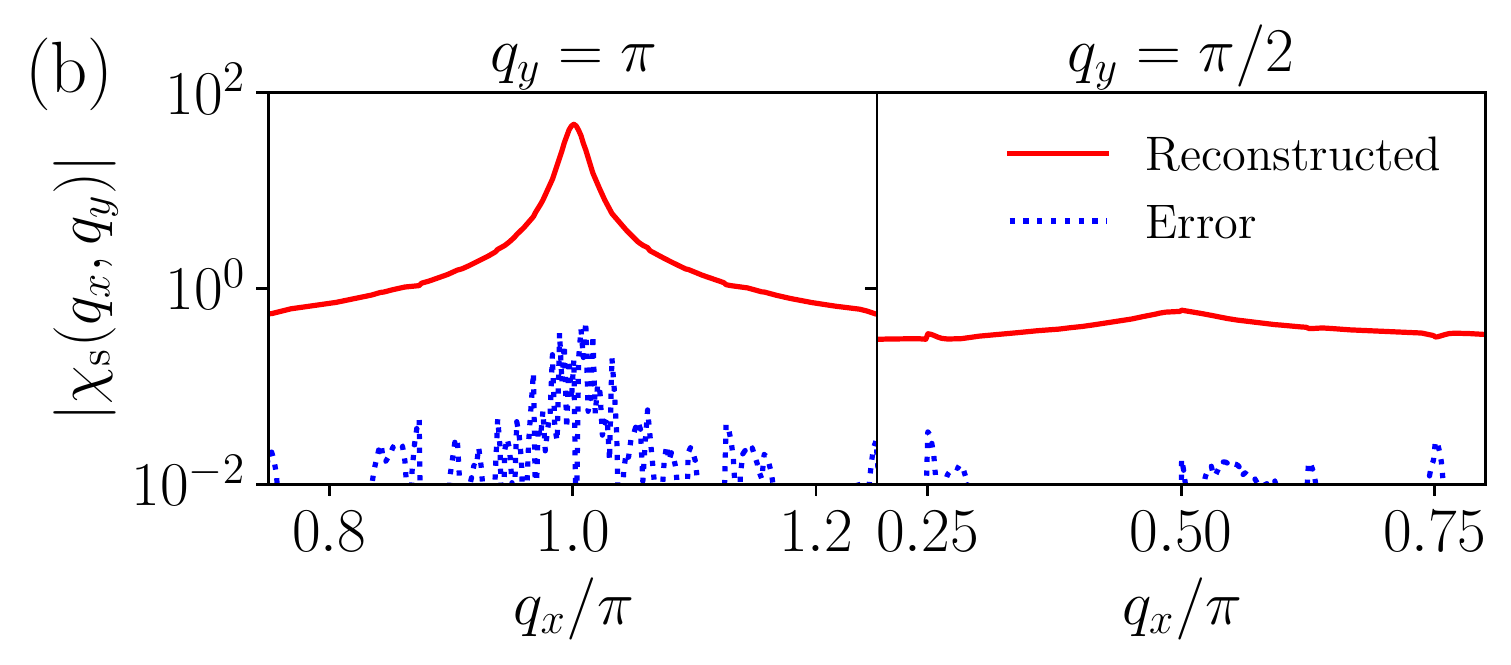}
          \includegraphics[width=1.0\columnwidth]{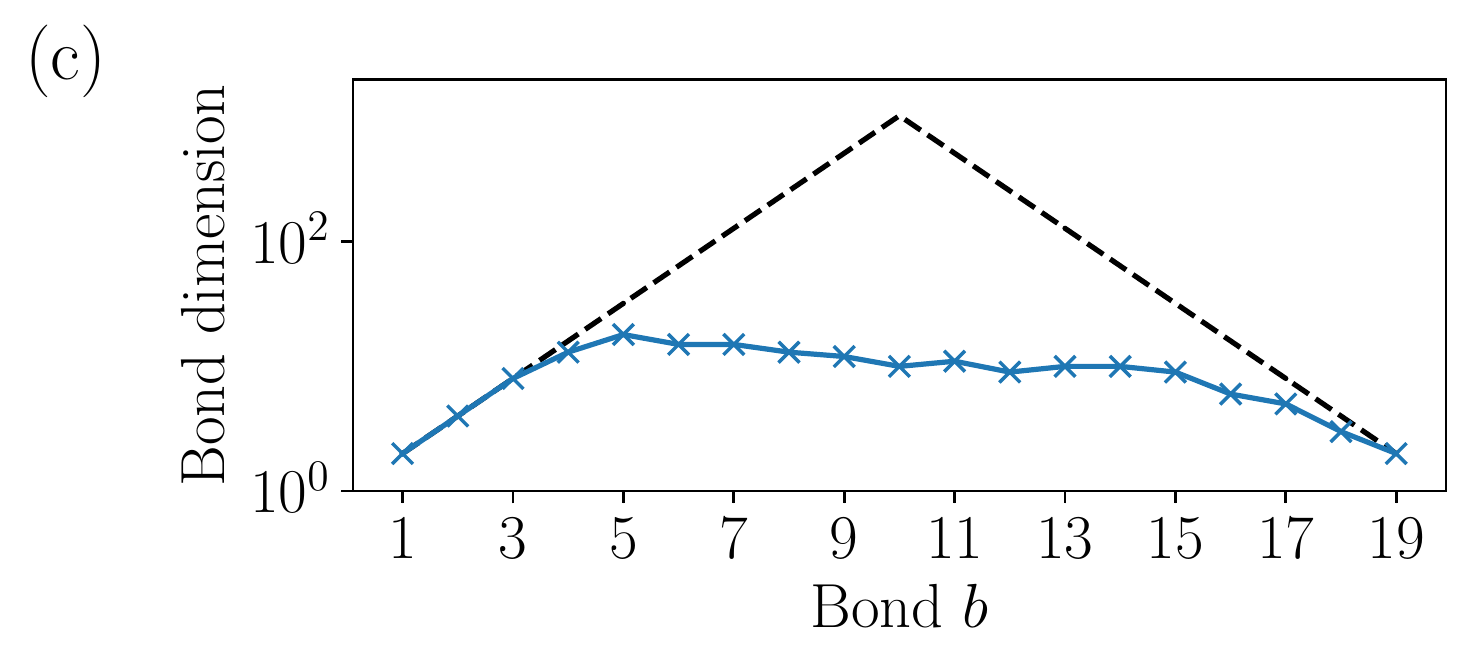}
          \caption{(a) Spin susceptibility of the 2D Hubbard model at zero frequency and an enlarged plot. (b) Reconstructed data and error at $q_y=\pi$ and $q_y=\pi/2$. (c) Bond dimensions of the MPS.
          The dashed line indicates the maximum bond dimensions in maximally entangled cases.
          {The compression rate is 269.97.}
          }
          \label{fig:flex-chi}
\end{figure}

We move on to the analysis of spin susceptibility.
Figure~\ref{fig:flex-chi}(a) shows the intensity map of the spin susceptibility at zero Matsubara frequency.
One can see a sharp peak at ${\bf q} = (\pi, \pi)$, reflecting the proximity to the AF phase at zero temperature.
There are additionally weaker signals on the diagonals.
Figure~\ref{fig:flex-chi}(b) compares the original data and the reconstructed ones along $q_y = \pi$ and $q_y = \pi/2$, respectively.
The cutoff was set to $\epsilon = 10^{-10}$
and the bond dimensions are shown in Fig.~\ref{fig:flex-chi}(c).
The maximum value of the bond dimensions is only around 20. Still, the compressed data can reproduce the sharp peak and the smaller features.
For more general cases with multiple peaks, which could happen for geometrically frustrated magnets, patching may help.

\subsection{Real frequency data}
\begin{figure}
    \centering
    \includegraphics[width=0.95\columnwidth]{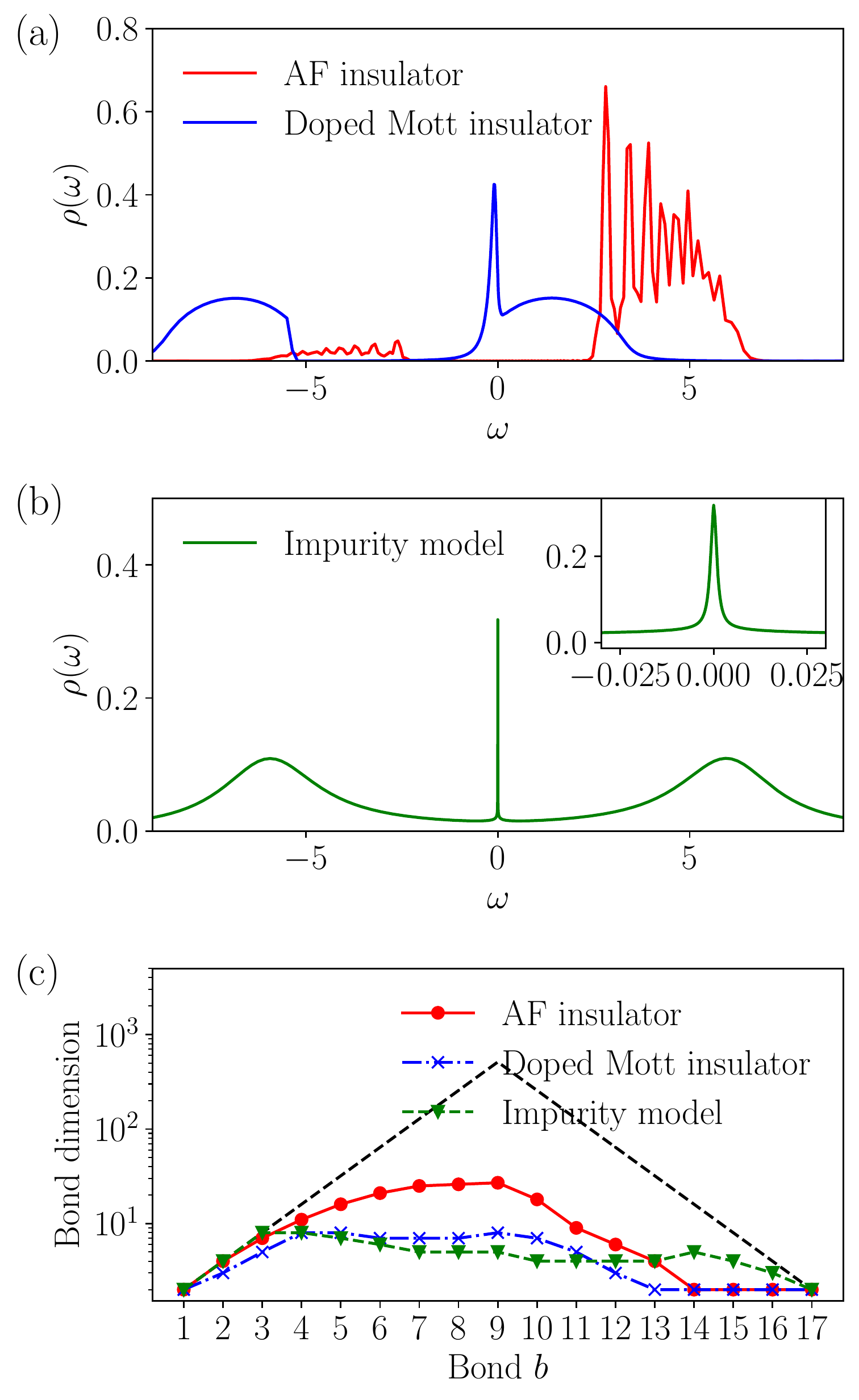}
    \caption{
    {
    Compression of the real-frequency spectral functions of an antiferromagnetic (AF) insulator, a doped Mott insulator, and an impurity model. See the main text for a more detailed description of the models. 
    (a) Spectral functions of the AF insulator and the doped Mott insulator,
    (b) spectral function of the impurity model (a zoom of the low frequency region is shown in the inset), (c) bond dimensions for $\epsilon=10^{-8}$ and $R=18$.
    The compression rates are 40.3, 280.7, 318.1, respectively.
    }
    }
    \label{fig:realfreq}
\end{figure}

{
As a next example we discuss the compressibility of real-frequency local spectral functions $\rho(\omega) = -\frac{1}{\pi} \Im G(\omega+\ii 0^+)$ of correlated systems, where electronic correlations lead to the emergence of fine structures or exponentially small energy scales.
Figure~\ref{fig:realfreq}(a) shows the results for an antiferromagnetic insulating ($U=8$ and $T=1/13$) and an 11\% doped Mott insulating ($U=8$, $\mu = U/2$+2.7, and $T=1/30$) state of the single-orbital Hubbard model~\eqref{eq:Hubbard} on the Bethe lattice with infinite coordination number. 
The unit of energy is the quarter of the bandwidth of the free system ($U=0$). We solved the Hubbard model using the real-time dynamical mean-field theory (DMFT)~\cite{GeorgesRMP, Aoki2014RMP} and the non-crossing approximation (NCA)~\cite{Eckstein2010b} and performed a Fourier transform to obtain the spectra.
In panel (b), we additionally show the spectral function for the single-orbital Anderson impurity model~\cite{Hewson} at half-filling  ($U=16$ and $T=0$) with constant hybridization set to $1$. In this case, the non-interacting density of states is a Lorentzian with width $1$.
The impurity model was solved by an approximate real-frequency solver which reproduces the exponential Kondo scale at strong coupling~\cite{Janis2017, Janis2017a}.
}

{
As seen in Fig.~\ref{fig:realfreq}(a), the spectral function of the AF insulator exhibits sharp peaks in the Hubbard bands originating from spin-polaron excitations~\cite{Sangiovanni2006}, while
the spectral function of the doped Mott insulator shows a sharp quasiparticle peak at $\omega=0$.
The spectral function of the impuirty model, shown in Fig.~\ref{fig:realfreq}(b), features a much sharper peak whose energy scale is smaller than the bandwidth by several orders of magnitude~\cite{Hewson}.
}

{
Figure~\ref{fig:realfreq}(c) shows the bond dimensions of MPSs constructed with $\epsilon=10^{-8}$ and $R=18$.
It is clearly seen that all the three spectral functions are QTT-compressible.
For the two spectral functions with a single sharp peak, the bond dimension decreases after the first few bonds, indicating energy separation.
The spectral function of the AF insulator, with multiple physical features, requires a larger bond dimension.
}

\begin{figure*}
    \centering
    \includegraphics[width=0.9\textwidth]{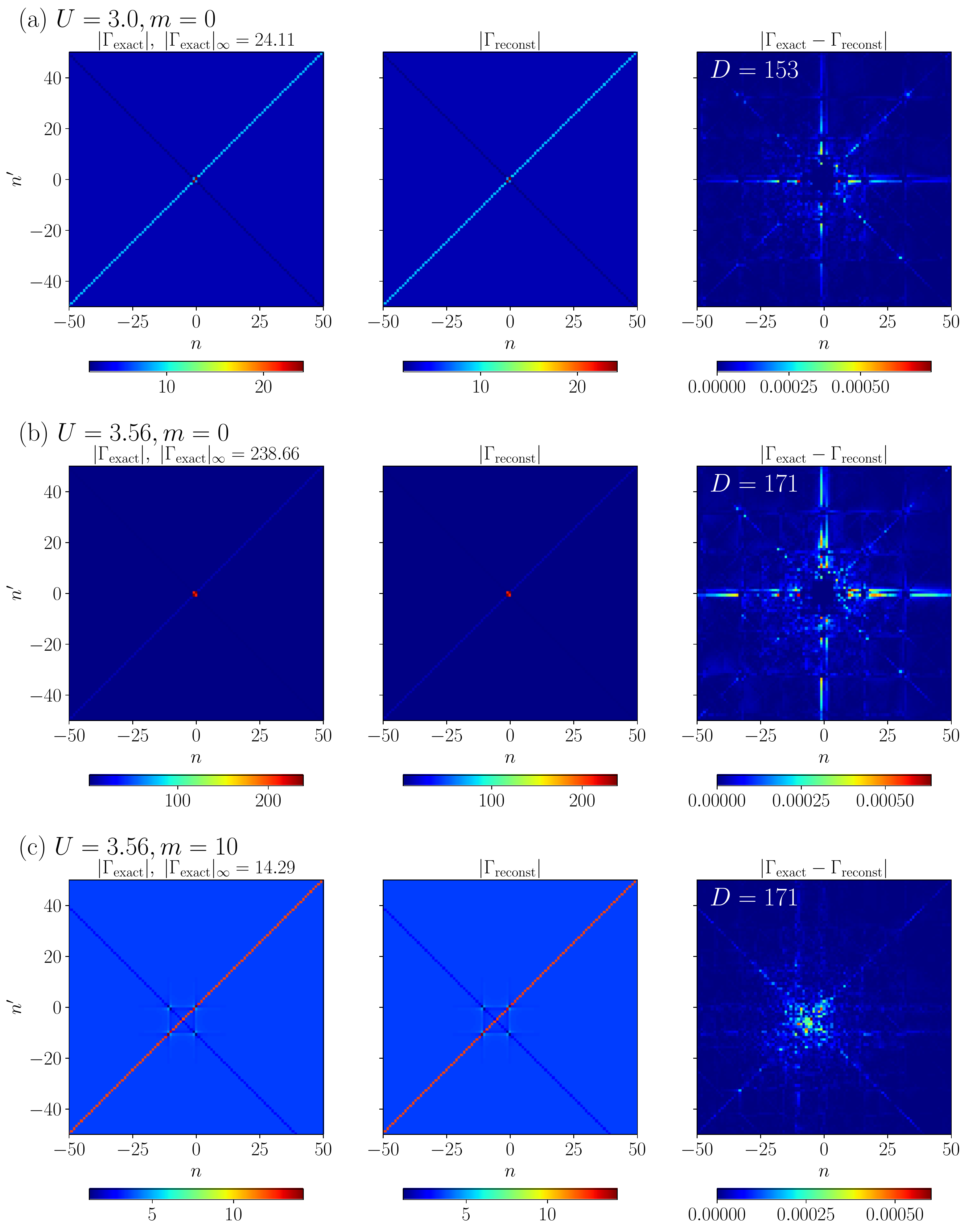}
    \caption{
        Irreducible vertex function $\Gamma_d$ of the Hubbard atom for two values of $U$ and for a given bosonic frequency $\frac{2m\pi}{\beta}$ ($\beta=1$).
        The middle panels show the reconstructed data from MPSs ($\epsilon=10^{-14}$).
        The right panels show the error in the reconstructed data.
        Note that the entire three-frequency dependence was fitted by a single MPS.
    }
    \label{fig:atom-3d-vertices}
\end{figure*}

\subsection{Hubbard atom}\label{sec:hubbardatom}
In this section, we extend the application of the QTT representation to three-frequency objects, namely two-particle vertex functions. We demonstrate the compactness of the representation by compressing vertex functions of the Hubbard atom for which the exact analytic forms are known~\cite{Thunstroem:PRB18}.

The two-particle Green's function in the so-called particle-hole (ph) frequency notation~\cite{RohringerRMP} is defined as
\begin{align}
    G^{(2)}_{\sigma_1 \sigma_2 \sigma_3 \sigma_4}(\iv,\iv';\iw) &= \int_0^{\beta}\!\!\dd\tau_1d\tau_2d\tau_3\; e^{-\ii\nu\tau_1}e^{\ii(\nu+\omega)\tau_2}\nonumber\\*
     \times \;\; e^{-\ii(\nu'+\omega)\tau_3}& \langle T_\tau c^{\phantom \dagger}_{\sigma_1}(\tau_1) c^\dagger_{\sigma_2}(\tau_2) c^{\phantom \dagger}_{\sigma_3}(\tau_3) c^\dagger_{\sigma_4}(0)\rangle,
\end{align}
where $\sigma_1 \ldots \sigma_4$ are spin indices 
and $\nu$, $\nu'$ are fermionic Matsubara frequencies, and $\omega$ is a bosonic Matsubara frequency.

$G^{(2)}$ can be decomposed into  so-called disconnected parts (products of one-particle Green's functions) and the connected part, which is a product of four Green's functions and the two-particle vertex $F$
\begin{equation}
\begin{split}
    &   G^{(2)}_{\sigma_1 \sigma_2 \sigma_3 \sigma_4}(\iv,\iv';\iw) = \beta  G_{\sigma_1}(\iv)  G_{\sigma_3}(\iv') \delta_{\omega,0} \delta_{\sigma_1,\sigma_2}\delta_{\sigma_3,\sigma_4}\\
    & \quad-\ \beta G_{\sigma_1}(\iv) G_{\sigma_2}(\iv+\iw) \delta_{\nu,\nu'} \delta_{\sigma_1,\sigma_4}\delta_{\sigma_2,\sigma_3} \\
    & \quad+\  G_{\sigma_1}(\iv) G_{\sigma_2}(\iv+\iw)  F_{\sigma_1 \sigma_2 \sigma_3 \sigma_4}(\iv,\iv';\iw)\\
       & \quad\times\ G_{\sigma_3}(\iv'+\iw)  G_{\sigma_4}(\iv').
\end{split}
\label{eq:F}
\end{equation}
The vertex $F$ is a sum of two-particle reducible and irreducible diagrams. Reducibility at the two-particle level is not uniquely defined and we need to specify in which channel the irreducible diagrams are not reducible. We choose the particle-hole channel here. The vertex $F$ is related to the irreducible diagrams collected in the irreducible vertex $\Gamma^\PH$ through the Bethe-Salpeter equation (BSE) in the ph channel 
\begin{align}
    &F_{\sigma_1 \sigma_2 \sigma_3 \sigma_4}(\iv, \iv';\iw) = \Gamma^\PH_{\sigma_1 \sigma_2 \sigma_3 \sigma_4}(\iv, \iv' ;\iw)  \nonumber\\
    & + \; \frac{1}{\beta^2} \underset{\sigma',\sigma''}{\sum_{\nu'', \nu'''}} \Gamma_{\sigma_1 \sigma_2 \sigma' \sigma''}^\PH(\iv,\iv'';\iw)    X_{\sigma' \sigma''}^{0,\PH}(\iv'',\iv''';\iw)\nonumber \\
    & \times \; F_{\sigma' \sigma'' \sigma_3 \sigma_4}(\iv''', \iv';\iw),
\end{align}
with
\begin{align}
    X_{\sigma \sigma'}^{0,\PH}(\iv,\iv';\iw) &= \beta G_{\sigma}(\iv')G_{\sigma'}(\iv' + \iw) \delta_{\nu \nu'}.
\end{align}
In the SU(2) symmetric case, which we consider here, the spin dependence can be diagonalized by introducing linear spin combinations known as density (d) and magnetic (m) channels. The BSE then takes the following form
\begin{align}
    &F_{d/m}(\iv, \iv';\iw) = \Gamma_{d/m}(\iv, \iv' ;\iw)  \nonumber\\
    & + \; \frac{1}{\beta^2} \sum_{\nu'',\nu'''} \Gamma_{d/m}(\iv,\iv'';\iw)  X^{0}(\iv'',\iv''';\iw)\nonumber \\
    & \times \; F_{d/m}(\iv''', \iv';\iw),\label{eq:bse-SU2}
\end{align}
where we dropped the $\PH$ superscript as well as the spin indices of the bare susceptibility $X^0$  (it is spin diagonal and equal for both spins in this case).

The frequency dependence of two-particle vertex functions is complicated due to the presence of sharp features that do not decay for large fermionic frequencies. Particularly challenging is the numerical treatment of irreducible vertices in the atomic limit due to the presence of divergencies~\cite{Schaefer:PRL13,Chalupa2018}.  At half-filling $\Gamma_d$ is known to diverge at $\beta U\simeq \{3.627$, $5.127$, $10.884$, $12.19$, $18.138$, $19.23, \ldots\}$. In the vicinity of these vertex divergences, the numerical treatment is challenging. In the following, we show the compression of the atomic irreducible vertex in the density channel $\Gamma_d$ for half-filling.

The left panels of Fig.~\ref{fig:atom-3d-vertices} show $\Gamma_d(\iv,\iv';\iw)$ computed for several values of $U$ and a fixed bosonic frequency $\frac{2m \pi}{\beta}$ (with $m=0$ and $m=10$).
We take $\beta=1$ throughout this subsection.
At $U=3$ and zero bosonic frequency $m=0$ [Fig.~\ref{fig:atom-3d-vertices}(a)], the main structure consists of diagonal lines extending to high fermionic frequencies.
Figure~\ref{fig:atom-3d-vertices}(b) shows the vertex $\Gamma_d$ for zero bosonic frequency near a divergence point (the divergence occurs at $U\simeq 3.627$). One can see that $\Gamma_d$ is dominated by large values at low frequencies.
At a finite bosonic frequency, as shown in Fig.~\ref{fig:atom-3d-vertices}(c), the vertex is not so strongly peaked, but additional box-like structures appear at low frequencies.

We decompose the vertex function on a grid of size $2^R \times 2^R \times 2^R$ with $R=10$ and cutoff $\epsilon=10^{-14}$.
The middle panels of Fig.~\ref{fig:atom-3d-vertices} show the reconstructed data from the MPSs.
The errors in the reconstructed data are shown in the right panels of Fig.~\ref{fig:atom-3d-vertices}.
It is clearly visible that the MPSs can describe all the complex structures in the three-frequency space.

\begin{figure}
    \centering
    \includegraphics[width=0.99\columnwidth]{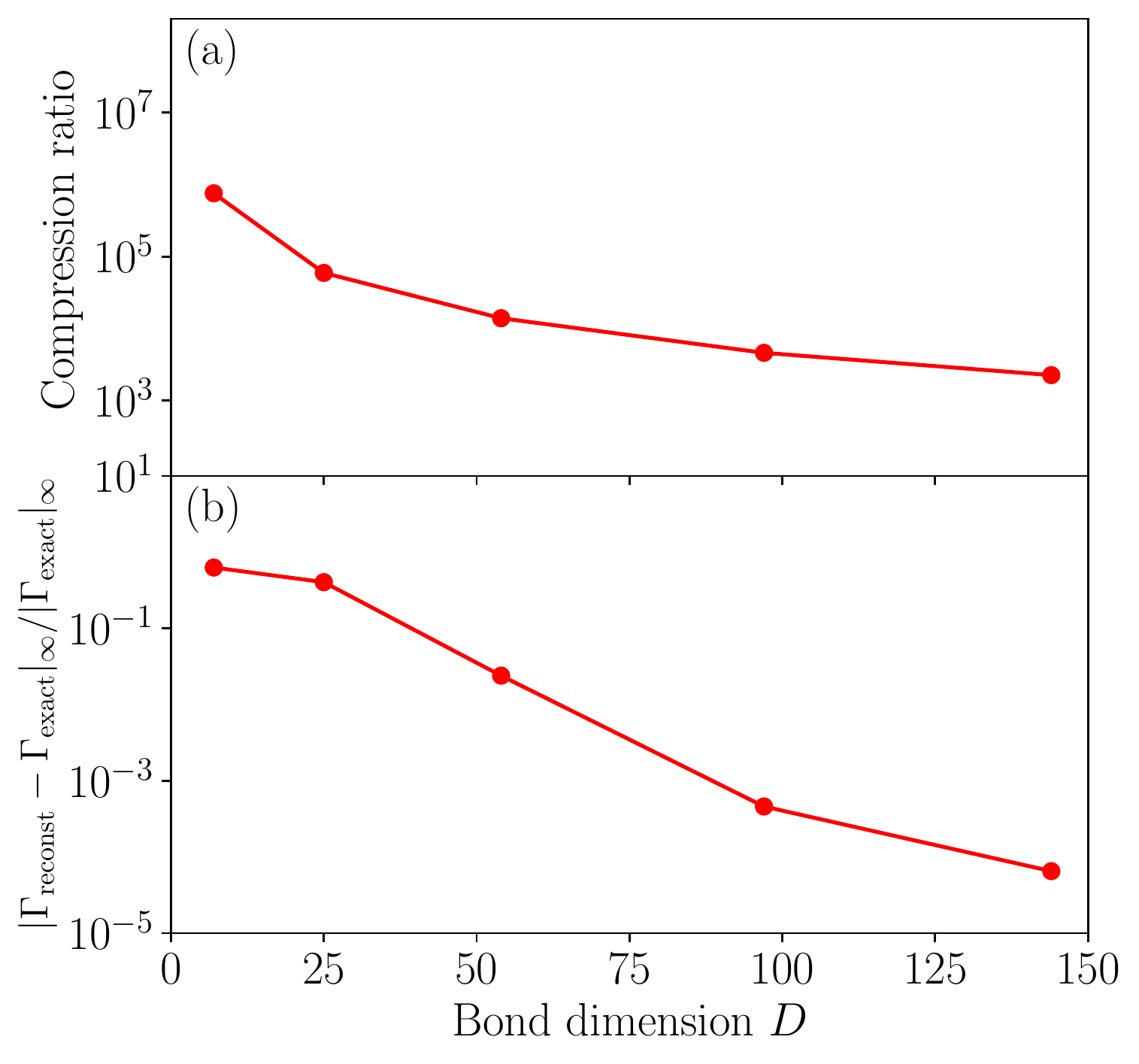}
    \caption{
    Dependence of (a) the bond dimension and (b) the compression rate on the error (accuracy) for $\Gamma_d$ of the Hubbard atom at $U=3$ ($\beta=1$). We compressed the data on a $2^R\times 2^R \times 2^R$ grid with $R=10$ for cutoff $\epsilon=10^{-6}$, $10^{-8}$, $10^{-10}$, $10^{-12}$ and $10^{-14}$.
{The symbol $|\cdots|_\infty$ denotes the maximum norm of a tensor, which is the maximum of the absolute values of its elements.}
    }
    \label{fig:atom-vertex-3d-scaling}
\end{figure}

Figure~\ref{fig:atom-vertex-3d-scaling} shows the dependence of the compression rate and the error on the bond dimension $D$.
We performed the compression for several different values of the cutoff $\epsilon$.
The compression rate is defined as the ratio of the number of elements in the MPS tensors and that of the original data.
The error roughly decays exponentially with increasing $D$.
Achieving the accuracy of $|\Gamma_\mathrm{reconst} - \Gamma_\mathrm{exact}|/|\Gamma_\mathrm{exact}|_\infty < 10^{-4}$ requires a bond dimension slightly larger than $100$. The compression rate is beyond $10^3$ even in this case.

\begin{figure}
    \centering
    \includegraphics[width=0.95\columnwidth]{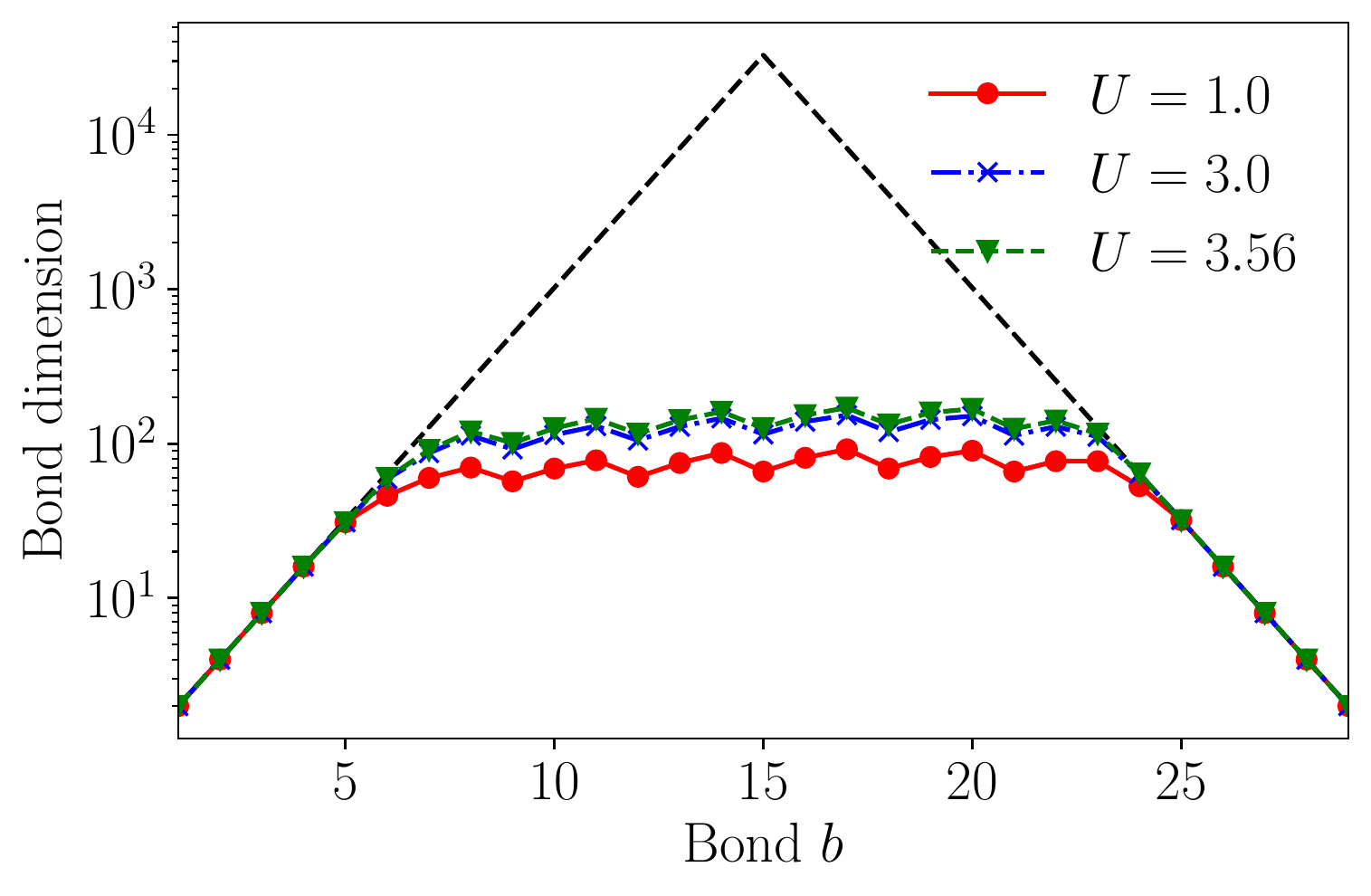}
    \caption{
    Bond dimensions of MPSs representing the vertex function $\Gamma_d$ with cutoff $\epsilon=10^{-14}$.
    The dashed line indicates the maximum bond dimensions in maximally entangled cases.
    }
    \label{fig:gamma-3d-linkdims}
\end{figure}

We now take a closer look at the bond dimensions.
Figure~\ref{fig:gamma-3d-linkdims} shows the bond dimensions along the chain.
After the first few bonds, the bond dimension stays almost constant, indicating a separation between different length scales.
The non-decaying bond dimension can be attributed to the non-decaying structures in the frequency space (with no high-frequency cutoff). In order to see that let us, for simplicity, consider 2D data at zero bosonic frequency [see Fig.~\ref{fig:atom-3d-vertices}(a)].
To simplify the discussion, we model the non-trivial diagonal structures by an identity matrix, $A_{ij} = \delta_{ij}$ of size $2^R \times 2^R$.
All the $2^R$ singular (eigen)values of the identity matrix are 1.
Thus, this matrix is not compressible by SVD.
Mapping to qubits yields
\begin{align}
    A_{(i_1 i_2 \cdots j_R)_2, (j_1 j_2 \cdots j_R)_2} &= \prod_{b=1}^R \delta_{i_b,j_b},
\end{align}
which indicates that the matrix can be represented exactly as an MPS with a bond dimension of 1.

\subsection{2P quantities from DFT+DMFT calculations}\label{sec:CeB6}
To demonstrate the compression in the case of multiple spin-orbital indices, we analyze the 2P response functions of a realistic multi-orbital model.
In a recent study~\cite{CeB6} where one of us was involved, the multipolar ordering in the $f$-electron compound CeB$_6$ was investigated using dynamical mean-field theory (DMFT) combined with density functional theory (DFT).
In this subsection, we will analyze the 2P data from this state-of-the-art DFT+DMFT calculation.

Ref.~\cite{CeB6} constructed a tight-binding model from DFT calculations and considered local correlation effects within DMFT.
The effective impurity model involves six local degrees of freedom of the $j=5/2$ multiplet, which was solved by the Hubbard-I approximation, i.e., exact diagonalization without hybridization.
The local interaction was set to $U = 6.2$ eV and the Hund's coupling to $J_\mathrm{H} = 0.8$ eV.

For a converged self-consistent solution of DMFT, they computed the multipolar susceptibility $\chi(\bm{q})$ in the $\PH$ channel at zero bosonic frequency through BSE.
First, they computed the local generalized susceptibility $X_\mathrm{loc}$ by exact diagonalization on a fermionic-frequency mesh of size $N_w \times N_w$. Then they computed the irreducible vertex $\Gamma$ by solving the local BSE.
By solving the lattice BSE with $\Gamma$, the multipolar susceptibility $\chi(\bm{q})$ was obtained.
For technical details, we refer to Ref.~\cite{CeB6}.

In this subsection, we analyze two important quantities: the local generalized susceptibility $X_\mathrm{loc}$ and the multipolar susceptibility $\chi(\bm{q})$.
The local generalized susceptibility is related to the local 2P Green's function by
\begin{align}
    X^\mathrm{loc}_{m_1 m_2, m_3 m_4}(\iv, \iv') &= \frac{1}{\beta} G^{(2)}_{m_2 m_1, m_4 m_3}(\iv, \iv'; \iw=0) \nonumber \\ & -  G_{m_2,m_1}(\iv)G_{m_4,m_3}(\iv'),
\end{align}
where $m_1,m_2,m_3,m_4$ stand for the eigenvalues of $j_z$, namely, $m=-5/2$, $-3/2$, $\cdots$, $+5/2$.
Introducing combined indices $I\equiv (m_1m_2)$ and $J\equiv (m_3m_4)$, we express this quantity as $X^\mathrm{loc}_{IJ}(\iv, \iv')$.
The combined indices are defined in row major order: $I=1, 2, \cdots, 36$ corresponds to $(m_1m_2)=(1,1),(1,2),\cdots, (6,6)$.

\begin{figure}
    \centering
    \includegraphics[width=0.85\columnwidth]{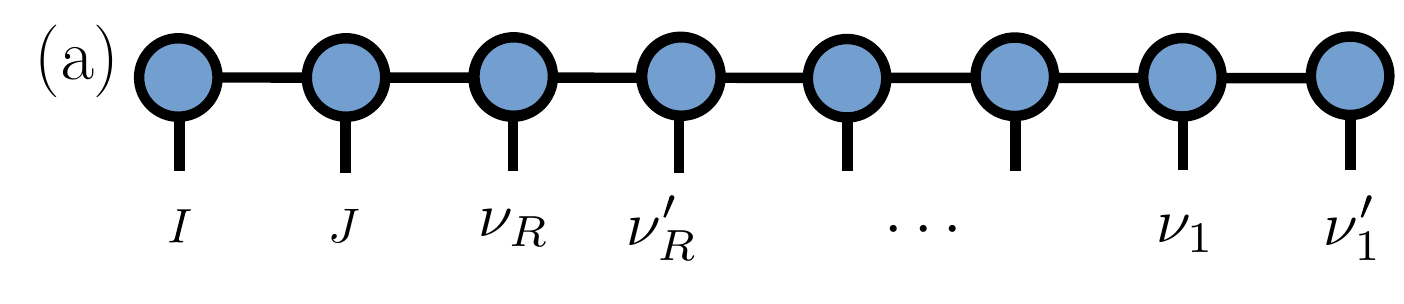}
    \includegraphics[width=0.85\columnwidth]{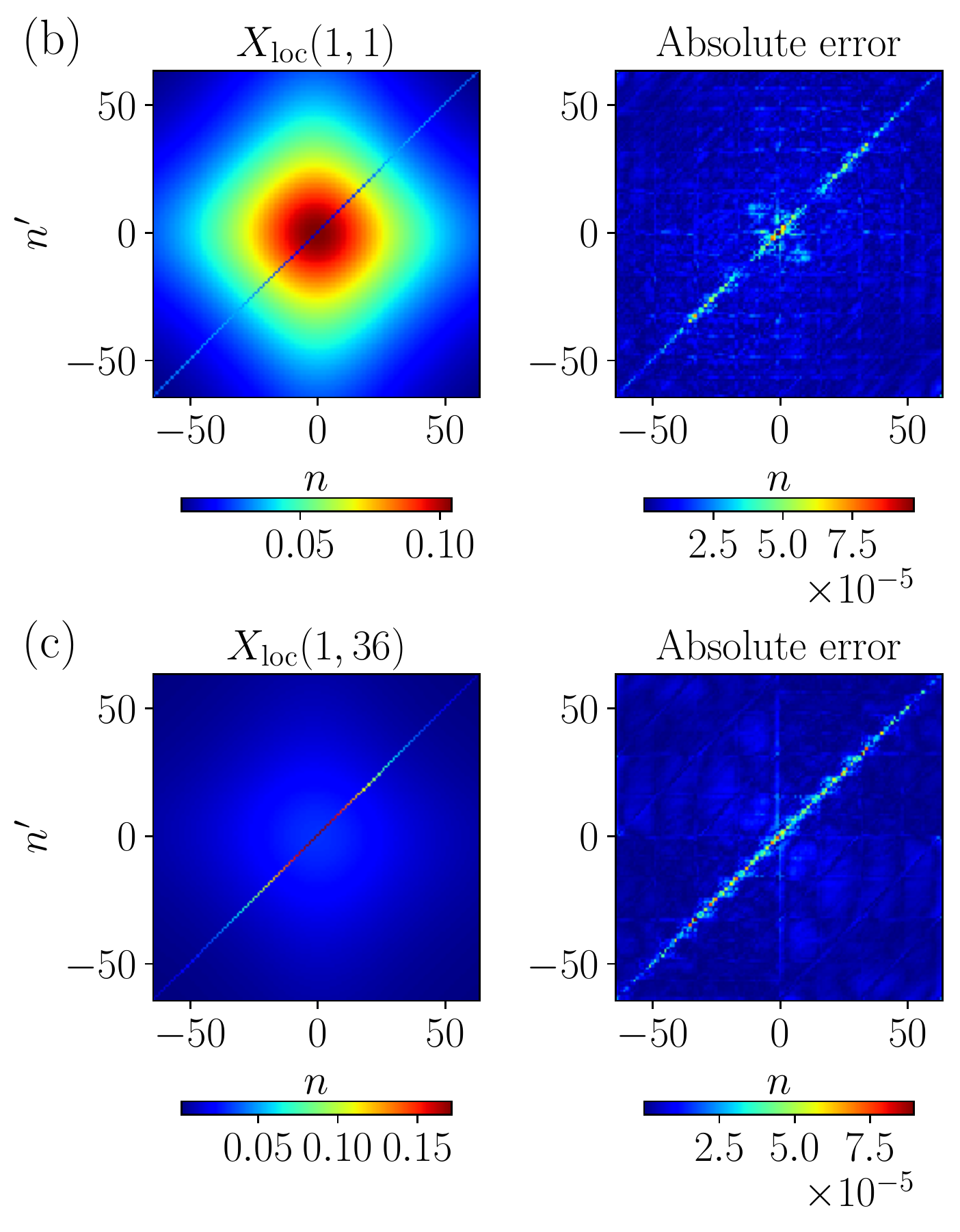}
    \includegraphics[width=0.85\columnwidth]{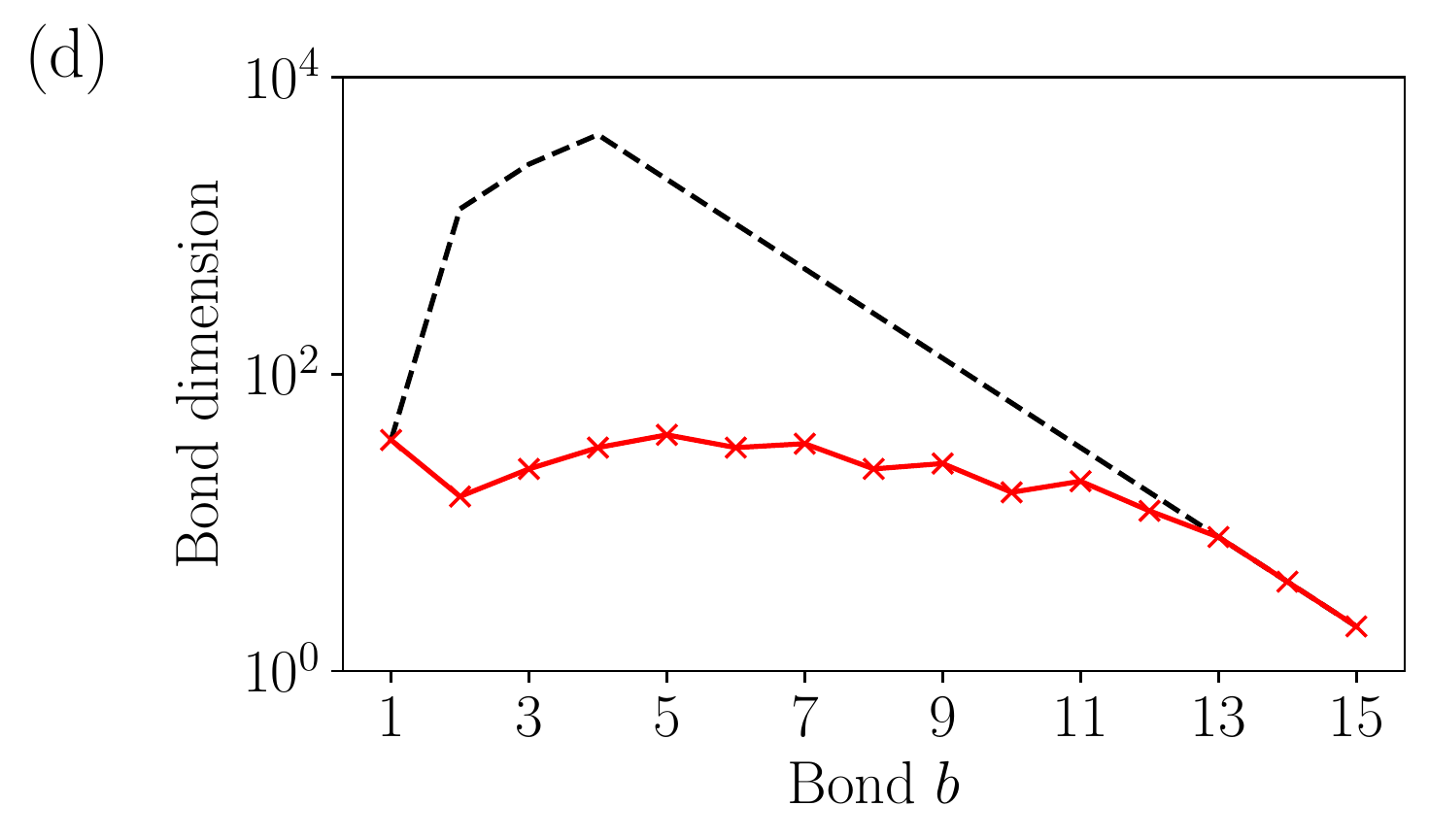}
    \caption{
        Local generalized susceptibility of CeB$_6$ computed by DFT+DMFT.
        (a) {MPS}, (b)/(c) intensity map of the reconstructed $X_\mathrm{loc}$ and error for $\epsilon=10^{-8}$, (d) bond dimensions along {the TT}.
        In (d), the dashed line indicates the maximum bond dimensions in maximally entangled cases.
        {The estimated compression rate is around $608$.}
    }
    \label{fig:CeB6-Xloc}
\end{figure}

Figure~\ref{fig:CeB6-Xloc}(a) illustrates the MPS used for compressing the local generalized susceptibility. A fermionic frequency $\nu$ is encoded as $\nu + N_w/2 = (\nu_1 \nu_2 \cdots \nu_R)_2$.
The combined indices for the $j=5/2$ multiplet, $I$ and $J$, are not decomposed further in this study.
Figures~\ref{fig:CeB6-Xloc}(b) and (c) show the results for $(I,J) = (1,1)$ and $(1, 36)$, respectively. The left panels plot the reconstructed data from the MPS constructed with $\epsilon=10^{-8}$ and $R=7$ ($N_w=128$), while the
right panels show the error.
In Fig.~\ref{fig:CeB6-Xloc}(b), one sees broad structures in addition to the diagonal line at $\nu=\nu'$.
A similar but weak broad structure is also seen in Fig.~\ref{fig:CeB6-Xloc}(c).
All these structures can be described with high accuracy, within an error of $<10^{-4}$.

Figure~\ref{fig:CeB6-Xloc}(d) shows the dimensions of {the MPS}. The bond dimension is highest at $b=1$ between the indices of $I$ and $J$. After showing another small local maximum around $b=5$, the bond dimension slowly decreases.
The estimated compression rate is around $608$.

We now analyze the multipolar susceptibility $\chi(\bm{q})$ obtained by solving the lattice BSE. The $\chi(\bm{q})$ is defined as
\begin{align}
    \chi_{m_1 m_2,m_3 m_4}(\bm{q}) \equiv \int_0^{\beta}d\tau \langle O_{m_1 m_2}(\bm{q},\tau) O_{m_4 m_3}(-\bm{q}) \rangle.
\end{align}
Here, the argument $\tau$ stands {for the imaginary time of} the Heisenberg operator.
The operator $O_{mm'}(\bm{q})$ is the Fourier transform of the local density operator $O_{mm'}(i) = \hat{f}_{im}^{\dag} \hat{f}_{im'}^{\phantom\dag}$, where $\hat{f}_{im}^{\dag}$ and $\hat{f}_{im}$ are the creation and annihilation operators for the $f$ electrons, respectively.

\begin{figure}
    \centering
    \includegraphics[width=0.85\columnwidth]{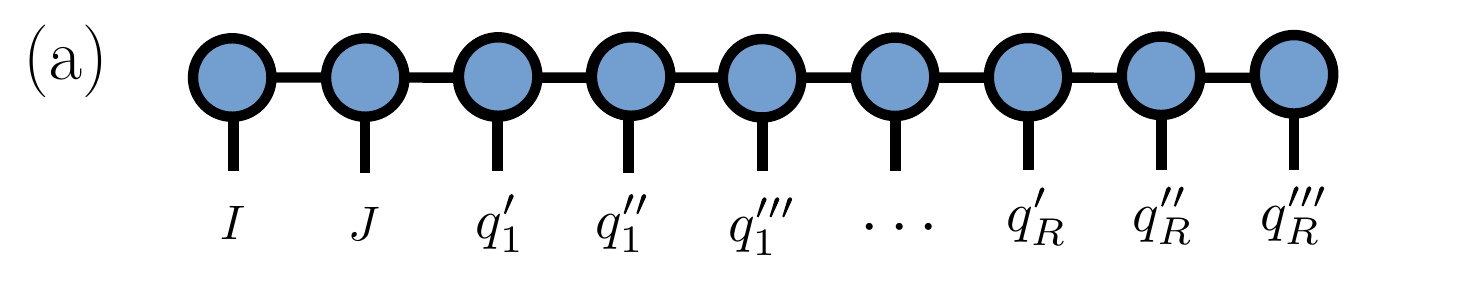}
    \includegraphics[width=0.85\columnwidth]{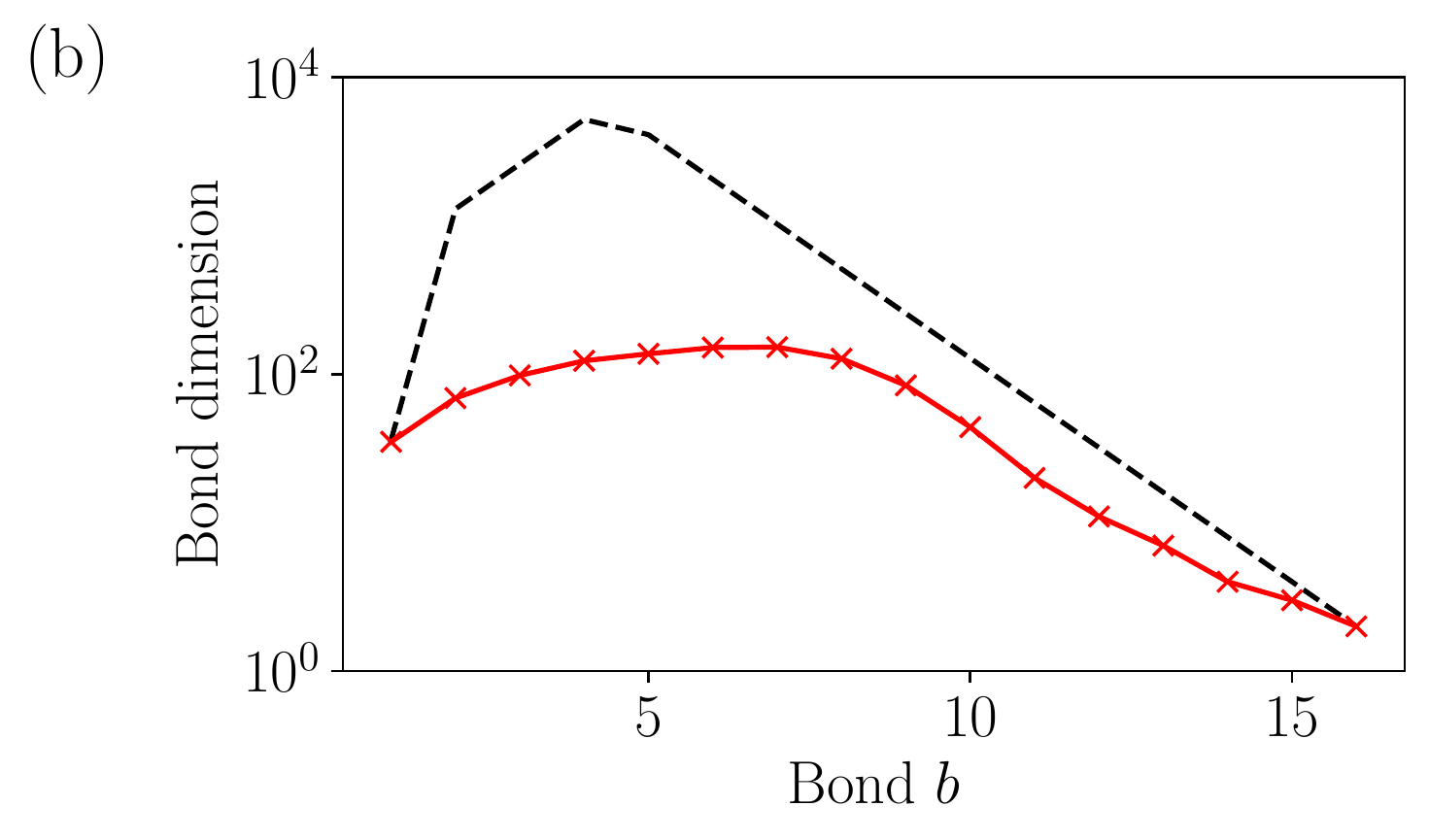}
    \caption{
        Multipolar susceptibility of CeB$_6$ computed by DFT+DMFT.
        (a) {MPS}, (b) bond dimensions along {the TT} for $\epsilon=10^{-8}$.
        In (b), the dashed line indicates the maximum bond dimensions in maximally entangled cases.
        {The estimated compression rate is around $134$.}
    }
    \label{fig:CeB6-chi}
\end{figure}

We analyze the data of size $36 \times 36 \times 32 \times 32 \times 32$, where the number of the momentum grid points is $32^3$.
A momentum grid point is denoted by three integers as $\bm{q} = (q', q'', q''')$ and 
we use {the MPS} illustrated in Fig.~\ref{fig:CeB6-chi}(a).
Since the multiplet indices correspond to the shortest (most relevant) length scale, it is a natural choice to place them on the left edge of {the MPS}.

Figure~\ref{fig:CeB6-chi}(b) shows the bond dimensions for $\epsilon=10^{-8}$ and $R=5$. The bond dimension of {the MPS} is slightly larger than 100. One can see a plateau behavior around $b=5$, supporting the length scale separation.
The estimated compression rate is around $134$.
Although this number is already impressive, it may even be underestimated: The $b$ dependence of the bond dimension in the right half of the chain may indicate that the mesh size was not large enough.

\begin{figure*}
  \centering
  \includegraphics[width=\textwidth]{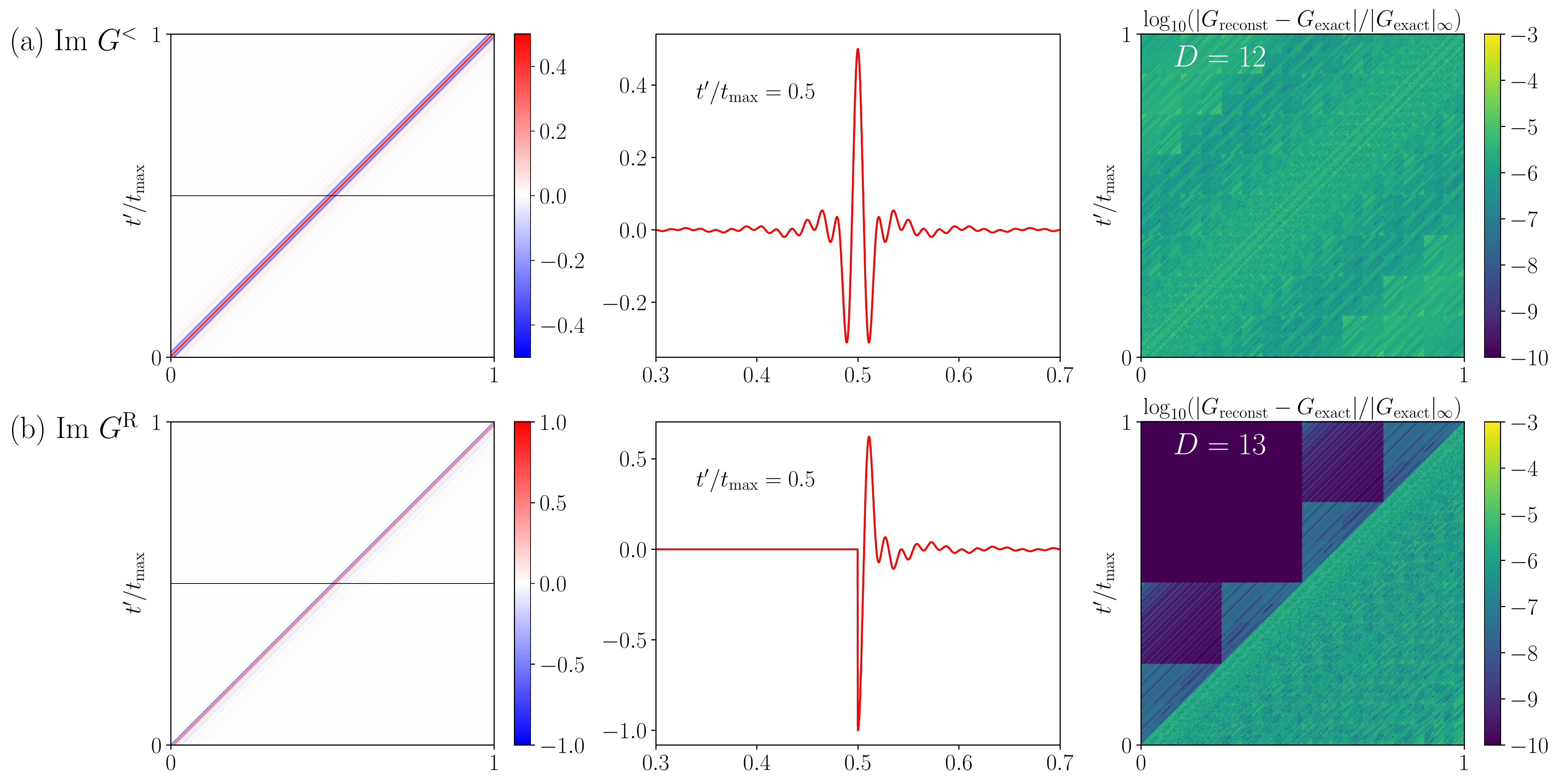}
  \includegraphics[width=\textwidth]{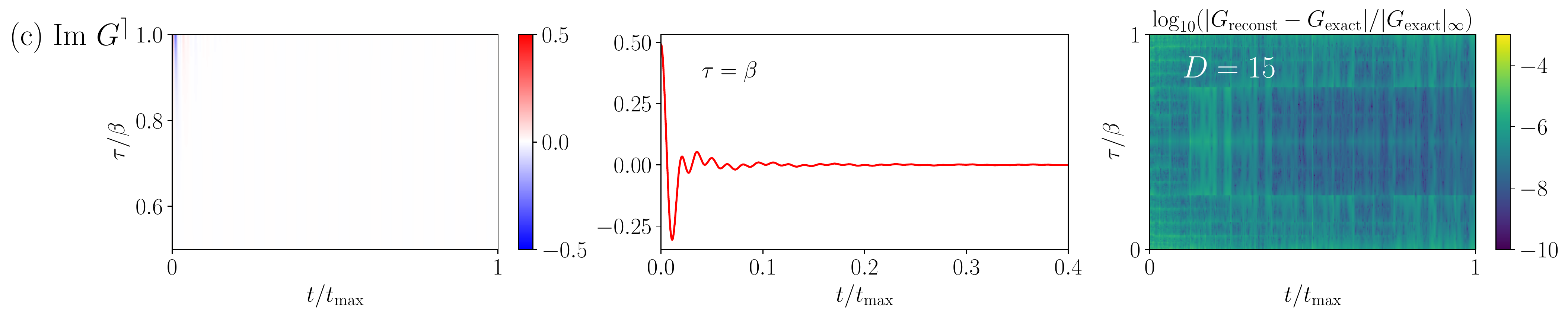}
  \vspace{-2em}
  \caption{Real-time and mixed Green's functions computed for the equilibrium paramagnetic system. Exact data for the imaginary parts of (a) the lesser, (b) the retarded, and (c) the left-mixing component are shown in the left panels. Cuts through these functions at $t'/t_\mathrm{max}=1/2$ or $\tau/\beta=1$ [holizontal lines in the left panels] are shown in the middle panel.
  The right panels show the {logarithm of the} error for cutoff $\epsilon=10^{-10}$ together with the bond dimension $D$ automatically set by $\epsilon$.
  {The symbol $|\cdots|_\infty$ denotes the maximum norm of a tensor, which is the maximum of the absolute values of its elements.}
  }
  \label{fig:keldysh-eq}
\end{figure*}

\subsection{Nonequilibrium Green's functions}\label{sec:noneq}
After analyzing Matsubara Green's functions and vertices in the previous subsections, we move on to the analysis of real-time and mixed real/imaginary-time Green's functions of equilibrium and nonequilibrium systems. In nonequilibrium or real-time Green's function calculations, the Green's functions are often defined on the so-called L-shaped contour, which consists of the Matsubara branch and a real-time contour~\cite{stefanucci_nonequilibrium_2013,Aoki2014RMP}. Depending on the position of the creation and annihilation operators on this contour, one can define different components of the Green's functions. A complete characterization is obtained in terms of the retarded component ($G^R(t,t') = -i\theta(t-t') \langle \{\hat{c}(t), \hat{c}^\dagger(t') \} \rangle$), the lesser component ($G^<(t,t') = i \langle \hat{c}^\dagger(t')\hat{c}(t) \rangle$), the left-mixing component ($G^\rceil(t,\tau') = i \langle \hat{c}^\dagger(\tau')\hat{c}(t) \rangle$), and the previously defined Matsubara component. 

In nonequilibrium Green's function methods, the interacting Green's function on the L-shaped contour is typically obtained by solving Dyson equations (Kadanoff-Baym equations) on this contour \cite{stefanucci_nonequilibrium_2013,Aoki2014RMP,Schuler2020-tq}. A standard implementation based on an equidistant time discretization with $N_t$ time steps on the real-time axis and $N_\tau$ time steps on the imaginary time axis requires a computational time of $\mathcal{O}(N_t^3)$ and memory of $\mathcal{O}(N_t^2)$, assuming that $N_t \gg N_\tau$ ~\cite{Schuler2020-tq}. Here, we will address the problem of storing the nonequilibrium (real-time) Green's functions and show that these functions are highly compressible. 

\begin{figure*}
    \centering
    \includegraphics[width=\textwidth]{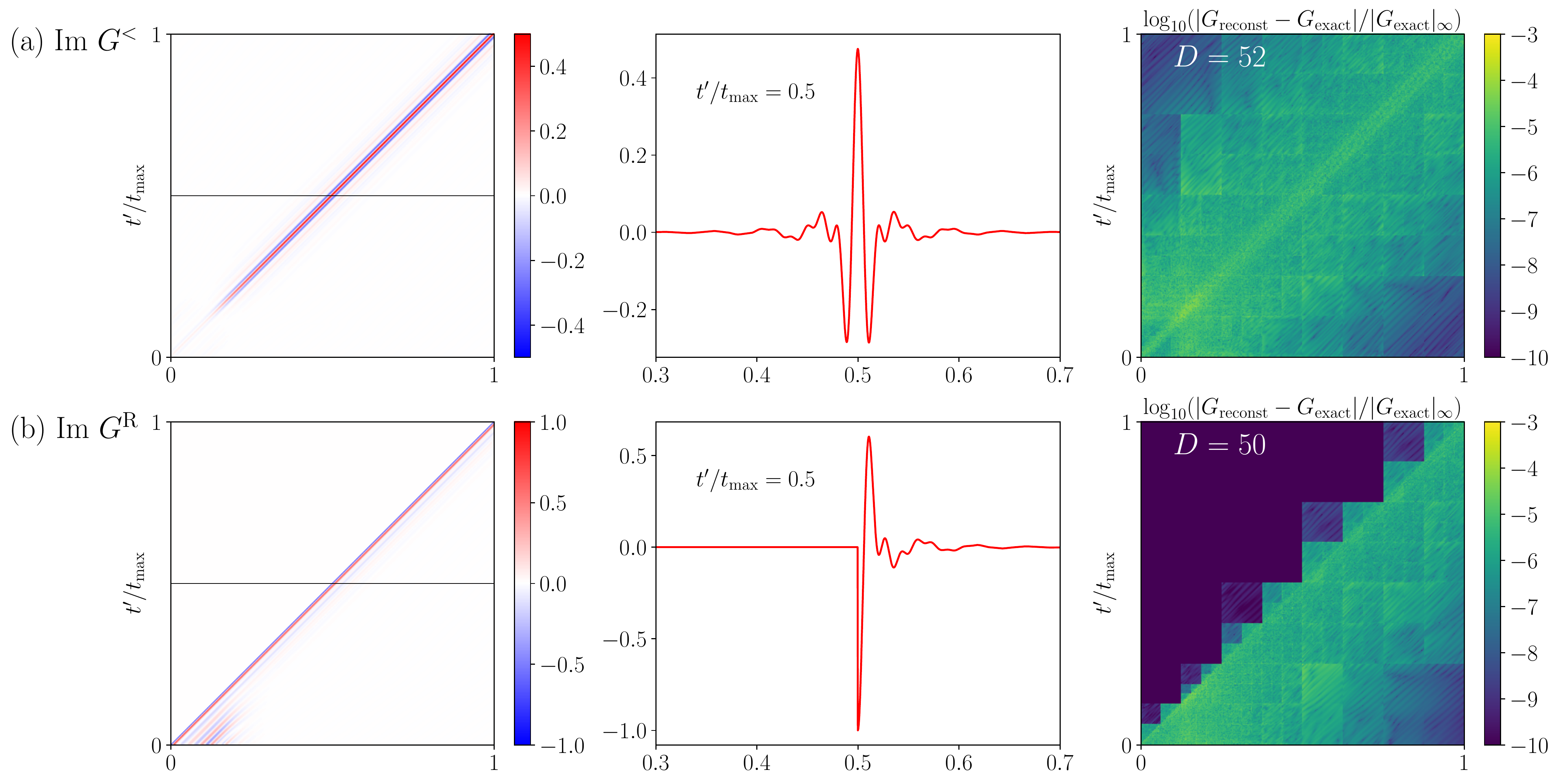}
    \includegraphics[width=\textwidth]{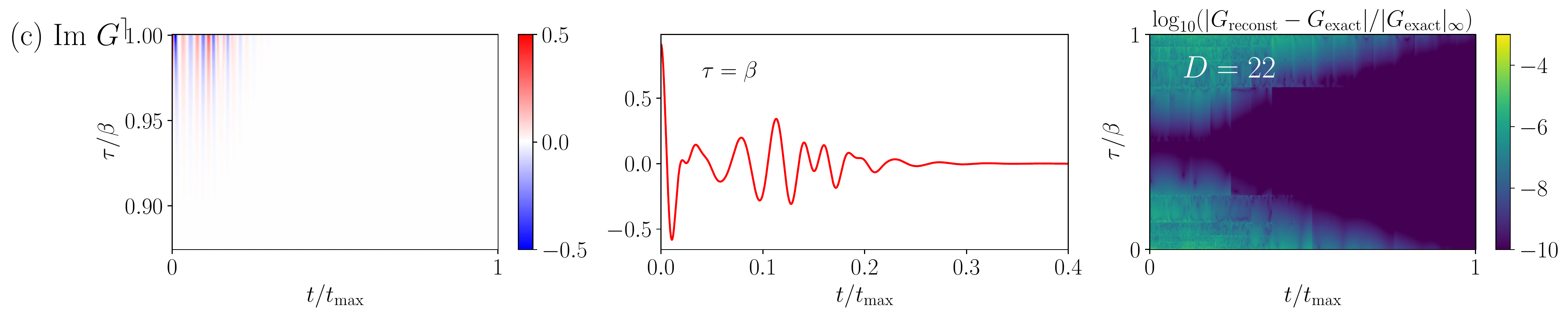}
    \vspace{-2em}
    \caption{
      Nonequilibrium Green's functions computed for the photo-excited antiferromagnetic system.
      See the caption of Fig.~\ref{fig:keldysh-eq} for the description of the panels.
    }
    \label{fig:keldysh-noneq2}
\end{figure*}

To illustrate {the QTT} compression in the nonequilibrium case, we focus on the single-band Hubbard model~\eqref{eq:Hubbard} in a time-dependent electric field
\begin{align} 
\mathcal{H}(t) = -\sum_{\langle ij\rangle} e^{i\phi_{ij}(t)} \hat{c}^\dagger_{i\sigma} \hat{c}_{j\sigma} + U \sum_i \hat{n}_{i\uparrow}\hat{n}_{i\downarrow}.
\end{align}
The electric field is included via a Peierls phase $\phi_{ij}$, which is the line-integral of the vector potential between the sites $i$ and $j$ \cite{Werner2018PRB}. 
We consider a half-filled system on the Bethe lattice, and calculate the Green's functions using the nonequilibrium (real-time) extension of  dynamical mean-field theory (DMFT) \cite{Aoki2014RMP}. 
Two representative cases will be analyzed: i) the paramagnetic Mott insulating system in equilibrium (see Fig.~\ref{fig:keldysh-eq}) and, ii) an initially  antiferromagnetic Mott insulating system which is excited with a short electric field pulse (see Fig.~\ref{fig:keldysh-noneq2}).
More specifically, we consider a Bethe lattice with infinite coordination number, which features a semi-circular density of states, and use the quarter of the bandwidth of the free system ($U=0$) as the unit of energy and $\hbar$ divided by the quarter of the bandwidth as the unit of time. We set the interaction to $U=6$ and use the non-crossing approximation to solve the effective impurity model in DMFT ~\cite{Eckstein2010b}. For calculation i), we choose the temperature $T=0.2$, while for ii), we use $T=0.05$, which is below the N\'eel temperature of the system. 
Setting $\hbar$, the bond length $a$, and the electron charge to unity, we choose the vector potential as $A(t) = \frac{E_0}{\Omega} F_{\rm G}(t,t_0,\sigma) \sin(\Omega(t-t_0))$ with $F_{\rm G}(t,t_0,\sigma) = \exp[-\frac{(t-t_0)^2}{2\sigma^2}]$.
The vector potential is related to the electric field $E(t)$ by $E(t)=-\partial_t A(t)$. We set $t_0=12,\sigma=3,\Omega=6,E_0=0.8$.

\begin{figure*}
\begin{minipage}[b]{0.48\textwidth}
    \centering
    \includegraphics[width=0.99\textwidth]{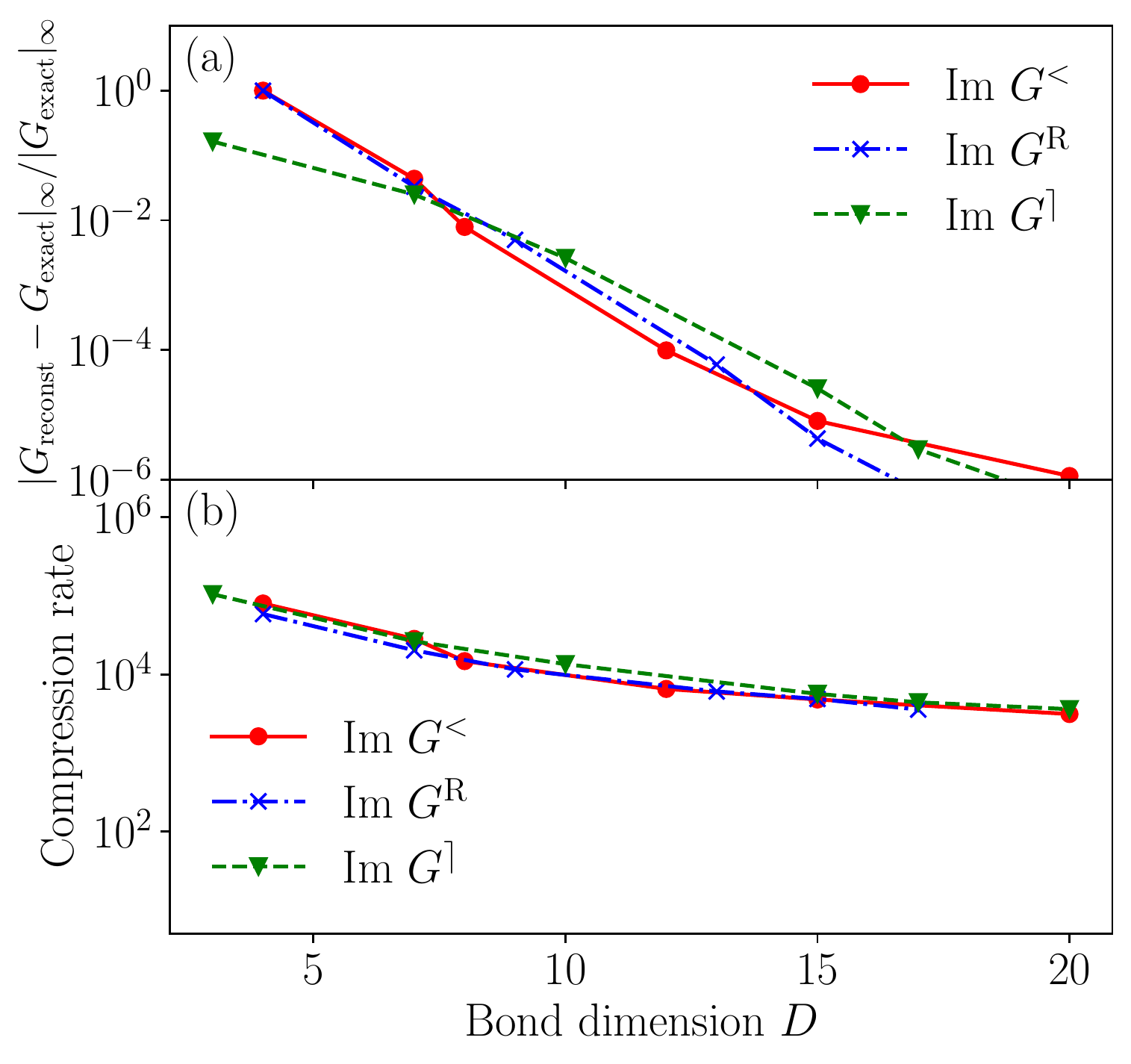}
\end{minipage}
\begin{minipage}[b]{0.48\textwidth}
    \centering
    \includegraphics[width=0.99\textwidth]{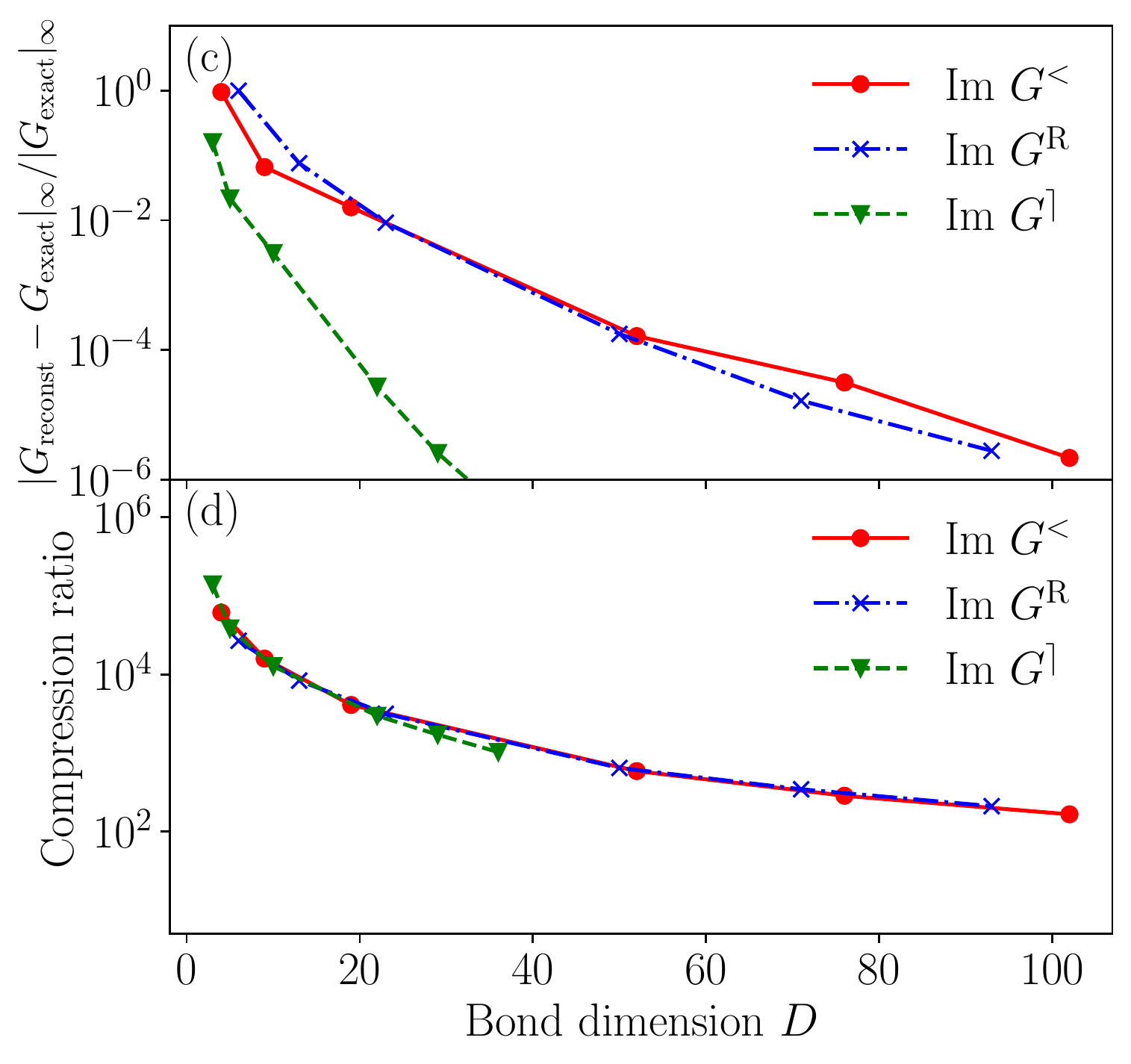}
\end{minipage}
\caption{Scaling of the relative accuracy [panels (a) and (c)] and compression rate [panels (b) and (d)] of the different components of the Green's function. The left panels are for the equilibrium case and the right panels for the nonequilibrium case. {The symbol $|\cdots|_\infty$ denotes the maximum norm of a tensor.}
}
\label{fig:keldysh-scaling}
\end{figure*}

The Green's functions are obtained using a time-stepping scheme which exploits the causal nature of the solution of the Kadanoff-Baym equations. We use $N_t=4096$ and $N_\tau=1024$, which means that the lesser and retarded components are stored on $(4096+1)^2/2=8.4\cdot 10^6$ grid points while the mixed component is stored on $(1024+1)\cdot (4096+1)=4.2\cdot 10^6$ grid points.
The left panels of Fig.~\ref{fig:keldysh-eq} and Fig.~\ref{fig:keldysh-noneq2}
show the imaginary parts of the interacting Green's function for the indicated components. The middle panels plot cuts at fixed $t'/t_\text{max}=0.5$ (indicated by a black line in the left panels) or $\tau/\beta=1$. We note that $\text{Im} G^<(t,t')$ is symmetric with respect to the diagonal $t=t'$, while $\text{Im} G^R(t,t')$ is nonzero only for $t>t'$ and features a jump of height 1 along the diagonal. The mixed component has a very different structure, since it connects to the Matsubara Green's function for $t\rightarrow 0$ and decays to zero for large $t$. We furthermore notice that the equilibrium lesser and retarded components are time translation invariant (i.e. they are functions of $t-t'$), while this is not the case for the pulse-excited system. The latter system features sharp peaks related to spin-polarons in the spectral function of the initial antiferromagnetic state~\cite{Sangiovanni2006}, and this leads to a slow decay in the retarded component away from the diagonal. At the same time, the weight in the lesser component (hole propagator) is suppressed since we plot the result for the minority-spin component. After the application of the pulse, the staggered  magnetization quickly melts, which leads to the disappearance of the spin polaron peaks and to half-filling ($\text{Im}G^<(t,t)=0.5$) for both spin components.

The right panels of Figs.~\ref{fig:keldysh-eq} and ~\ref{fig:keldysh-noneq2} demonstrate that despite the different qualitative features of the three components and the two distinct set-ups, {the QTT} compression scheme is capable of reproducing the Green's functions to high accuracy with modest bond dimensions. More specifically, in the case of the paramagnetic equilibrium system, a relative accuracy better than $|G_\text{reconstructed}-G_\text{exact}|_\infty/|G_\text{exact}|_\infty< 10^{-4}$ is achieved with bond dimensions $D~\sim 10$, while in the more challenging nonequilibrium case (with antiferromagnetic order in the initial state and a lack of translation invariance) a similar precision is reached with $D~\sim 50$. A noteworthy observation is that the compression scheme does not seem to encounter any difficulties in resolving the jump in the retarded component (compare the results for $\text{Im}G^<$ and $\text{Im}G^R$). While a representation in terms of average and relative times might look more natural in the case of the retarded component, the transformation to this representation can be achieved by {an MPO} of very small bond dimension, as discussed in Sec.~\ref{sec:argument}. 
Whether or not a reduction of the bond dimension can be realized by introducing variable transformations will be the subject of a separate study. 

To illustrate how the accuracy of the compressed Green's function improves with increasing bond dimension $D$, we plot in panels (a) and (c) of Fig.~\ref{fig:keldysh-scaling} the dependence of the maximum relative error on $D$ for the imaginary parts of the different components. The accuracy improves roughly exponentially with increasing $D$, both in the equilibrium case and in the nonequilibrium case. An interesting question is what these numbers imply for the memory requirement of {the QTT} representation and the compression rate. The results are shown in panels (b) and (d) of the figure. We see that in the equilibrium case, for bond dimension $D\sim 10$ and relative deviations smaller than $\sim 10^{-4}$, a compression rate of about $10^4$ is realized, which for example means that instead of $8.4\cdot 10^6$ data points for the lesser or retarded components, we need to store less than 1000 numbers. In the nonequilibrium case, the compression rate is lower, but still impressive. For $D\sim 50$, which again ensures relative deviations smaller than $\sim 10^{-4}$, the memory cost is reduced by approximately three orders of magnitude.  

\section{Computation}
\label{sec:computation}

In this section, we demonstrate how to perform basic operations for diagrammatic calculations in the compressed form.

\subsection{Fourier transform}
In this subsection, we discuss the Fourier transform between the Matsubara-frequency and imaginary-time domains.
As discussed in Sec.~\ref{sec:fft}, we precompute the {MPO} for the Fourier transform.
To test the numerical stability of the Fourier transform using {the MPO},
we consider the Green's function associated with a single pole,
\begin{align}
    G(\tau) &= - \frac{e^{-\tau \omega}}{1 + e^{-\beta \omega}},
\end{align}
where we take $\beta=100$ and $\omega=1$.
We first construct {an MPS} of bond dimension 1 for $G(\tau)$ using Eq.~\eqref{eq:pole-mps} with a given $R$.
Then, we apply the {MPO} to {the MPS} to obtain {an MPS} for $G(\iv)$, whose bond dimension is truncated using $\epsilon=10^{-20}$.

\begin{figure}
    \centering
    \includegraphics[width=0.925\columnwidth]{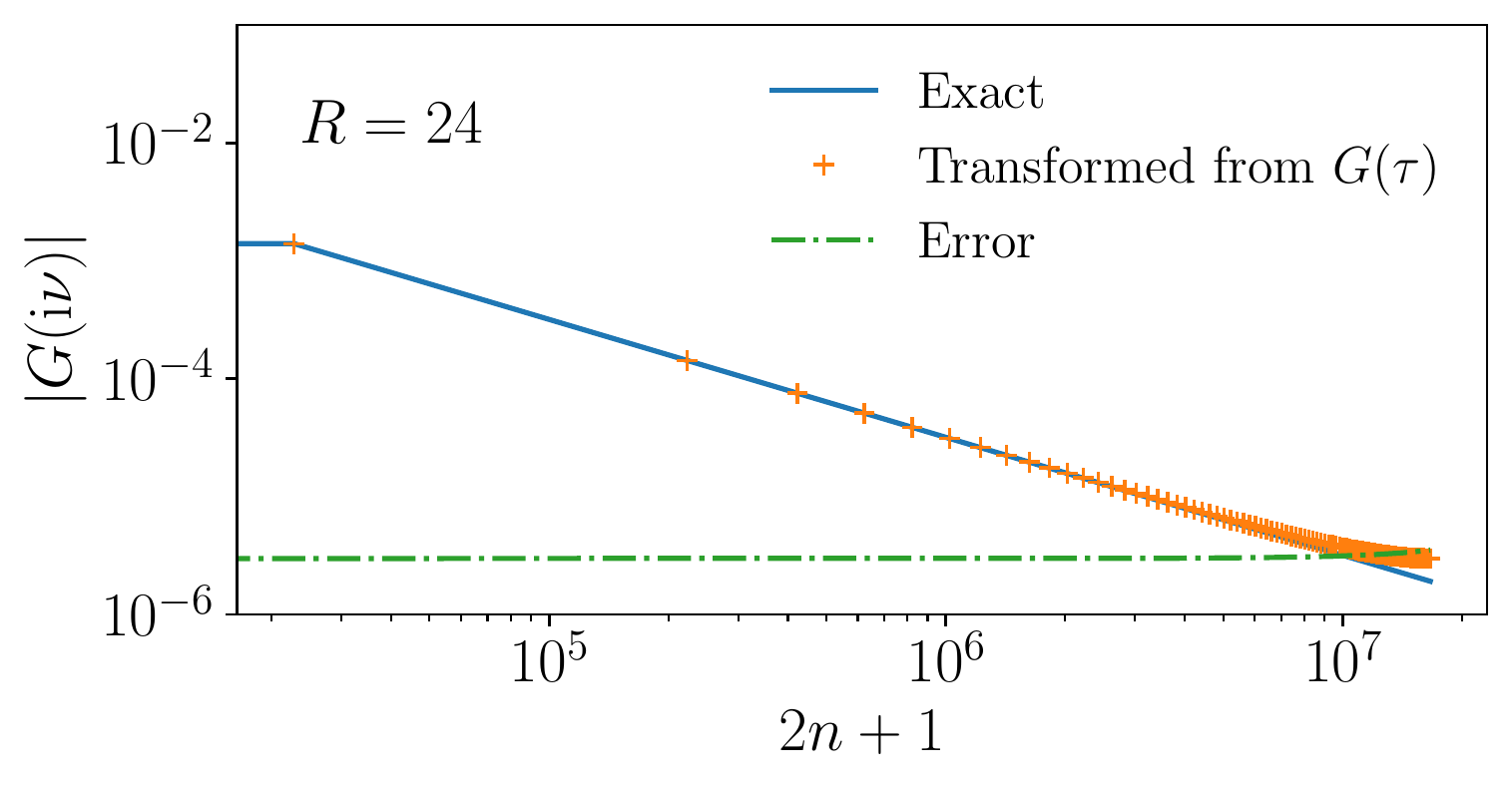}
    \vspace{-1.5em}
    \caption{
        Fast Fourier transform of the one-particle Green's function.
        $G(\iv)$ transformed from $G(\tau)$ [$\iv = \mathrm{i}(2n+1)\pi/\beta$].
        We plot the data for every $10^{5}$.
    }
    \label{fig:ft}
\end{figure}

Figure~\ref{fig:ft} shows the results for $R=24$, where we compare $G(\iv)$ reconstructed from {the MPS} to the exact values.
One can see that the error level is constant throughout the frequency mesh.
The error essentially originates from the discretization in $\tau$, which can be reduced exponentially by increasing $R$, as we will see later.

\begin{figure}
    \centering
    \includegraphics[width=0.925\columnwidth]{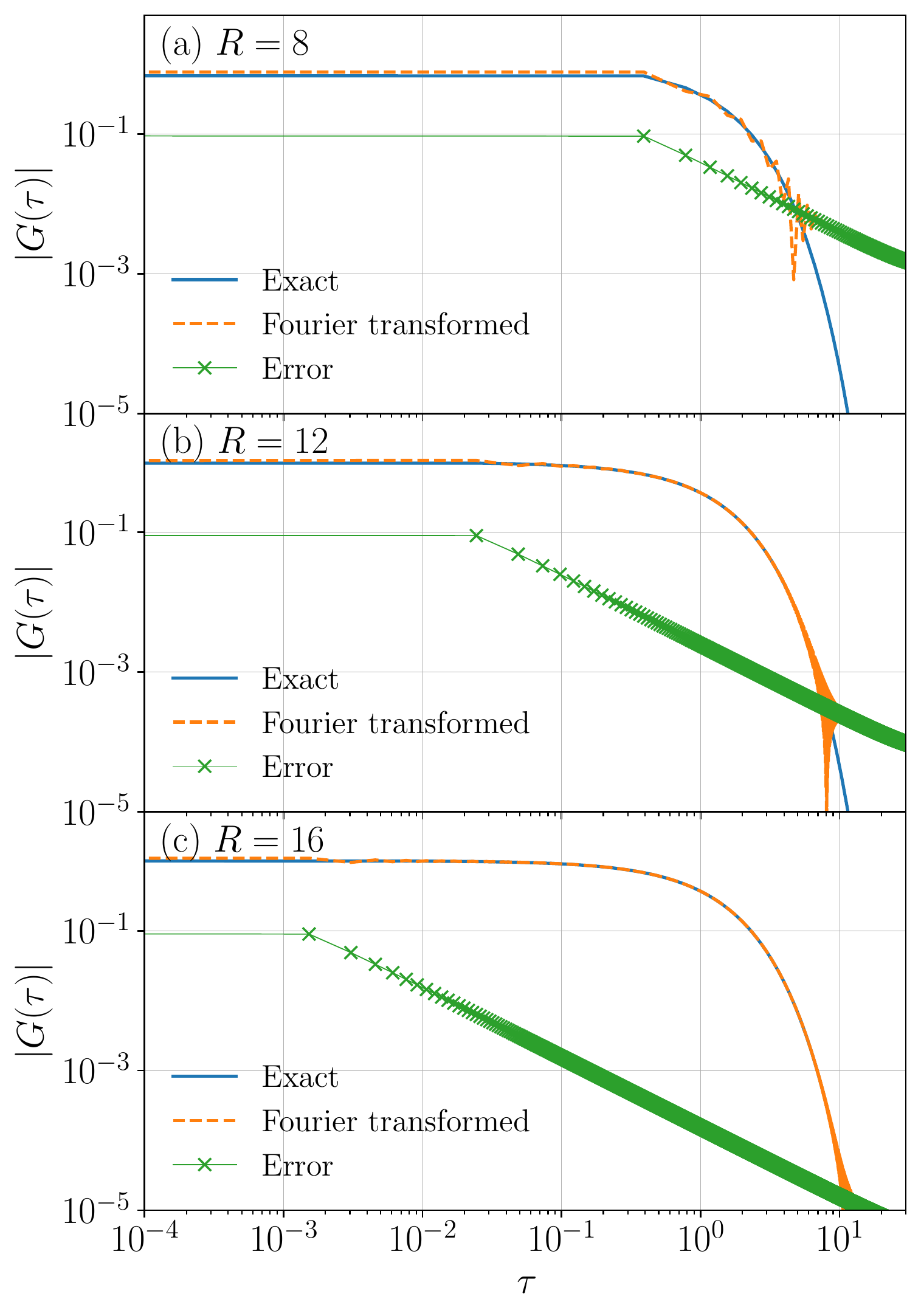}
    \vspace{-1.5em}
    \caption{
        Fourier transform of the Green's function to the $\tau$ domain for $R=8$, 12, 16.
        The other parameters are the same as in Fig.~\ref{fig:ft}.
    }
    \label{fig:ft-error}
\end{figure}

Next, we test the inverse Fourier transform from $G(\iv)$ to $G(\tau)$.
In practice, we decompose the numerical data of $G(\iv)$ on a mesh of size $2^R$ by SVD with $\epsilon=10^{-20}$.
Then, we apply the {MPO} of the inverse Fourier transform to {the MPS} of $G(\iv)$, yielding {an MPS} of $G(\tau)$.
Figure~\ref{fig:ft-error} shows the results for $R=8$, 12, 16.
The error around $\tau > \beta/2^R$ vanishes exponentially with increasing $R$, while
the error around $\tau=0$ stays almost constant.
The region with the large error vanishes exponentially in width with increasing $R$.

The large error at $\tau=0$ can be attributed to the truncation of the Matsubara sum: A discontinuity of $G(\tau)$ at $\tau=0$ cannot be reproduced by summing over a finite number of Matsubara frequencies.
To be more specific, for a finite $R$, the transformed $G(\tau=0)$ equals to $(G_\mathrm{exact}(\tau=0^+) + G_\mathrm{exact}(\tau=0^-))/2$, where $G_\mathrm{exact}$ is the exact Green's function.
If $G_\mathrm{exact}(\tau=0^+) - G_\mathrm{exact}(\tau=0^-) (\equiv \Delta) \neq 0$, the error at $\tau=0$ is larger than or equal to $|\Delta|/2$ for $R<\infty$.
This does not matter in practice since the error is localized in the exponentially narrow regions near $\tau=0$ and $\beta$.

\subsection{Dyson equation}
In this subsection, we describe how to solve the Dyson equation.
Without loss of generality, we restrict ourselves to 1D cases.
The Dyson equation can be expressed as
\begin{align}
    A(k) G(\iv, k) &= 1,\label{eq:dyson-diag}
\end{align}
where the linear operator $A(k)$ is defined as
\begin{align}
A(k) &\equiv \iv - \epsilon(k) - \Sigma(\iv, k).
\end{align}
For a fixed $\iv$, this equation can be expressed in compressed form as
\begin{align}
 A(k_1, \cdots, k_R) G(k_1,  \cdots, k_R) &= 1,
\end{align}
where
\begin{align}
 & A(k_1, \cdots, k_R) \nonumber\\
 & \equiv \iv~1_{k_1}\cdot(\cdots)\cdot 1_{k_R} - \epsilon(k_1, \cdots, k_R) - \Sigma(k_1, \cdots, k_R; \iv).\label{eq:A-mps}
\end{align}
Hereafter, we assume that the self-energy is given as {an MPS}.
The {MPS} for $\epsilon(k)$ can be constructed from the hopping matrix as follows.
The $\epsilon(k)$ ($k=0, \cdots, 2^R-1$) can be expressed as
\begin{align}
    \epsilon(k) &= \sum_{r=0}^{2^R-1} e^{\ii 2\pi k r/2^R} t_r,
\end{align}
where $t_r$ is the real-space hopping matrix.
In the present formalism,  the hopping ``matrix'' can be expressed as a $R$-way tensor of size $(2, 2, \cdots, 2)$.
For a tight-binding model, {an MPS} with a small bond dimension can be constructed explicitly for the hopping matrix as
\begin{align}
    \sum_{r'} t_{r'} T^{(1)}(r_R') \cdot (\cdots) \cdot T^{(R)}(r_1'),\label{eq:tr-mps}
\end{align}
where {the TT} tensor is defined as
\begin{align}
    (T^{(n)}(r_{R-n+1}'))_{a_{n-1}, a_n}^{r_{R-n+1}} &\equiv \delta_{r_{R-n+1}, r_{R-n+1}'}.
\end{align}
Note that the physical index is $r_n$ and $r_n'$ is an external tensor parameter.
The bond dimension of {the MPS} in Eq.~\eqref{eq:tr-mps} equals to or is smaller than the number of nonzero elements of $t_r$.
The bond dimension is only two for a nearest-neighbor 1D tight-binding model.
For a more complex hopping matrix, one may have to compress {the MPS} numerically.
The resultant {MPS} for the hopping matrix $t_r$ can be Fourier transformed numerically to $k$ space using the {MPO} of FFT.

Once {an MPS} for $A(k)$ is constructed, one can readily solve the Dyson equation ~\eqref{eq:dyson-diag} in QTT representation using a standard Krylov method.
In the following numerical demonstration, we transform {the MPS} of $A(k)$ to {an MPO} as described in Sec.~\ref{sec:elementwiseprod}.

\begin{figure}
    \centering
    \includegraphics[width=0.99\columnwidth]{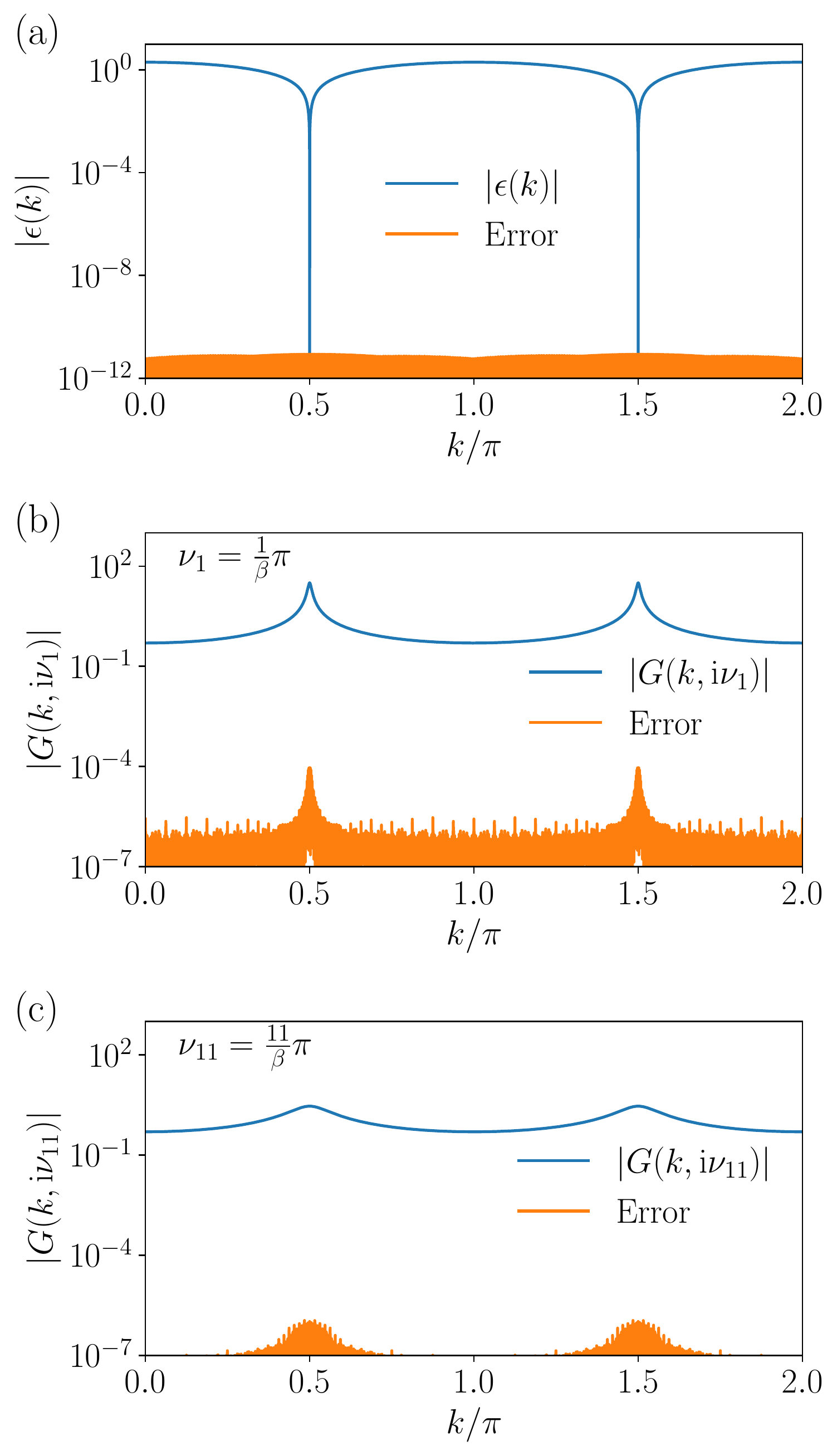}
    \caption{
    Solving the Dyson equation for the nearest-neighbor tight-binding model on the 1D lattice for $\beta=100$. (a) $\epsilon(k)$ computed by Fourier transform in TT form. (b) [(c)] Green's functions computed for $\nu=\frac{1}{\beta}\pi$ [$\nu=\frac{11}{\beta}\pi$].
    }
    \label{fig:dyson}
\end{figure}

As a simple case, we consider a nearest-neighbor tight-binding model on the 1D lattice. 
This case without self-energy is challenging because of sharp peaks in $G(k, \mathrm{i}\nu)$ at low frequencies and at the Fermi points.
Figure~\ref{fig:dyson}(a) shows $\epsilon(k)$ computed by the Fourier transform for $R=20$ without patching.
For constructing the {MPO} of the Fourier transform, we used the cutoff $\epsilon=10^{-25}$.
The resultant {MPO} for $\epsilon(k)$ has $D=2$.
The $\epsilon(k)$ was reconstructed on a grid of size $2^{20}=1048576$, which is compared with the exact result $\epsilon(k) = 2\cos(k)$. The noise level is constant over the whole interval.
The signal-to-noise ratio for $\epsilon(k)$ becomes worse around the Fermi points.
One can improve {MPSs} for $\epsilon(k)$ patch-wise to improve the signal-to-noise ratio.

Figures~\ref{fig:dyson}(b) and (c) show the Green's function computed by solving the Dyson equation in QTT representation for $\nu=\frac{1}{\beta}\pi$ and $\nu=\frac{11}{\beta}\pi$, respectively ($\beta=100$).
In particular, we used the generalized minimal residual method (GMRES)~\cite{Saad1986-ou} with cutoff $\epsilon=10^{-15}$ for truncating {MPSs} during the Krylov-subspace construction.
One can see that the reconstructed $G(k, \iv)$ matches the exact value accurately over the interval.
This proves the numerical stability of the present method.

The error becomes slightly larger around the Fermi points and at low frequencies, which can be attributed to the large signal-to-noise ratio in $\epsilon(k)$.
This issue becomes less serious at higher frequencies and presumably also in the presence of a self-energy.

\subsection{Bethe-Salpeter equation}\label{sec:bse}
We now solve the BSE for the Hubbard atom in {the TT} form.
We evaluate the RHS of Eq.~\eqref{eq:bse-SU2} using TT from the exact vertices and compare the resultant $F$ with the LHS (exact full vertex).
The evaluation of the RHS is done as follows for a fixed bond dimension $D$.

\begin{enumerate}
    \item Compute {MPSs} for $X^0$, $\Gamma$ and $F$ with the fixed bond dimension $D$.
    \item Compute {the MPS} of $\phi' \equiv X^0 F$.
    \item Compute {the MPS} of $\phi \equiv \beta^{-2} \Gamma \phi'$.
    \item Compute {the MPS} of $\Gamma + \phi$.
\end{enumerate}
At the end of Steps 2, 3, 4, we truncate the resultant {MPS} to the bond dimension $D$.
At the end of Step 4, we should have $\Gamma + \phi \simeq F$.

\begin{figure}
    \centering
    \includegraphics[width=0.99\columnwidth]{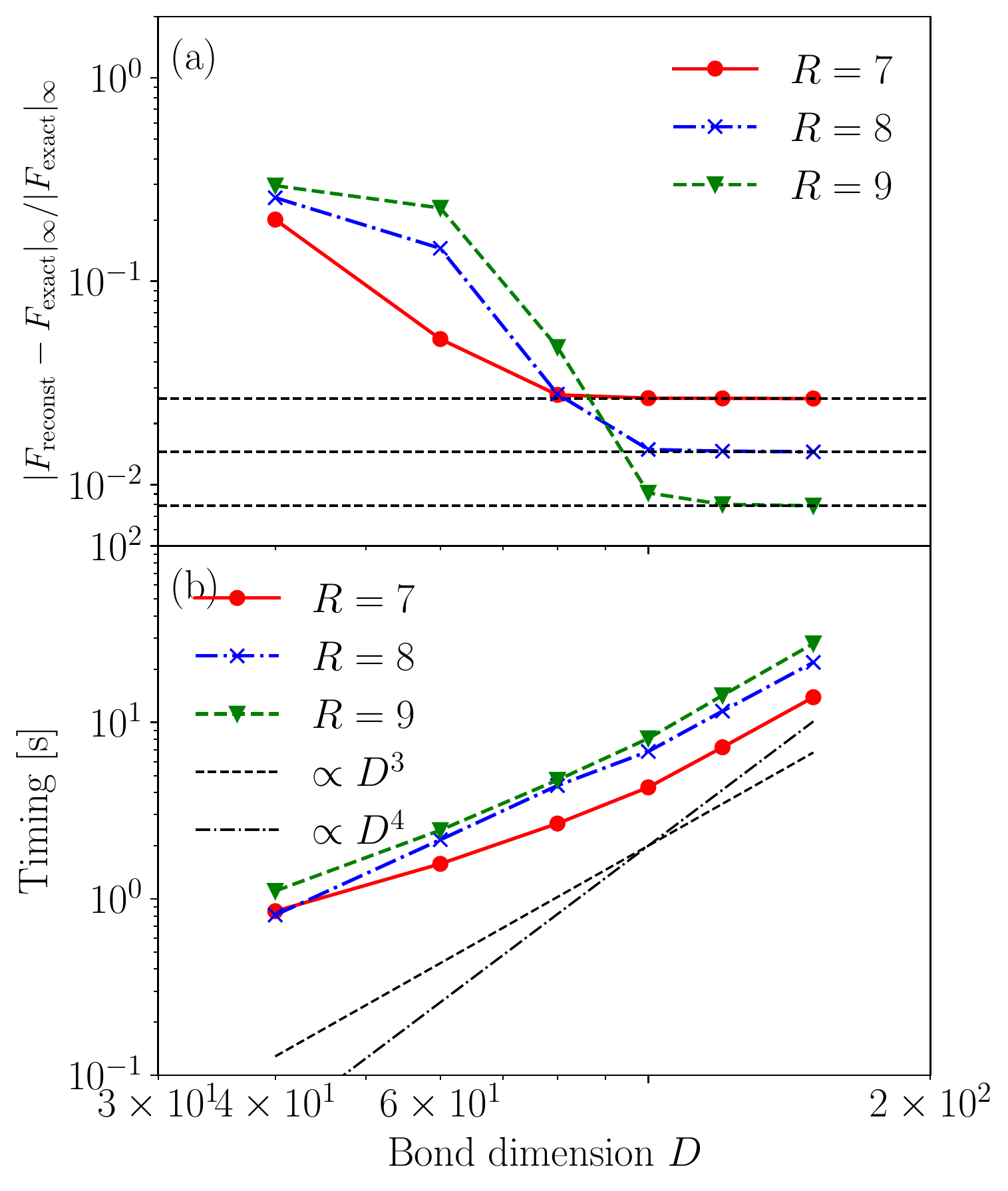}
    \caption{
        (a) Error in the reconstructed full vertex by solving the BSE, (b) timings of the one-shot solution of the BSE for $U=3$ and $R=7, 8, 9$.
        The BSE was solved effectively on a $2^R \times 2^R \times 2^R$ grid.
        The horizontal dashed line in (a) denotes the error level set by the finite-size effects of the grid (see the text).
        {The symbol $|\cdots|_\infty$ denotes the maximum norm of a tensor.}
    }
    \label{fig:atom-bse-scaling}
\end{figure}

Figure~\ref{fig:atom-bse-scaling} shows the results for $\beta U = 3$ and $R=9$ on a $2^R\times 2^R\times 2^R$ grid.
Figure~\ref{fig:atom-bse-scaling}(a) shows how the error in the result decays with increasing bond dimension $D$ for several values of $R$.
The error vanishes exponentially and eventually saturates due to the finite-size effect of the grid.
The finite-size error was estimated by performing the one-shot BSE calculation without compression directly on a fermionic-frequency mesh of size $2^R \times 2^R$ at each bosonic frequency using matrix multiplications.
Note that this finite-size error vanishes slowly as $\mathcal{O}(1/M^p)$ with the mesh size $M=2^R$ ($p=1$).
In the present approach, the finite-size error vanishes exponentially with $R$.

Figure~\ref{fig:atom-bse-scaling}(b) shows the timings of the present approach. We ran the code with 8 threads on an AMD EPYC 7702P 64-Core Processor. We performed the matrix multiplication using the fitting algorithm, whose computational cost scales as $\mathcal{O}(D^4)$.
One can see that the timings depend weakly on $R$ as expected.
The timings grow slightly slower than the expected scaling $\mathcal{O}(D^4)$.
This indicates that the bond dimensions are still too small to see the asymptotic scaling.

For a fixed temperature, the run time of the overcomplete IR method~\cite{Wallerberger2021-kv} scales as $\mathcal{O}(L^8)$, where $L\propto -\log \epsilon$.
For a fixed box size $R$, the run time of the present method grows only as $\mathcal{O}(L^4 N)$, where $L \propto -\log \epsilon_\mathrm{MPS}$, $N \propto -\log \epsilon_\mathrm{box}$,
$\epsilon_\mathrm{MPS}$ is the target accuracy for compressing {MPSs}, and $\epsilon_\mathrm{box}$ is the target accuracy for the finite-size error of the grid.
Thus, the present approach is asymptotically superior to the overcomplete IR method for high target accuracy.

\section{Summary and conclusion}
\label{sec:conclusions}
In summary, we proposed {a multi-scale space-time ansatz} for the correlation functions of quantum systems based on {quantics tensor trains (QTT)}. We numerically established the validity of the ansatz for the space-time dependence of correlation functions in various equilibrium and nonequilibrium systems. Furthermore, we proposed efficient algorithms for basic operations required for diagrammatic calculations. In particular, we showed that the Fourier transform can be performed using {a matrix product operator (MPO)}, with a small bond dimension.
Finally, we numerically demonstrated the computation of the Fourier transform, solution of the Dyson equation and evaluation of the Bethe-Salpeter equation.

This study will open a new route to efficient computations of quantum field theories using classical computers.
A possible application is the numerical integration of high-order perturbation series, which has been recently attacked by a low-rank tensor train approximation using tensor cross interpolation formula without the multi-scale ansatz~\cite{Fernandez2022}. 
It is of interest to test the efficiency of more sophisticated tensor networks, such as tree tensor networks and multi-scale tensor networks.

In {the present QTT approach}, diagrammatic calculations are mapped to standard operations {in the TT format}, which can be parallelized using many GPUs and/or CPUs.
{The QTT representation} is capable of treating high dimensional frequency and momentum domains in a straightforward way.

There are many possible applications of {QTT}.
In equilibrium calculations,
handling 2P quantities with three frequencies and momenta requires  huge computational  and memory costs.
This limits the application of sophisticated diagrammatic approaches at the 2P level and makes it unfeasible to address  low-temperature phenomena in real materials.
Examples include parquet(-like) equations ~\cite{Yang2009-ul, Tam2013-ez, Augustinsky2011-bq, Janis2017, Li2019,Pudleiner2019,Astretsov2020,Eckhardt2020,Krien2020a, Krien2020b, Krien2022} and the DW equation~\cite{Tazai2022,Kontani2022,Onari2016-sp}.
For the DFT+DMFT method, a challenging issue is the computation of 2P response functions~\cite{Hafermann2014,Shinaoka:2020ji,Kunes2020, Kunes2021, CeB6}.
Non-local extensions of DFT+DMFT~\cite{Toschi2011-iy, Otsuki2014,Valli2015-er, Rohringer2018-gt} require efficient treatment of BSE and parquet equations as well.
\textit{Ab initio} fRG~\cite{fRG_ab_initio} and downfolding beyond the constrained RPA method (cRPA)~\cite{Honerkamp2012-zc, Maier2012-bx, Kinza2015-up, Honerkamp2018-pr,Van_Loon2021-rw} are other interesting targets.

\textit{Ab initio} calculations where the bands span a wide energy range require an efficient treatment of internal degerees of freedom, which may be enabled by the {the QTT} representation.
Vertex corrections to the Migdal-Eliasberg equation~\cite{Schrodi2020-le} and \textit{GW}+BSE~\cite{Maggio2017} could also be addressed.

{
Recent work on the analytic structure of multi-point correlation functions~\cite{Kugler2021-gq,Lee2021-ho} is in principle orthogonal to the compression
strategies presented in this work.  It is however intriguing to explore synergies,
for example, by trying to represent the partial spectral function representations
using {MPSs} rather than the full object.
}

In nonequilibrium simulations, the huge memory cost of storing momentum-dependent two-time Green's functions has restricted lattice calculations based on the $\Sigma^{(2)}$ \cite{Tsuji2014}, FLEX \cite{Stahl2021} or two-particle self-consistent (TPSC) \cite{Simard2022} approaches to short times and coarse momentum resolutions. An interesting aspect of correlated nonequilibrium systems however is the emergence of distinct behaviors on different timescales \cite{Aoki2014RMP}, as exemplified by the concepts of prethermalization \cite{Berges2004,Moeckel2008} and nonthermal fixed points \cite{Berges2008,Tsuji2013}. The {QTT representation} essentially eliminates the memory bottleneck and it should enable new implementations of the diagrammatic equations which give access to slow dynamics and provide insights into the role of nonlocal correlations. 

While current implementations of nonequilibrium calculations rely on a time-stepping scheme~\cite{Schuler2020-tq}, it is rather cumbersome to combine this strategy with compressed representations of nonequilibrium Green's functions~\cite{Kaye2021-ly}. The routines discussed in this work suggest that it may be advantageous to give up the time-stepping and to implement the calculations on a fixed time contour using {the QTT} representations. How this affects the numerical stability and convergence properties of the simulations is an interesting subject for future studies. 

It is also interesting to explore the theoretical possibility of implementing diagrammatic calculations using a real quantum computer. This may allow to handle difficult cases leading to a large bond dimension with {QTT}. It however requires the implementation of non-unitary operations such as element-wise multiplication and convolutions using a unitary quantum circuit. This remains a challenging problem in quantum information theory.

{\it Note added}. While finalizing this manuscript, we became aware of an independent work~\cite{Jielun2022} where an upper bound of the bond dimension of the {MPO} for Fourier transform was analytically derived.

\begin{acknowledgments}
H.S. was supported by JSPS KAKENHI Grants No. 18H01158, No. 21H01041, No. 21H01003, and 23H03817, and JST PRESTO Grant No. JPMJPR2012, Japan.
Y.M. was supported by JSPS KAKENHI Grants No. JP20K14412 and No. JP21H05017 and by JST CREST Grant No. JPMJCR1901.
K.N. was supported by JSPS KAKENHI Grants No. JP21J23007. 
P.W. acknowledges support from ERC Consolidator Grant No. 724103. A.K. acknowledges support by Austrian Science Fund (FWF) through Projects No. P 32044 and P 36213.
H.S thanks K. Yoshimi, J. Otsuki, T. Koretsune, Y. Yanase, T. Okubo, M. Kitatani, Y. Nagai, W. Mizukami, Y. Yamaji, and N. Witt for the fruitful discussions.
H.S especially thanks J. Otsuki for providing the numerical raw data of his DFT+DMFT calculation.
We carried out part of the calculations using computer code based on \texttt{SparseIR.jl}~\cite{sparse-ir} and \texttt{ITensors.jl}~\cite{Fishman2022-mu} written in \texttt{Julia}~\cite{bezanson2017julia}.
H.S and M.W gratefully thank E. Miles Stoudenmire for fruitful discussions and his advice on implementing our code using \texttt{ITensors.jl}. M.W. and A.K. sincerely thank Jan Kune\v{s} for illuminating discussions. 
We used \texttt{NESSi}~\cite{Schuler2020-tq} to generate the real-time Green's function data.
\end{acknowledgments}

\appendix

\section{Matrix product states (MPS)}
\label{sec:mps}
\begin{figure*}
    \centering
    \includegraphics[width=0.95\textwidth]{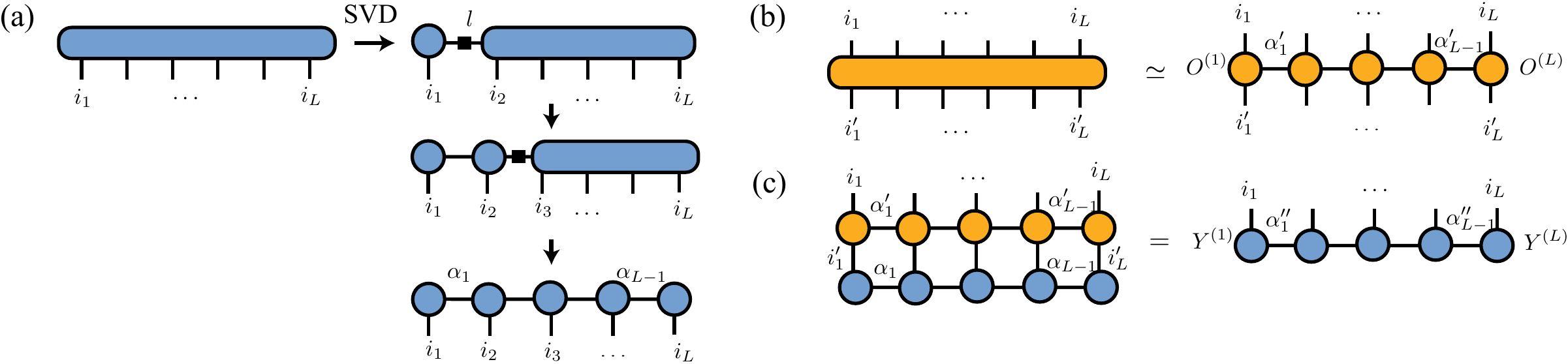}
    \caption{
    (a) Decomposing a tensor into a tensor train or matrix product state,
    by SVD, yielding {a matrix product state (MPS)} of the dependence on $i$. The filled square denotes a diagonal matrix consisting of singular values. The singular values are absorbed into the right singular matrix.
    {(b) Matrix product operator (MPO).}
    (c) Multiplication of {an MPO and an MPS}.
    We do not show the dummy virtual bonds at the edges for simplicity.
    }
    \label{fig:mps}
\end{figure*}
Here we provide a brief overview of tensor trains, which we call {\em tensor train states} to disambiguate them from operators introduced in the next section.
For a comprehensive review, we refer the reader to Refs.~\onlinecite{Schollwock2011-eq,Oseledets2011-uj}.

Let $A$ be an $n_1 \times \cdots \times n_L$-tensor, i.e., an object with $L$ indices $i_1, \ldots, i_L$. We will write this as $A(i_1, i_2, \ldots, i_L)$. {A matrix product state (MPS)} is an approximation of the single $L$-way tensor $A$ by a contraction of $L$ three-way tensors $A^{(1)}, \ldots, A^{(L)}$:
\begin{align}
    & A(i_1, i_2, \cdots, i_L)\nonumber\\
    & \quad\simeq \sum_{\alpha_0=1}^{D_0} \cdots \sum_{\alpha_L=1}^{D_L}
    A^{(1)}_{i_1,\alpha_0\alpha_1}
    A^{(2)}_{i_2,\alpha_1\alpha_2}
    \cdots
    A^{(L)}_{i_L,\alpha_{L-1}\alpha_L} \nonumber\\
    & \quad\equiv A^{(1)}_{i_1} \cdot (\cdots) \cdot A^{(L)}_{i_L},
    \label{eq:mps}
\end{align}
where $A^{(l)}$ is now an auxiliary $n_l \times D_{l-1} \times D_l$ tensor.
Aside from the outer or physical indices $i_1, \ldots, i_L$, we have
introduced dummy or virtual indices $\alpha_0, \ldots, \alpha_L$.
The index $\alpha_l$ thereby forms a ``bond'' between the tensors $A^{(l-1)}$
and $A^{(l)}$, hence its bound $D_l$ is called bond dimension.
(By definition, $D_0 = D_L = 1$.) The bond dimension $D$ of the {MPS}
is defined as the largest bond dimension of its constituents, $D = \max_l D_l$.  With the outer indices held fixed, the virtual indices chain the tensors $A^{(l)}$ into a matrix product, which informs the condensed
notation $A^{(1)}_{i_1} \cdot (\cdots) \cdot A^{(L)}_{i_L}$ for the {MPS}.

Any tensor can be decomposed into {an MPS}.
As illustrated in Fig.~\ref{fig:mps}(a), we can reshape the original tensor $A$ into an $n_1 \times (n_2 \cdots n_L)$ matrix:
\begin{equation}
    A(\i_1, \underbrace{i_2, i_3 \ldots, i_L}_J) \equiv A(i_1, J).
    \label{eq:A_matrix}
\end{equation}
By means of a singular value decomposition (SVD), we can detach the first tensor $A^{(1)}$
from $A$:
\begin{equation}
    A(i_1, J) = \sum_{\ell=1}^{D'_1} (U)_{i_1\ell} s_\ell (V^\dagger)_{\ell J} \equiv A^{(1)}_{i_1}\cdot A'(i_2,\ldots,i_L),
    \label{eq:A_SVD}
\end{equation}
where $s_1 \ge s_2 \ge \ldots \ge 0$.  Iterating Eqs.~(\ref{eq:A_matrix}) and (\ref{eq:A_SVD}) on $A'$, we obtain {an MPS}~(\ref{eq:mps}).

The bond dimension $D'_k$ obtained thusly is given by $D'_k = \min(n_1\cdots n_k, n_{k+1}\cdots n_K)$,
i.e. it grows exponentially as we move from the edges toward the center of the {MPS}.  Fortunately, in many interesting cases, different indices $i_k$ are not strongly entangled, making $D_k$ a very
loose bound.  To utilize this, we approximate $A$ in Eq.~(\ref{eq:A_SVD}) with its low-rank approximation
$\tilde A$:
\begin{equation}
    A(i_1, J) \approx \tilde A(i_1, J) = \sum_{\ell=1}^{D_1} (U)_{i_1\ell} s_\ell (V^\dagger)_{\ell J}
    \label{eq:A_trSVD}
\end{equation}
where $D_1 \le D'_1$, i.e., we simply discard the smallest singular values in Eq.~(\ref{eq:A_SVD}).  The error of this approximation is usually taken to be with respect to the Frobenius norm:
\begin{equation}
    \epsilon = \frac{|| A - \tilde A ||^2_F}{||A||^2_F} = \frac{\sum_{\ell=D_1+1}^{D'_1} s^2_\ell}{\sum_{\ell=1}^{D'_1} s^2_\ell},
    \label{eq:epsilon}
\end{equation}
and can be shown to be optimal for a given rank $D'_1$ by virtue of the SVD.
(We follow common MPS convention and express the approximation error in terms of
the {\em squared} deviation.)

Equations~(\ref{eq:A_trSVD}) and (\ref{eq:epsilon}) now provide us with a way to construct a (lossily) compressed
form of the {MPS}.  We start with an error bound $\epsilon$ and optionally a maximum bond dimension $D$.
Instead of the SVD (\ref{eq:A_SVD}), we use the truncated SVD (\ref{eq:A_trSVD}) to detach the first
tensor $A^{(1)}$ from $A$, ensuring that the approximation error (\ref{eq:epsilon}) stays below
our error bound.  We then iterate (``sweep'') through the indices to construct the truncated {MPS}.
The {MPS} obtained this way is optimal and we will refer to it as {an MPS} in its {\em canonical} form. 
{An MPS} in canonical form cannot be compressed further without sacrificing further accuracy.

The above procedure is not only useful in constructing the {MPS}, but also
for ``recompressing'' {an MPS} of bond dimension $D'$ into one of bond dimension $D \le D'$.
This is necessary because intermediate results arising from, e.g., element-wise products of two {MPSs},
are not canonical, i.e., these computations yield {an MPS} with a large bond dimension but with
large redundancy.  The recompression algorithm proceeds along the lines of Eq.~(\ref{eq:A_trSVD}) and (\ref{eq:epsilon}), but is slightly more involved, which is why we do not detail it here.
We just remark that truncating the bond dimension from $D^\prime$ to $D$ costs $\mathcal{O}(n D^{\prime 3}L)$ computational time for $D^\prime \gg D$, where $n$ is the maximum dimension of physical indices.

\section{Matrix product operator (MPO)}
\label{sec:mpo}

{Matrix product operators (MPOs)} are a natural generalization of the {MPS} concept to operators.
{An MPO} is a decomposition of a $n_1 \times n_1' \times \cdots \times n_L \times n_L'$ ($2L$-way) tensor $O$ into a product of $L$ four-way tensors $O^{(1)}, \ldots, O^{(L)}$:
\begin{align}
    & O^{i_1, \cdots, i_L}_{i_1', \cdots, i_L'}\nonumber\\
    & \quad\simeq \sum_{\alpha_0=1}^{D_0} \cdots \sum_{\alpha_L=1}^{D_L}
    O^{(1)}_{i_1i'_1,\alpha_0\alpha_1}
    O^{(2)}_{i_2i'_2,\alpha_1\alpha_2}
    \cdots
    O^{(L)}_{i_Li'_L,\alpha_{L-1}\alpha_L}
    \nonumber\\
    & \quad\equiv O^{(1)}_{i_1i'_1} \cdot (\cdots) \cdot O^{(L)}_{i_Li'_L},
    \label{eq:mpo}
\end{align}
where $O^{(l)}$ is now an auxiliary $n_l \times n'_l \times D_{l-1} \times D_l$ tensor,
$\alpha_0, \ldots, \alpha_L$ are again the bond indices, $D_0,\ldots, D_L$ the bond dimensions,
$D_0 = D_L = 1$, and $\cdot$ again is shorthand for the contraction
along the bond dimension.  We illustrate Eq.~(\ref{eq:mpo}) in Fig.~\ref{fig:mps}(b).

Crucially, {an MPO}~(\ref{eq:mpo}) can be applied to {an MPS}~(\ref{eq:mps})
``tensor-by-tensor'':
\begin{align}
    & \sum_{i_1', \cdots, i_L'} O^{i_1, \cdots, i_L}_{i_1', \cdots, i_L'} X(i_1', \cdots, i_L')\nonumber\\
    &\quad = \sum_{i_1', \cdots, i_L'} [O^{(1)}_{i_1i_1'} X^{(1)}_{i_1'}]
             \cdot (\cdots) \cdot 
             [O^{(L)}_{i_Li_L'} X^{(L)}_{i_L'}]
    \nonumber\\
    &\quad \equiv Y^{(1)}_{i_1} \cdot (\cdots) \cdot Y^{(L)}_{i_L},
    \label{eq:mpo-times-mps}
\end{align}
where $\cdot$ now indicates contraction over the bond indices of both $O$ and $X$.
We illustrate Eq.~(\ref{eq:mpo-times-mps}) in Fig.~\ref{fig:mps}(c).

Eq.~(\ref{eq:mpo-times-mps}) shows that the result tensor $Y$ can be expressed in {MPS} form.
However, its bond dimension is as large as $D' \times D$, where $D'$ and $D$ are the bond dimensions of $O$ and $X$, respectively.
The resultant {MPS} is not in canonical form and thus can be compressed significantly.
The compression by SVD would cost $\mathcal{O}(D^3 (D')^3)$ operations, which is usually inefficient.
The multiplication and subsequent compression can be done simultaneously using the density-matrix method or the fitting method~\cite{Stoudenmire2010-ya} where we avoid creating an intermediate {MPS} with a large bond dimension. In the present study, we use the fitting method.
The computation time scales as $\mathcal{O}(D^5)$ and $\mathcal{O}(D^4)$ for these two algorithms, respectively, when $D'\simeq D$.

In solving diagrammatic equations, one may have to evaluate
element-wise products
$C(t,t') = A(t,t')B(t,t')$
or
convolutions
$C(t,t') = \int \dd t'' A(t,t'')B(t'',t')$.
As we describe in Secs.~\ref{sec:elementwiseprod} and \ref{sec:matmul}, these operations on two {MPSs} can be recast into {an MPO--MPS} multiplication.

One can perform many operations in {the MPS} form.
For instance, one can add two {MPSs} with bond dimensions $D_1$ and $D_2$, where the resultant {MPS} has bond dimension $D_1 + D_2$. The resultant {MPS} can be compressed by SVD efficiently, at a computational cost which scales as $\mathcal{O}((D_1 + D_2)^3)$.
Thus, addition is usually computationally cheap compared to multiplication.
The same applies to the addition of two {MPOs}.

\section{Fast Fourier transform}\label{appendix:fft}
\label{app:fft}
We are attempting to construct {matrix product operator (MPO)} $\DFT^{k_1\ldots k_R}_{x_R\ldots x_1}$ for the discrete Fourier transform:
\begin{align}
    f(k) &= \sum_{x=0}^{2^R-1} \exp(\frac{2\pi\ii}{2^R}kx) f(x), \label{eq:dft}\\
    \hat F^{(1)}_{k_1} &\cdots \hat F^{(R)}_{k_R} =
        \sum_{\{x_r\}} \DFT^{k_1\ldots k_R}_{x_R\ldots x_1}
        F^{(R)}_{x_R} \cdots F^{(1)}_{x_1}.
    \label{eq:dft-mpo}
\end{align}

We start with the definition of the discrete Fourier transform on a grid of size $2N$:
\begin{equation}
    \hat f(k) = \sum_{x=0}^{2N-1} \exp(\frac{2\pi\ii}{2N}kx) f(x),
    \label{eq:fft}
\end{equation}
where $f(x)$ is the discrete real-space signal, $\hat f(k)$
is the corresponding momentum space function, and $x, k \in \{0, \ldots, 2N-1\}$.
The standard Cooley--Tukey algorithm reduces a discrete Fourier transform (\ref{eq:fft})
of size $2N$ to two discrete Fourier transforms of size $N$:
\begin{equation}
    \hat f(k + \kappa N)
        = \sum_{\xi=0}^1 \ee^{\frac{\pi\ii}N \xi(k + \kappa N)}
            \sum_{x=0}^{N-1} \exp(\frac{2\pi\ii}{N}kx) f(2x+\xi),
    \label{eq:cooley-tukey}
\end{equation}
where now $x, k \in\{0, \ldots, N-1\}$ and $\kappa, \xi \in \{0, 1\}$.
Iterating Eq.~(\ref{eq:cooley-tukey}) yields the fast Fourier transform
(FFT) for problem sizes $2N=2^R$ for some integer $R$.

Let us again expand $f(x)$ and $\hat f(k)$ as {MPSs}, which
we reproduce here:
\begin{subequations}%
\begin{align}
    \hat f(k) &= \hat F^{(1)}_{k_1}\cdot \hat F^{(2)}_{k_2} \cdot (\cdots) \cdot \hat F^{(R)}_{k_R}, \\
    f(x) &= F^{(R)}_{x_R}\cdot F^{(R-1)}_{x_{R-1}} \cdot (\cdots) \cdot F^{(1)}_{x_1},
\end{align}
\label{eqs:tensors}%
\end{subequations}
where $x=(x_1\cdots x_R)_2$ and $k=(k_1\cdots k_R)_2$ are again the bits of $x$ and $k$, respectively, and $\cdot$ denotes the contraction of the 
matrices $f$ or $\hat f$ along the bond dimension.

Note again that the order of tensors is reversed in $f$ with respect to $\hat f$.
This circumvents
the necessity of bit reversals in traditional FFT.  Using our convention instead, we
can identify $\xi = x_R$ and $\kappa = k_1$ in Eq.~(\ref{eq:cooley-tukey}),
allowing us to perform the FFT from ``left to right'' in both $f$ and $\hat f$.
{Empirically, we found that including the bit reversal adds a large amount of entanglement, which is why its avoidance is critical in this case.}

We are attempting to construct {an MPO} for the discrete FT in Eq.~\eqref{eq:dft-mpo}.
Imposing the tensor structure (\ref{eqs:tensors}) on the discrete FT (\ref{eq:cooley-tukey}),
we obtain the following recurrence for the {MPO} (\ref{eq:dft-mpo}):
\begin{equation}
    \DFT^{k_1\ldots k_R}_{x_R\ldots x_1}
    = \prod_{r=1}^{R} \exp(\frac{2\pi\ii}{2^r} x_R k_r)
    \DFT^{k_2\ldots k_R}_{x_{R-1}\ldots x_1}.
    \label{eq:fft-recursive}
\end{equation}
Unwinding the recurrence (\ref{eq:fft-recursive}), we obtain:
\begin{equation}
    \DFT^{k_1\ldots k_R}_{x_R\ldots x_1} =
        \prod_{r=1}^R \prod_{s=1}^r \exp(\frac{2\pi\ii}{2^s} x_r k_s),
    \label{eq:fft-expl}
\end{equation}
in other words, simply a collection of phase shifts applied
whenever some bits in both $k$ and $x$ are set.

\begin{figure}[b]
    \includegraphics{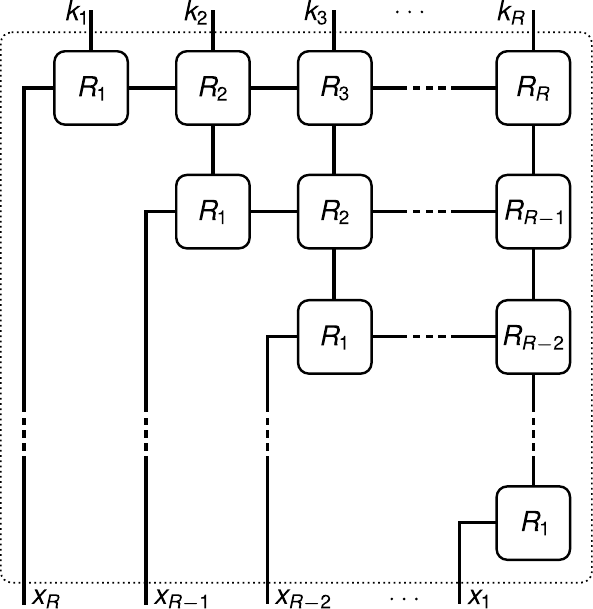}
    \caption{
        {MPO} for the discrete Fourier transform (\ref{eq:dft-mpo}) of {an MPS}.
    }
    \label{fig:fft-tn}
\end{figure}

In order to write the DFT as a tensor network, we introduce a set of phase shift tensors:
\begin{equation}
    \vcenter{\hbox{\includegraphics{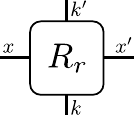}}}
    \equiv \exp\left(\frac{2\pi\ii}{2^r} xk\right) \delta_{xx'} \delta_{kk'},
    \label{eq:Phiphi}
\end{equation}
and also three- and two-legged versions where removing a leg
corresponds to removing the associated dependency and delta function.
Using Eq.~(\ref{eq:Phiphi}), we can rewrite the {MPO} (\ref{eq:fft-expl}) in its tensor network form, depicted in Fig.~\ref{fig:fft-tn}.
Conceptually, this diagram clarifies the structure of the FFT---leveraged in high-performance
libraries such as FFTW---as a network of simple $2\times 2$ kernels together with a permutation
of inputs.  More importantly, it can be applied efficiently to {an MPS} layer-by-layer with intermediate compression steps.
Alternatively, one can construct {an MPO} for the whole steps recursively.
The essentially idential quantum circuit was already derived in Ref.~\cite{Yoran2007-ek,Holzapfel2015-ru} in the context of quantum information theory.

\section{Matsubara-frequency mesh}
\label{app:mesh}
We define the grid points as $(2n+\xi)\pi T$ ($n=-N/2, -N/2+1, \cdots, N/2-1$), where $\xi=0,1$ for bosons and fermions, respectively.
This choice has the advantage that all the frequencies are sorted in ascending order.
For this convention, the {MPO} for the Fourier transform from imaginary times to Matsubara frequencies is given by
\begin{align}
\hat G_{n'} &= G(\iv_n) \nonumber \\
&= \int_0^\beta d \tau e^{\ii \nu_{n'+N/2} \tau} G(\tau) \\
&\approx \frac{\beta}{2^R} \sum_{m=0}^{2^R-1} e^{\ii \pi \frac{2n'm}{2^R}} e^{\ii\pi\frac{-2^R+\xi}{2^R} m} G_m,
\end{align}
where $\tau_m = \frac{\beta}{2^R} m$ ($m=0, 1, \cdots, 2^R-1$),
$n'= n + 2^{R-1}~(= 0, 1, \cdots, 2^R-1)$.
By using $m=(m_1 m_2 \cdots m_R)_2$ and $\theta \equiv \pi\frac{-2^R+\xi}{2^R}$, we obtain {an MPO} for the transform:
\begin{align}
\frac{\beta}{2^{2R}} \mathcal{F}^{-1} \mathcal{P},
\end{align}
where the phase-rotation layer $\mathcal{P}$ is given by
\begin{align}
\mathcal{P} &= \mathrm{U1}(2^{R-1}\theta) \cdot \mathrm{U1}(2^{R-2}\theta) \cdot (\cdots) \cdot \mathrm{U1}(\theta)
\end{align}
with the single-qubit rotation
\begin{align}
\mathrm{U1}(\alpha) &\equiv
\begin{pmatrix}
1  &  0 \\
0  &  e^{\ii\alpha}
\end{pmatrix}.
\end{align}
The Fourier transform $\mathcal{F}$ is defined in Eq.~\eqref{eq:fft}.

\bibliography{main,nogaki,noneq,shinaoka}

\end{document}